\documentclass[epj]{svjour}
\usepackage{latexsym}
\usepackage{graphics}
\usepackage{amssymb}
\usepackage{graphicx, color}
\usepackage{amsmath}
\usepackage{mathrsfs}
\usepackage{bm}
\usepackage{ulem}

\begin{document}

\title{Probing exotic phenomena at the interface of nuclear and particle physics with the electric dipole moments of diamagnetic atoms:
A unique window to hadronic and semi-leptonic CP violation}
\author{N. Yamanaka\inst{1,2}
\and B. K. Sahoo\inst{3} 
\and N. Yoshinaga\inst{4}
\and T. Sato\inst{5} 
\and K. Asahi\inst{5,6} 
\and B. P. Das\inst{6}\thanks{\emph{Email address:} bpdas.iia@gmail.com} }         

\offprints{}       

\institute{iTHES Research Group, RIKEN, Wako, Saitama 351-0198, Japan 
\and Complex Simulation Group, School of Biomedicine, Far Eastern Federal University, Vladivostok, 690950 Russia
\and Atomic, Molecular and Optical Physics Division, Physical Research Laboratory, Ahmedabad-380009, India 
\and Graduate School of Science and Engineering, 255 Shimo-Okubo, Sakura-ku, Saitama City, Saitama 338-8570, Japan 
\and Nishina Center, RIKEN, 2-1 Hirosawa, Wako-shi, Saitama, 351-0198 Japan 
\and Department of Physics and International Education and Research Center of Science, Tokyo Institute of Technology, 2-12-1 Ookayama Meguro-ku, Tokyo 152-8550, Japan
}

\date{Received: date / Revised version: date}

\abstract{
The current status of electric dipole moments of diamagnetic atoms which involves the synergy between atomic 
experiments and three different theoretical areas -- particle, nuclear and atomic is reviewed. 
Various models of particle physics that predict CP violation, which is necessary for the existence of such electric dipole moments, are presented. These include the standard model of particle physics and 
various extensions of it. 
Effective hadron level combined charge conjugation (C) and parity (P) symmetry violating interactions are derived taking into consideration different ways in which a nucleon interacts with other nucleons as well as with electrons. 
Nuclear structure calculations of the CP-odd nuclear Schiff moment are discussed using the shell model and other theoretical approaches. 
Results of the calculations of atomic electric dipole moments due to the interaction of the nuclear Schiff moment with the electrons and the P and time-reversal (T) symmetry violating tensor-pseudotensor electron-nucleus are elucidated using different relativistic many-body theories. 
The principles of the measurement of the electric dipole moments of diamagnetic atoms are outlined. 
Upper limits for the nuclear Schiff moment and tensor-pseudotensor coupling constant are obtained combining the results of atomic experiments and relativistic many-body theories. 
The coefficients for the different sources of CP violation have been estimated at the elementary particle level for all the diamagnetic atoms of current experimental interest and their implications for physics beyond the standard model is discussed. 
Possible improvements of the current results of the measurements as well as quantum chromodynamics, nuclear and atomic calculations are suggested.
}
\maketitle

\section{Introduction}
\label{intro}

The important predictions of the standard model (SM) of particle physics  \cite{SM1,SM2,SM3}  have been verified largely due to the remarkable
advances in accelerator technology \cite{higgsatlas,higgscms}. A number of ingenious high energy experiments are currently underway to 
search for new phenomena beyond the SM.  Many of these experiments are being  performed using the Large Hadron Collider (LHC) at the TeV 
scale. A complementary approach to search for new physics beyond the SM is characterized by non-accelerator low energy 
precision tests of fundamental physics. It involves measuring observables and comparing the experimental results with the predictions 
of the SM. This is an indirect approach to new physics beyond the SM, but the observation of rare or forbidden phenomena is an 
indubitable proof of the existence of a new theory. Although conclusions reached by such an approach may in some case not be as specific 
in identifying the underlying fundamental theory as in the direct high energy physics approach, its sensitivity to new physics may well 
exceed the energy of collider experiments.

The combined charge conjugation (C) and parity (P) symmetry (CP) violation is considered to have relevance to the huge discrepancy from the SM prediction which is observed 
in the matter-antimatter asymmetry of the Universe \cite{sakharov}, and is currently an issue of primary importance 
in elementary particle physics \cite{khriplovich,bigibook,roberts}. CP violation has been studied in various physical systems,
but has so far been observed only in the $K$ \cite{christenson} and $B$ mesons \cite{abe,aubert,aaij,alvarez}, 
in which cases the  experiments are in agreement with predictions of the SM. 
In the SM, it arises from the complex phase of the Cabibbo-Kobayashi-Maskawa (CKM) matrix \cite{K-M,Jarlskog}.
It is well known that this phase cannot generate excess of matter over antimatter in the early Universe  \cite{farrar,huet,dine}.
It is therefore imperative to find one or several new sources of CP violation beyond the SM. A variety of studies on CP violation including experiments to observe the electric dipole moments (EDMs) of different systems have lent themselves to searches for new physics beyond the SM \cite{purcell,hgedm1987,rosenberry,regan,baker,griffith,hudson,baron,bishof,graner,Pendlebury}.

A non-degenerate physical system can possess a permanent EDM due to violations of P and time-reversal (T) symmetries 
\cite{ramsey1,fortson}. T violation implies the CP violation as a consequence of the CPT theorem \cite{luders}. An atom 
could possess an EDM due to the possible existence of (i) the electron EDM ($d_e$) (ii) P and T violating (P,T-odd) electron-nucleus 
interactions and (iii) the hadronic CP violation. EDMs of open shell (paramagnetic) atoms arise primarily due $d_e$ and the P,T-odd 
electron-nucleus scalar-pseudoscalar (S-PS interaction,  but the dominant contributions to the EDMs of closed-shell (or diamagnetic) atoms 
come from the hadronic CP violation and the electron-nucleus tensor-pseudotensor (T-PT) interaction. Atomic EDMs are sensitive to new physics 
beyond the standard model (BSM) and can probe CP violating interactions corresponding to mass scales of tens of TeV or 
larger \cite{bernreuther,barr,pospelovreview,ramsey}. The results of atomic EDM experiments and theory currently constrain various extensions 
of the SM. 
Experiments are underway to improve the limits of EDMs in paramagnetic (open-shell) \cite{weiss,heinzen,harada} and diamagnetic (closed-shell)
atoms \cite{furukawa-xe,inoue-xe,rand,tardiff,fierlinger,schmidt,yoshimi}. Their results in combination with state of the art 
theoretical calculations can extract various CP violating coupling constants 
at the elementary particle level via the hadronic, nuclear and atomic theories \cite{khriplovich,pospelovreview,heedmreview,ginges,Fukuyama,dzubareview,engel}.

It is necessary at this stage to emphasize the importance of the study of EDMs of the diamagnetic atoms. 
Many low energy observables used in the precision tests of fundamental physics, including EDMs of the paramagnetic atoms, are sensitive to
limited sectors (e.g. leptonic, hadronic, Higgs, etc) of a particular particle physics model. However, the EDMs of diamagnetic atoms arise 
from new physics in multiple sectors of a variety of extensions of the SM, since the hadronic sector opens up many possible scenarios for 
CP violation at the elementary level (quark EDMs, quark chromo-EDMs, gluon chromo-EDMs, quark-quark (q-q) interactions, 
etc.). This means that one experimental constraint cannot in principle determine the unknown coupling constants of the models.
Unraveling new physics beyond the SM in the context of EDMs of diamagnetic atoms is equivalent to  finding  the values for the couplings of 
new interactions that are solutions of a set of coupled equations obtained from experiments on atomic EDMs. The number of systems for EDM 
experiments must be at least equal to the number of coupling constants in order to uniquely determine those constants; assuming that 
uncertainties associated in all the results are of similar order. It is therefore desirable 
to perform EDM experiments on a number of different diamagnetic atoms.

\begin{figure*}
\begin{center}
\resizebox{16.0cm}{!}{\includegraphics{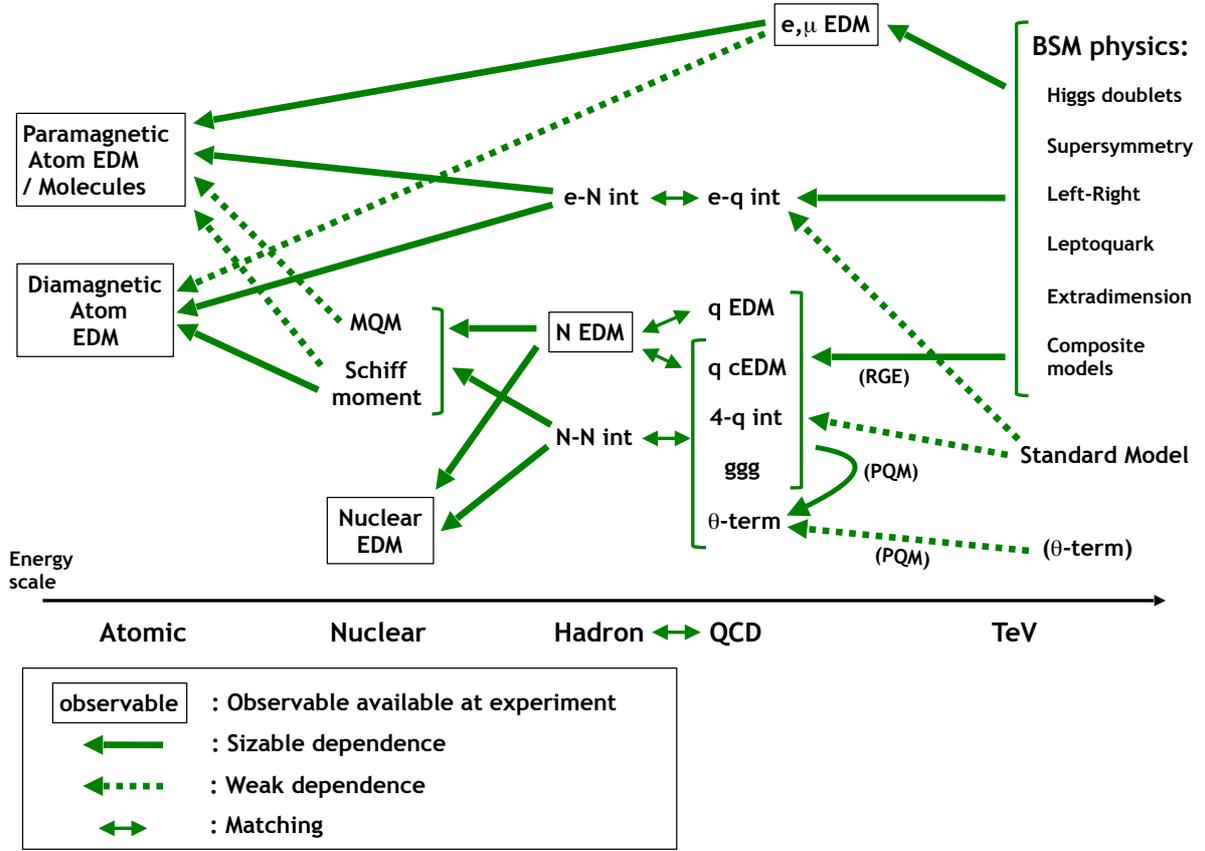}}
\caption{\label{fig:flow_diagram} 
Flow diagram of the dependence of the elementary level P,CP-odd processes on the EDMs of composite systems, whose EDMs can be measurable.
``RGE" means renormalization group evolution and ``PQM" means Peccei-Quinn mechanism.
} 
\end{center}
\end{figure*}

The experimental limit on the EDM of mercury atom ($^{199}$Hg) has improved several times since the first measurement in 1987 \cite{hgedm1987}, and it is
currently the lowest limit reported for the EDM of any system ($d_{\rm Hg}< 7.4 \times 10^{-30}e$ cm) \cite{graner}.
Improvements are expected in the EDM measurements of other diamagnetic systems such as the Xe and Ra in the near future.
However, since the EDMs of the diamagnetic atoms depend on many fundamental sectors, considerable theoretical effort has to be put in relating these EDMs to new physics beyond the SM (see Fig. \ref{fig:flow_diagram}).
In particular, the atomic and nuclear level many-body physics as well as the nonperturbative effects of quantum chromodynamics (QCD) contribute to the theoretical
uncertainties in the determination of their sensitivity to fundamental theories. Recent advances in the atomic and nuclear many-body as well as QCD calculations using numerical methods have reduced these uncertainties, but further progress  is necessary in this direction.

The focus of this review article is the recent advances in the EDMs of the diamagnetic atoms which arise predominantly from the nuclear 
Schiff moment (NSM) \cite{schiff} 
and CP violating electron-nuclear interaction. 
The former arises from CP violating nucleon-nucleon (N-N) interactions and EDMs of nucleons, which in turn originate from
CP violating quark level.
The latter is fundamentally due to the CP violating electron-quark (e-q) interactions.
We shall summarize our current understanding of physics beyond the SM that has been obtained by combining the results of experiment as well 
as atomic theory, nuclear theory and QCD relevant in the evaluation of the EDMs of diamagnetic atoms.
The theoretical uncertainty in the determination of these EDMs is the combined uncertainties resulting from the calculations in these 
three different theories.
It is therefore important to identify the large sources of errors in extracting the CP violating couplings at the particle 
physics level from the EDM experimental data.

The article is organized in the following manner:
Sec. \ref{sec:particlephysics} covers CP violations at the particle physics level that are suitable for the kind of atomic 
EDM that is considered in this review. The derivation of hadron level effective CP-odd interactions are then presented in Sec. 
\ref{sec:hadron_physics}. Sec. \ref{Sec:Nuclear structure calculation} deals with the NSM and the nuclear structure issues involved in its 
calculation. Different features of relativistic many-body theories which are necessary to calculate the EDMs of diamagnetic atoms are 
presented in Sec. \ref{sec:1}. An introduction to the principles of the measurement of EDMs of diamagnetic atoms and the current status of 
the search for EDMs of these atoms are given in Sec. \ref{sec:experiment}.
We summarize the effect of CP-odd interactions at the particle physics level on the EDMs of diamagnetic atoms in Sec. \ref{sec:discussion},
and analyze the candidates for BSM physics which can be constrained. Finally, our concluding remarks regarding the search 
for the EDMs of diamagnetic atoms are made in Sec. \ref{sec:summary}.

\section{Sources of P and CP violations in particle physics\label{sec:particlephysics}}

In this section, we describe the physics of CP violation at the level of elementary particle physics.
First, we present the relevant CP violating operators, and then show that the SM contribution to them is small.
We then briefly review several motivated candidates of new theories beyond SM.
We also introduce the Peccei-Quinn mechanism which is almost mandatory to resolve the problem of too large QCD $\theta$-term.
We finally see the procedure to renormalize the CP-odd operators from the elementary physics to the hadronic scale,
to pass on to the hadron level analysis.

\subsection{\label{sec:quark-gluon_cpv}CP violating operators after integration of heavy degrees of freedom}

After integrating out heavy new physics particles of BSM, the Higgs boson and massive electroweak gauge bosons (and eventually the top quark), 
we are left with an infinite number of operators which form the quark and gluon
level effective interactions. As the coupling constants of those interactions are suppressed by the power of the energy scale of new physics,
operators with the lowest mass dimension are important in the physics of strong interaction.
Here we list the CP violating operators generated at the elementary level up to mass dimension six, which are relevant in 
the physics of the EDM of atoms:
\begin{itemize}
\item
$\theta$-term:
\begin{equation}
{\cal L}_{\theta} = \frac{g_s^2}{64 \pi^2} \bar \theta \, \epsilon^{\mu \nu \rho \sigma} G_{\mu \nu}^a G_{\rho \sigma}^a 
.
\label{eq:theta-term}
\end{equation}

\item
Fermion EDM:
\begin{equation}
{\cal L}_{\rm EDM} = -\frac{i}{2} d_f \, \bar \psi \sigma_{\mu \nu} \gamma_5 \psi F^{\mu \nu}
,
\end{equation}
where $\psi$ denotes the electron or the quark and also it follows $\bar \psi = \gamma_o \psi^{\dagger}$.

\item
quark chromo-EDM:
\begin{equation}
{\cal L}_{\rm cEDM} = -\frac{i}{2} d^c_q \, g_s  \bar \psi_q \sigma_{\mu \nu} t_a \gamma_5 \psi_q G_a^{\mu \nu}
,
\end{equation}
where $\psi_q$ is the field operator of the quark $q$. 

\item
Weinberg operator:
\begin{equation}
{\cal L}_w = \frac{1}{6} w 
f^{abc} \epsilon^{\alpha \beta \gamma \delta} G^a_{\mu \alpha } G_{\beta \gamma}^b G_{\delta}^{\mu,c}
,
\label{eq:weinberg_operator}
\end{equation}
where $f^{abc}$ is the $SU$(3) structure constant of the Lie algebra. 

\item
P, CP-odd or equivalently P,T-odd 4-quark interactions:
\begin{eqnarray}
{\cal L}_{\rm 4q}
&=&
\sum_{q} 
\Bigl[
C^q_4 \bar q q \, \bar q i\gamma_5 q + C^q_5 \bar q \sigma^{\mu \nu} q \, \bar q i \sigma_{\mu \nu} \gamma_5 q
\Bigr]
\nonumber\\
+ &&
\frac{1}{2}
\sum_{q\neq q'} 
\Bigl[
2 \tilde  C_1^{q'q} \bar q' q' \, \bar q i\gamma_5 q
+2 \tilde C_2^{q'q} \bar q'_\alpha q'_\beta \, \bar q_\beta i\gamma_5 q_\alpha \nonumber \\ + &&
 \tilde C_3^{q'q} \bar q' \sigma^{\mu \nu} q' \, \bar q i \sigma_{\mu \nu} \gamma_5 q
+  \tilde C_4^{q'q} \bar q'_\alpha \sigma^{\mu \nu} q'_\beta \, \bar q_\beta i \sigma_{\mu \nu} \gamma_5 q_\alpha
\Bigr] ,
\ \ \ 
\label{eq:p,cp-odd_4-quark_interaction}
\end{eqnarray}
where the color indices $\alpha$ and $\beta$ were explicitly written when the color contraction is not taken in the same fermion bilinear.

\item
P, CP-odd or equivalently P,T-odd e-q interactions:
\begin{eqnarray}
{\cal L}_{eq} &=& 
-\frac{G_F}{\sqrt{2}} \sum_{q} [
C_{eq}^{\rm SP} \bar qq \, \bar e i \gamma_5 e
+C_{eq}^{\rm PS} \bar q i\gamma_5 q \, \bar e e \nonumber \\ &&
+\frac{1}{2} C_{eq}^{\rm T} \epsilon^{\mu \nu \rho \sigma} \bar q \sigma_{\mu \nu} q \, \bar e \sigma_{\rho \sigma} e
 ] \,
,
\label{eq:p,cp-odd_e-q_interaction}
\end{eqnarray}

\end{itemize}
where superscripts SP, PS, and T denote the scalar-pseudoscalar (S-PS), pseudoscalar-scalar (PS-S), and T-PT e-q interactions, respectively.

We must note that these effective interactions are defined at some energy scale.
In perturbative evaluations, they are usually given at the energy scale where the new particle BSM is integrated out (typically at the
TeV scale).

\subsection{The SM contribution\label{sec:smcpv}}

Let us start with the SM contribution to the elementary level CP violation \cite{SM1,SM2,SM3}.
Apart from the strong $\theta$ term, CP-violation comes from the Kobayashi-Maskawa phase \cite{K-M} in the form of Jarlskog invariant \cite{Jarlskog}. 
The standard form of Cabibbo-Kobayashi-Maskawa (CKM) matrix is given by
\small
\begin{eqnarray}
V
\equiv
 \left(
  \begin{array}{ccc}
   c_{12} c_{13}
   & s_{12} c_{13}
   & s_{13} e^{-i\delta}\\
   - s_{12} c_{23} - c_{12} s_{23} s_{13} e^{i\delta}
   & c_{12} c_{23} - s_{12} s_{23} s_{13} e^{i\delta}
   & s_{23} c_{13}\\
   s_{12} s_{23} - c_{12} c_{23} s_{13} e^{i\delta}
   & -c_{12} s_{23} - s_{12} c_{23} s_{13} e^{i\delta}
   & c_{23} c_{13}
  \end{array}
 \right) 
,
\nonumber
\end{eqnarray}
and the Jarlskog invariant is
\begin{eqnarray}
J_{CP}
\equiv
 \left|
  \Im( V_{\alpha j} V_{\beta j}^\ast V_{\alpha k}^\ast V_{\beta k} )
 \right|
=
 s_{12} s_{23} s_{13} c_{12} c_{23} c_{13}^2 \sin\delta.
\label{Jarlskog}
\end{eqnarray}
Here $\Im$ implies an imaginary part and $s_{ij}=\sin\theta_{ij}$ and $c_{ij}=\cos\theta_{ij}$. 
This combination of CKM matrix elements is the minimal requirement to generate CP violation.

\begin{figure}[h]
\begin{center}
\resizebox{6.0cm}{!}{\includegraphics{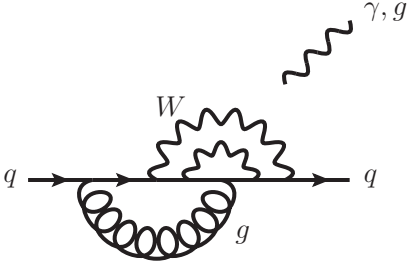}}
\caption{\label{fig:dEDM_SM3loop} 
Example of a diagram contributing to the EDM (chromo-EDM) of light quark at the three-loop level in the SM \protect\cite{czarnecki}. 
The external electromagnetic (or color) field, denoted by the isolated wavy line, is to be inserted in all possible propagators of 
electrically charged (colored) particles.
} 
\end{center}
\end{figure}

The CP violation in the SM therefore requires at least two $W$ boson exchanges.
For the quark EDM and the chromo-EDM, the two-loop level contribution is also known to vanish due to the GIM mechanism \cite{GIM,donoghuesmedm,shabalin1,shabalin2}, 
and the leading order one is given by at the three-loop level \cite{czarnecki} (see Fig. \ref{fig:dEDM_SM3loop}).
Their effect on the nucleon EDM is around $d_N \sim 10^{-35} e$ cm, much smaller than the present
experimental limit of that of the neutron ($d_n < 10^{-26} e$ cm) \cite{baker,Pendlebury}.

The EDM of the electron is also generated by the CP phase of the CKM matrix.
This effect starts from the four-loop level, and its value is $d_e \sim 10^{-44} e$ cm \cite{pospelovsmelectronedm,booth,pospelovsmatomicedm}.
We must note that the effect of the CP phase of the neutrino mixing matrix is negligible due to the small neutrino
mass. If the neutrinos are Majorana fermions the effect of additional CP phases can generate the electron EDM from the two-loop level, and a
larger value will be allowed for $d_e$ \cite{Archambault,hemfv1,hemfv2,novales}.

Purely gluonic CP-odd processes such as the $\theta$-term or the Weinberg operator are also known to be very small. The $\theta$-term generated
by the CKM phase is $\bar \theta \sim 10^{-17}$ \cite{ellissmtheta,khriplovichtheta}, which yields a nucleon EDM of $|d_N| \sim 10^{-33} e$ cm.
The Weinberg operator gives an even smaller nucleon EDM, of order $10^{-40}e$ cm \cite{smweinbergop}.

\begin{figure}[h]
\begin{center}
\resizebox{8.0cm}{!}{\includegraphics{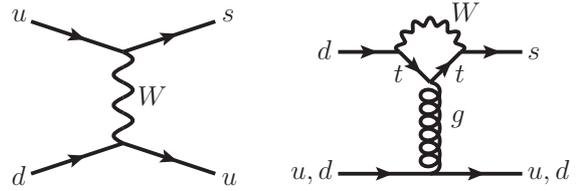}}
\caption{\label{fig:tree_penguin} 
Tree level $|\Delta S|=1$ $W$ boson exchange diagram (left) and the penguin diagram (right).
} 
\end{center}
\end{figure}

In the strongly interacting sector, the most widely accepted leading hadronic CP violation due to the CP phase of the CKM matrix is generated by the long distance effect.
The long distance contribution of the CKM phase arises from the interference between the tree level strangeness violating $|\Delta S|=1$ $W$ boson exchange process and the penguin diagram (see Fig. \ref{fig:tree_penguin}), which forms the Jarlskog invariant (\ref{Jarlskog}).
From a naive dimensional analysis, the nucleon and nuclear EDMs are estimated as
$d \sim O(\frac{\alpha_s}{4\pi} G_F^2 J \Lambda_{\rm QCD}^3 ) \sim 10^{-32}e$ cm, which is larger than the contribution from the short distance
processes (quark EDM, chromo-EDM, Weinberg operator, etc). Previous calculations of the nucleon EDM are in good
agreement with this estimations \cite{ellissmedm,nanopoulossmedm,Deshpande,gavelasmedm,smneutronedmkhriplovich,eeg,hamzaouismedm,smneutronedmmckellar,mannel,seng}.

The CP violating effects in the SM exhibit an EDM well smaller than the experimental detectability, and a large room
is left for the discovery of new source of CP violation BSM.

\subsection{Sources of CP violation from BSM physics}

\begin{figure*}
\begin{center}
\resizebox{14.0cm}{!}{\includegraphics{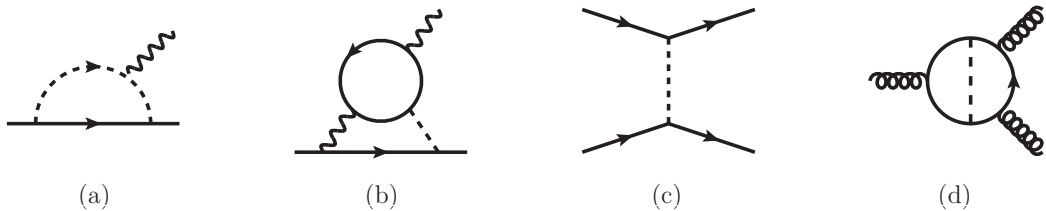}}
\caption{\label{fig:elementary_cpv}
Leading contribution of the new physics of BSM to the effective interaction at the TeV scale.
(a) One-loop level fermion EDM, (b) Barr-Zee type diagram, (c) CP-odd four-fermion interaction, (d) Weinberg operator.
The external wavy lines of (a) and (b) are either a photon or a gluon, and the internal one of (b) is either a photon, weak gauge bosons,
or gluon.
} 
\end{center}
\end{figure*}

In many scenarios of BSM, large EDMs are predicted, because of higher order contributions that can arise 
at the one- or two-loop levels. These contributions are overwhelmingly exceed over the loop suppressed SM contribution. 
In Fig. \ref{fig:elementary_cpv}, we present the typical lowest order CP violating processes of BSM contributing to the EDMs at  
the elementary level. In this subsection, we would like to elaborate several such well motivated candidates of BSM which can generate 
EDMs.

\subsubsection{Higgs doublet models\label{sec:2HDM}}

The Higgs boson was recently discovered \cite{higgsatlas,higgscms}, but the detailed Higgs potential is still
unknown. There are currently many well-motivated extensions of the Higgs sector BSM.
The most well-known one is the two-Higgs doublet model (2HDM), and extensive studies have been performed 
\cite{barr-zee,weinbergop,dicus,chang2hdm1,leigh,chang2hdm2,mahanta,kao2hdm,barger,bowser-chao,buras2hdm,brod,sinoue2hdm,cheung2hdm,jung2hdmedm,tabe,bian,chenedm,eeghiggs}.

As the Higgs boson has a small coupling with light fermions, the one-loop level fermion EDM and the CP-odd four-fermion interactions are suppressed in 2HDM \cite{barrmasiero,e-nint}.
The leading contribution to the elementary level CP violation contributing to the EDM is the two-loop level Barr-Zee type diagram \cite{barr-zee} [Fig. \ref{fig:elementary_cpv} (b)], enhanced by the large Yukawa coupling of the top quark of the inner loop.
The Barr-Zee type diagram contribution to the EDM of SM fermion can be written as \cite{barr-zee}
\begin{equation}
d_f
=
\frac{Q_f e \alpha_{\rm em}}{48 \pi^3 m_t} 
\left[
(Y_f^{(+)} Y_t^{(+)} ) (f+g)
-(Y_f^{(-)} Y_t^{(-)} ) (f-g)
\right]
,
\label{eq:barr-zeefermionedm}
\end{equation}
where
\begin{eqnarray}
f
&\equiv &
\frac{m_t^2}{2m_H^2} \int_0^1 dx \frac{1-2 x (1-x) }{ x(1-x) -m_t^2 / m_H^2}
\ln \frac{x(1-x)}{m_t^2 / m_H^2}
,
\\
g
&\equiv &
\frac{m_t^2}{2m_H^2} \int_0^1 dx \frac{1 }{ x(1-x) -m_t^2 / m_H^2}
\ln \frac{x(1-x)}{m_t^2 / m_H^2}
,
\end{eqnarray}
and $Y_f^{(\pm )}$ and $Y_t^{(\pm )}$ are the Yukawa couplings relating the lightest Higgs boson ($m_H = 125$ GeV) with the fermion $f = e, u ,d$ and the top quark, respectively.
The first (second) term of Eq. (\ref{eq:barr-zeefermionedm})
is generated 
by the vacuum expectation value $\langle \phi_i^0 \phi_1^{0*} \rangle$ ($\langle \phi_i^0 \phi_1^{0} \rangle$), where $\phi_i$
is the Higgs doublet interacting with the up-type quark ($i=1$) or the down-type fermion ($i=2$, for the electron or
down-type quarks).
These vacuum expectation values strongly depend on the Higgs potential.
Those couplings are obtained from the diagonalization of the Higgs doublets.

In diamagnetic atoms, the most important CP violating process is the quark chromo-EDM:
\begin{equation}
d^c_q
=
\frac{g_s \alpha_s}{32 \pi^3 m_t} 
\left[
(Y_q^{(+)} Y_t^{(+)} ) (f+g)
-(Y_q^{(-)} Y_t^{(-)} ) (f-g)
\right]
.
\label{eq:barr-zeequarkchromoedm}
\end{equation}
With $Y_q^{(\pm)} \sim 10^{-6}$, we have $d^c_f \sim Y_t^{(+)} \times 10^{-25}$ cm.
We must note that the Weinberg operator (\ref{eq:weinberg_operator}) is also generated in the 2HDM [see Fig. \ref{fig:elementary_cpv} (d)] \cite{weinbergop,dicus}.
Its typical value is $w \sim 10^{-10}$ GeV$^{-2}$, with $m_H = 125$ GeV.
We will see in later sections that this contribution is subleading for the nucleon EDM.

\subsubsection{Supersymmetric (SUSY) models\label{sec:susy}}

As the next attractive model for BSM physics, we have the minimal supersymmetric standard model (MSSM) \cite{haber,gunionmssm,martinmssm}.
The MSSM contains several phenomenological interactions which generically possess CP phases.
In the most simplified parametrization, the Higgs bilinear $\mu$-term
\begin{equation}
{\cal L}_\mu =  
e^{i \theta_\mu} \mu^2 H_u \cdot H_d 
,
\label{eq:mu-term}
\end{equation}
from the superpotential, which is required to give mass to higgsinos, 
and the supersymmetry breaking sfermion trilinear interactions 
\begin{eqnarray}
{\cal L}_A 
&=& 
e^{i \theta_A} A_u \tilde u_R H_u \cdot \tilde Q_L
+e^{i \theta_A} A_d \tilde d_R H_d \cdot \tilde Q_L
\nonumber\\
&&+e^{i \theta_A} A_e \tilde e_R H_d \cdot \tilde L_L
+{\rm h.c.}
,
\end{eqnarray}
are CP violating.
Here the dot denotes the $SU(2)_L$ inner product.
For the sfermion trilinear interactions, we often assume a flavor diagonal one, with a common CP phase $\theta_A$.
This assumption is due to the strong constraints on flavor changing neutral current from phenomenology \cite{ramsey,gabbiani}.

Under this MSSM Lagrangian, the fermion EDM appears at the one-loop level \cite{barrmasiero,ellismssmedm,buchmullermssmedm,polchinski,aguilamssmedm,nanopoulosmssmedm,duganmssmedm,nathmssmedm,kizukurimssmedm1,kizukurimssmedm2,inuimssmedm,mssmreloaded,bingli} 
[see Fig. \ref{fig:elementary_cpv} (a)].
The electron EDM and the quark EDM, in the simplified parametrization of MSSM where masses of all the supersymmetric particles as well as $|\mu|$ are given by $M_{\rm SUSY}$, are given by \cite{pospelovreview}
\begin{eqnarray}
d_e
&\approx &
\frac{e m_f}{16 \pi^2 M_{\rm SUSY}^2} 
\left( 
\frac{5 g_2^2 + g_1^2}{24} \sin \theta_\mu \tan \beta 
+\frac{g_1^2}{12} \sin \theta_A 
\right)
, \ \ \ \ 
\\
d_q
&\approx &
 \frac{Q_q e m_f}{16 \pi^2 M_{\rm SUSY}^2} \frac{2 g_s^2}{9} \left( \sin \theta_\mu [ \tan \beta ]^{-2 Q_q + \frac{1}{3}} - \sin \theta_A \right)
,
\label{eq:susy_qedm}
\end{eqnarray}
 respectively, with $Q_q$ is the electric charge of the quark $q$, and $g_1$, $g_2$ and $g_s$ are the couplings of the
$U(1)_Y$ and $SU(2)_L$ gauge theories and QCD, respectively. The quark chromo-EDM is similarly given by
\begin{eqnarray}
d^c_q
&\approx &
 \frac{ g_s m_f}{16 \pi^2 M_{\rm SUSY}^2} \frac{5 g_s^2}{18} \left( \sin \theta_\mu [ \tan \beta ]^{-2 Q_q + \frac{1}{3}} - \sin \theta_A \right)
, \ \ \ \ 
\label{eq:susy_cedm}
\end{eqnarray}
where $\tan \beta \equiv \frac{v_d}{v_u}$ is the ratio between the vacuum expectation values of the up-type and down-type Higgs fields.
As for the Higgs doublet models, we also see here a dependence on $\tan \beta$.
By assuming $M_{\rm SUSY} = O({\rm TeV})$ and $\theta_\mu, \theta_A , \beta =O(1)$, the MSSM contribution to the EDMs of the fermions and 
the chromo-EDMs of quarks at the scale $\mu = 1$ TeV [$\alpha_s (\mu = 1\, {\rm TeV}) \approx 0.09$] become $d_e = O(10^{-27})e$ cm,
$d_q = O(10^{-25})e$ cm and $d^c_f = O(10^{-25})$ cm\footnote{Unfortunately such models, except for the few predictive
ones, have so many undetermined parameters and we should be careful about under what assumptions such and such predictions have been made.
On this point one can refer to Ref. \cite{F-A} for more detailed clarification.}. 

To conceive natural scenarios in MSSM, it is often assumed that the first and the second generations have no sfermion trilinear interactions.
In such a case, the leading order CP violation are given by the two-loop level effect, namely 
the Barr-Zee type diagrams [Fig. \ref{fig:elementary_cpv} (b)] \cite{west,kadoyoshi,chang,pilaftsis1,chang2,mssmrainbow1,demir,feng1,feng2,lihiggshiggsino,mssmrainbow2,carena,nakai} and the Weinberg operator [Fig. \ref{fig:elementary_cpv} (d)] \cite{daimssm,arnowitt1,arnowitt2}.
We must note that the Barr-Zee type diagram and the four-fermion interaction are enhanced when $\tan \beta$ is large \cite{demir,fischler,falk,lebedevmssm,pilaftsishiggsmssm}.
Global analyses with constrained supersymmetric parameters by the Grand unification theory (GUT) strongly constrain CP
phases \cite{mssmunify1,mssmunify2,mssmunify3,mssmunify4,mssmunify5,mssmunify6}.

Another natural supersymmetric scenario is the split SUSY model \cite{splitsusy1,splitsusy2}, relying on the GUT.
In this case, the sfermions are much heavier than the gauginos, and one-loop level diagrams, which must contain sfermions, are suppressed.
The Barr-Zee type diagram with chargino inner loop therefore becomes dominant \cite{splitsusy3,splitsusy4,dhuria,sarellis}.

The SUSY model can be extended with additional interactions, with several motivations. The first 
possibility is to take into account additional soft supersymmetry breaking terms, in particular the flavor violating ones which are not 
forbidden by any symmetries or by other experimental constraints. The flavor non-diagonal soft breaking terms can generically have CP phases.
This extension was motivated by the deviation of the CP violating $B \to \phi K_s$ decay \cite{grossman,barbieri} suggested by Belle experiment \cite{belle}. 
The effects of those flavor violating terms on the EDM are however large, and it was found that the EDM experimental data can strongly constrain their CP phases \cite{hisanoshimizu1,hisanoshimizu2,endomssmflavor,chomssmflavor,hisanonagai1,hisanonagai2,hisanonagai3,Altmannshofer}.

Another possible way to extend the MSSM is to add new interactions in the superpotential.
The scenario on these lines is the next-to-minimal supersymmetric standard model (NMSSM) which considers an additional scalar superfield in the Higgs sector \cite{nmssm}.
This model can dynamically generate the $\mu$-term (\ref{eq:mu-term}) and circumvent the problem of $\mu$-term.
It is also motivated by the difficulty to explain the appearance of the light Higgs boson in the simple parametrization of the MSSM.
 In the NMSSM, the EDMs of fermions do not become large \cite{nmssmedm}. If we further enlarge the superpotential by adding
new local gauged terms (BLMSSM) \cite{blmssm1,blmssm2}, the fermion EDMs can become large, and the CP phases will be
strongly constrained by the current experimental data \cite{blmssmedm}.
The EDM is even more enhanced if we also allow the R-parity violation, where baryon and lepton numbers are not conserved \cite{rpvreview1,rpvreview2,rpvreview3,rpvreview4}.
If we neglect the one-loop level fermion EDM which is only generated in the presence of soft breaking bilinear R-parity violating interaction \cite{keum1,keum2,choirpv,chiourpv}, the leading CP violation processes are the Barr-Zee type diagram \cite{godbole,abelrpv,changrpv,rpvedm1,rpvedm2,rpvedm3,rpvedm4} and the CP-odd four-fermion interaction \cite{rpvedm1,rpv4f1,faessler1,faessler2}.
The majority of CP phases of the R-parity violating couplings are strongly constrained by the current EDM experimental data.

 Obviously, the SUSY extensions allow larger observable EDMs as the number of parameters increases.
This fact does not depend on whether we have extended the superpotential or the soft supersymmetry breaking interaction.
The supersymmetric SM is an excellent example of new physics which contributes to the EDM of composite systems through various elementary level CP-odd operators.
Current EDM experimental data strongly constrain the CP phases of models with large degree of freedom.
In the analysis of theories and models which have a large parameter space, it was often assumed that only a small numbers of couplings are active, and the effect of the others were neglected.
We however have to note that cancellations may occur among supersymmetric CP phases \cite{susycancel1,susycancel2,susycancel3,susycancel4,susycancel5,susycancel6,susycancel7,susycancel8,susycancel9,susycancel10,susycancel11,susycancel12,susycancel13,yamanakabook,linearprogramming}.
In that case, still large CP phases may be allowed, and they may be relevant in the ongoing EDM experiments.

\subsubsection{Left-Right symmetric models\label{sec:LRSM}}

The Left-Right symmetric models contain an additional gauge theory which couples to the right-handed fermions of SM \cite{left-right1,left-right2,left-right3}.
An $SU(3)_c \times SU(2)_L \times SU(2)_R \times U(1)_{B-L}$ gauge group is assumed to be spontaneously broken at some high energy scale, and gives the SM as an effective theory below it.
Phenomenologically, a mixing of $W$ boson with a heavier $W_R$ boson is possible.
The mass of additional weak gauge boson is constrained by LHC experiment, and the current lower bound is a few TeV \cite{LRLHC1,CMSW'1,ATLASW',CMSW'2,CMSW'3}.

In low energy effective theory, we obtain a 4-quark interaction with the structure $(V-A) \times (V+A)$:
\begin{eqnarray}
{\cal L}_{\rm LR} 
&=&
i {\rm Im} (\Xi )
\left[
\bar u_R \gamma_\mu d_R \cdot \bar d_L \gamma^\mu u_L
-\bar d_R \gamma_\mu u_R \cdot \bar u_L \gamma^\mu d_L
\right]
\nonumber\\
&=&
\frac{{\rm Im} (\Xi )}{12}
\bigl[
2 (\bar qq \cdot \bar q i\gamma_5 \tau_z q
-\bar q \tau_z q \cdot \bar q i\gamma_5 q )
\nonumber\\
&&\hspace{4em}+ 
3 (\bar q t_a q \cdot \bar q i\gamma_5 \tau_z t_a q
-\bar q t_a \tau_z q \cdot \bar q i\gamma_5 t_a q )
\bigr]
,
\ \ \ \ \ 
\label{eq:4-q_lagrangian}
\end{eqnarray}
where the coupling constant $\Xi$ scales as $O(m_{W_R}^{-2})$.
The terms in the last line are the color octet four-quark interaction, with $t_a$ the generator of the $SU(3)_c$ group.
If $\Xi$ has a CP phase, the EDM is induced in hadronic systems \cite{LRedm1,LRedm2,LRedm3,LRzhang1,LRzhang2,LRxu,eft6dim,LRLHC2,dekens}.
It is important to note that the above four-quark interaction breaks both the chiral and isospin symmetries \cite{eft6dim}.
This property is useful in estimating the leading CP-odd hadron level effective interaction generated by it \cite{sengisovector}.
Moreover, the effective interaction (\ref{eq:4-q_lagrangian}) is generated at the scale $\mu = m_W$, where the $W$ boson is integrated out.

\subsubsection{Models with vectorlike fermion}

The vectorlike fermions are spin $\frac{1}{2}$ particles which have the same gauge charges for their left- and right-handed components \cite{Aguilar}.
They are not constrained by the analysis of the Higgs boson in collider experiments, as it was for extensions with extra generations of chiral fermions \cite{djouadi}.
This class of models are attractive since those particles are often relevant in extensions of SM with composite sectors \cite{dimopouloscomposite,kaplancomposite,peterson} or extradimensions \cite{contino,hewette6}.

As a model independent feature, the vectorlike fermions may mix with SM fermions, but those processes are strongly constrained by the flavor changing neutral current \cite{Aguilar,vectorlikefcnc1,vectorlikefcnc2,vectorlikefcnc3,vectorlikefcnc4,vectorlikefcnc5,vectorlikefcnc6,vectorlikefcnc7,vectorlikefcnc8,vectorlikefcnc9,vectorlikefcnc10,vectorlikefcnc11}.
Regarding more model dependent aspects, additional dynamically generated bosons may accompany vectorlike fermions, such as the Higgs bosons, Kaluza-Klein particles, or higher energy resonances, and their interactions with SM fermions may generate EDM at the one-loop level.
This process is also strongly constrained by phenomenology \cite{vectorlikefcnc8,liaockm,Iltan,chang5d,agashe,nishiwaki,Kalinowski}.
Under those constraints, the vectorlike fermions may appear in the intermediate states connected only by the exchange of gauge bosons \cite{fan}.
The leading CP violating process is therefore the Weinberg operator \cite{choi}.

The contribution of the Weinberg operator in the vectorlike fermion models can be written as
\begin{equation}
w_{\rm VF}
=
-\sum_i^{N_F} \frac{g_s \alpha_s Y_i Y_{Pi} }{(4\pi)^3 M_i^2} h(M_i, m_{H'})
,
\label{eq:vectorlikeweinbergop}
\end{equation}
with $ h(M, m_{H'}) \equiv \frac{M^4}{2} \int_0^1 dx \int_0^1 du \frac{u^3 x^3 (1-x)}{[ M^2 x(1-ux) + m_{H'}^2 (1-u) (1-x) ]^2}$.
Here we have assumed a boson $H'$ which couples to  $N_F$ vectorlike fermions with mass $M_i$ ($i=1,\cdots ,N_F$).
In the limiting case $M \gg m_{H'}$, we have $h(M, m_{H'}) \approx \frac{1}{16}$ \cite{mssmreloaded}.
In technicolor theories, an effective $WW \gamma$ interaction is generated by a similar mechanism \cite{appelquist1}.

\subsubsection{Leptoquark models}

The leptoquarks are bosons which couple to both leptons and baryons, and often appears in scenarios with GUT. Those which violate the baryon number are strongly constrained by the proton decay, but those which conserve lepton
and baryon numbers are allowed up to the constraints from the LHC experiments \cite{leptoquarkcms1,leptoquarkcms2,leptoquarkcms3,leptoquarkcms4,leptoquarkatlas1,leptoquarkatlas2}, and their interaction can be probed using low energy precision tests \cite{davidson}.
The simplest interaction of the scalar leptoquark is given as
\begin{equation}
{\cal L}_{\rm LQ} 
=
\sum_{i,j} \varphi (\lambda_{ij} \bar Q_{L i} \cdot e_{Rj} +\lambda'_{ij} \bar u_{R i} \cdot L_{Lj})
+{\rm h.c.}
,
\end{equation}
where $\varphi$ is the leptoquark field, and the indices $i,j$ denote the flavor.

If the couplings $\lambda$ and $\lambda'$ have relative CP phases, the EDM will be induced in atomic systems. The leading CP violation is 
 given by the one-loop level fermion EDM \cite{barrmasiero,geng} and the CP-odd e-q interaction
[see Eq. (\ref{eq:p,cp-odd_e-q_interaction})] \cite{e-nint,hee-Nedm,herczegleptoquark}. For the atomic system, the latter is especially 
important, since it contributes to the tree level. The Leptoquark model is one of the rare models which
contribute to the T-PT CP-odd e-N interaction [the term with $C_{\rm T}$ in Eq. (\ref{eq:p,cp-odd_e-q_interaction})].

\subsection{Renormalization group evolution (RGE)}

In the usual discussion of particle physics, the effect of BSM physics is calculated at some high energy scale, much higher than that of the strong interaction $\mu \gg \Lambda_{\rm QCD} \sim 200$ MeV.
On the other hand, their matching with the hadronic effective interaction is done at the hadron scale, we must evolve the 
Wilson coefficients of elementary level interactions down to the hadronic scale.
In this Sub-sec., we first present the 
RGE of purely hadronic CP-odd operators, and then that of CP-odd e-q interactions, which do not mix 
with each other.

\subsubsection{RGE of strong CP-odd operators\label{sec:quarkRGE}}

The effective CP-odd Lagrangian and their Wilson coefficients are given as
\begin{eqnarray}
{\cal L}_{eff}
&=&
\sum_{i=1,2,4,5} \sum_q C_i^q (\mu) O_i^q (\mu)
+ C_3 (\mu) O_3 (\mu) \nonumber \\ &&
+ \sum_{i=1,2} \sum_{q\neq q'} C_i^{q' q} (\mu) \tilde O_i^{q'q} (\mu) \nonumber \\ &&
+ \frac{1}{2} \sum_{i=3,4} \sum_{q\neq q'} C_i^{q' q} (\mu) \tilde O_i^{q'q} (\mu)
,
\label{eq:wilsoncoefficients}
\end{eqnarray}
with 
\begin{eqnarray}
O_1^q 
&=&
-\frac{i}{2} m_q \bar q Q_q e \sigma_{\mu \nu} F^{\mu \nu} \gamma_5 q
,
\label{eq:o1q}
\\
O_2^q 
&=&
-\frac{i}{2} m_q \bar q g_s \sigma_{\mu \nu} G^{\mu \nu}_a t_a \gamma_5 q
,
\label{eq:o2q}
\\
O_3 
&=&
-\frac{1}{6} g_s f_{abc} G_{\mu \nu ,a} G^{\nu}_{\ \rho , b} G_{\alpha \beta , c} \epsilon^{\rho \mu \alpha \beta}
,
\label{eq:o3}
\\
O_4^q 
&=&
\bar q q \, \bar q i \gamma_5 q
,
\label{eq:o4}
\\
O_5^q 
&=&
\bar q \sigma^{\mu \nu} q \, \bar q i \sigma_{\mu \nu} \gamma_5 q
,
\label{eq:o5}
\\
\tilde O_1^{q' q} 
&=&
\bar q' q' \, \bar q i \gamma_5 q
,
\label{eq:o1'}
\\
\tilde O_2^{q' q} 
&=&
\bar q'_\alpha q'_\beta \, \bar q_\beta i \gamma_5 q_\alpha
,
\label{eq:o2'}
\\
\tilde O_3^{q' q} 
&=&
\bar q' \sigma^{\mu \nu} q' \, \bar q i \sigma_{\mu \nu} \gamma_5 q
,
\label{eq:o3'}
\\
\tilde O_4^{q' q} 
&=&
\bar q'_\alpha \sigma^{\mu \nu} q'_\beta \, \bar q_\beta i \sigma_{\mu \nu} \gamma_5 q_\alpha
,
\label{eq:o4'}
\end{eqnarray}
where the color indices $\alpha$ and $\beta$ were explicitly written when the color contraction is not taken in the same fermion bilinear.
The summation of the quark $q$ for the above operators must be taken for the relevant flavor at the renormalization scale chosen (e.g. $q=u,d,s$ for $\mu = 1$ GeV).

We note that the above CP-odd operators are defined in a scale where the Higgs boson, massive electroweak gauge bosons and the top quark are integrated out.
Their Wilson coefficients therefore also involve the BSM CP-odd effect related with those particles at a higher energy scale through the RGE \cite{brod,dekens1,brod2,cirigliano2,cirigliano3}.
In this section, we do not treat them explicitly, but consider their effects through the renormalized Wilson coefficients at the electroweak scale as the initial 
condition.

The evolution of the Wilson coefficients is dictated by the renormalization group equation, which mixes the CP-odd operators when the scale is changed.
It is given by the following differential equation
\begin{equation}
\frac{d}{d \ln \mu} {\mathbf C} ( \mu )
=
\hat \gamma^T (\alpha_s) {\mathbf C} ( \mu )
.
\label{eq:renor_eq}
\end{equation}
The anomalous dimension matrix is given by
\begin{equation}
\hat \gamma
=
\hat{Z}^{-1}
\frac{d}{d \ln \mu}
\hat{Z}
,
\end{equation}
with $\hat{Z}$ the renormalization matrix. 
By integrating (\ref{eq:renor_eq}) with the initial condition at the scale of new physics $\mu' = M_{\rm NP}$, 
we have
\begin{equation}
{\mathbf C} ( \mu )
=
\hat{U} ( \mu , \mu' = M_{\rm NP})
{\mathbf C} (  \mu' = M_{\rm NP})
,
\end{equation}
where
\begin{equation}
\hat{U} ( \mu , \mu' =  M_{\rm NP})
=
T_g
\exp
\int_{g(M_{\rm NP})}^{g(\mu)} dg' \frac{\hat \gamma^T (g')}{\beta (g')}
,
\end{equation}
with the strong coupling $g \equiv \sqrt{4\pi \alpha_s}$, and the coupling ordered product operator $T_g$.
The anomalous dimension matrix and the beta function $\beta (g)$ are expanded in terms of the QCD coupling as
\begin{eqnarray}
\hat \gamma (g)
&=&
\hat \gamma^{(0)} 
+ \hat \gamma^{(1)} 
+\cdots
,
\\
\beta (g)
&=&
-\beta_0 \frac{g^3}{16 \pi^2}
-\beta_1 \frac{g^5}{(16 \pi^2)^2}
+\cdots
.
\end{eqnarray}

Let us see the leading logarithmic order contribution.
The leading order coefficient of the beta function is $\beta_0 = \frac{11}{3}n_c - \frac{2}{3} n_f$
with the color number $n_c =3$.
The anomalous dimension matrix $\hat \gamma^{(0)}$, depending on $n_f$, is expressed in terms of submatrices as \cite{tensorrenormalization1,braaten1,boyd,braaten2,dineweinbergop,tensorrenormalization2,degrassi,yang,dekens1}
\begin{equation}
\hat \gamma^{(0)}
=
\left(
\begin{array}{ccc}
\frac{\alpha_s}{4 \pi} \hat \gamma_s & \hat 0 & \hat 0 \cr
\frac{\alpha_s}{(4 \pi )^2} \hat \gamma_{sf} & \frac{\alpha_s}{4 \pi} \hat \gamma_f & \hat 0 \cr
\frac{\alpha_s}{(4 \pi )^2} \hat \gamma'_{sf}  & \hat 0 & \frac{\alpha_s}{4 \pi} \hat \gamma_f \cr
\end{array}
\right)
,
\end{equation}
where $\hat 0$ is the null matrix with arbitrary dimension, and 
\begin{eqnarray}
\hat \gamma_s
&=&
\left(
\begin{array}{ccc}
8 C_F & 0 & 0 \cr
8 C_F & 16 C_F - 4 n_c & 0 \cr
0 & 2 n_c & n_c + 2 n_f + \beta_0 \cr
\end{array}
\right)
,
\\
\hat \gamma_f
&=&
\left(
\begin{array}{cc}
-12 C_F +6 & \frac{1}{n_c} -\frac{1}{2} \cr
\frac{48}{n_c} + 24 & 4 C_F +6  \cr
\end{array}
\right)
,
\\
\hat \gamma'_f
&=&
\left(
\begin{array}{cccccc}
-12 C_F & 0 & 0 & 0 & \frac{1}{n_c} & -1 \cr
-6 & \frac{6}{n_c} & 0 & 0 & -\frac{1}{2} & c_1 \cr
0 &0 & -12 C_F & 0 & \frac{1}{n_c} & -1 \cr
0& 0& -6 & \frac{6}{n_c} & -\frac{1}{2} & c_2 \cr
\frac{24}{n_c} & -24 & \frac{24}{n_c} & -24 & 4C_F & 0 \cr
-12 & c_3 & -12 & c_4 & 6 & c_5 \cr
\end{array}
\right) 
,
\\
\hat \gamma_{sf}
&=&
\left(
\begin{array}{ccc}
4 & 4 & 0 \cr
-32 n_c -16 & -16 & 0  \cr
\end{array}
\right)
,
\\
\hat \gamma'_{sf}
&=&
\left(
\begin{array}{ccc}
0 & 0 & 0 \cr
0 & 0 & 0 \cr
0 & 0 & 0 \cr
0 & 0 & 0 \cr
-16 n_c \frac{m_{q'}}{m_q} \frac{Q_{q'}}{Q_q} & 0 & 0  \cr
-16 \frac{m_{q'}}{m_q} \frac{Q_{q'}}{Q_q} & -16 \frac{m_{q'}}{m_q}  & 0  \cr
\end{array}
\right)
,
\end{eqnarray}
where $C_F = 4/3$, $c_1=-C_F + \frac{1}{2n_c}$, $c_2= -C_F + \frac{1}{2n_c}$, 
 $c_3=-24 C_F + \frac{12}{n_c}$, $c_4=-24 C_F + \frac{12}{n_c}$, and $c_5=-8C_F -\frac{6}{n_c}$. 

Let us show the results for three explicit cases with the initial condition $\mu = M_{\rm NP} = 1$ TeV.
For the quark EDM, there is no mixing with other operators.
If only the quark EDM is dominant at the initial scale, we have
\begin{equation}
\frac{d_q (\mu =  \mu_{\rm had}) }{ d_q (\mu = M_{\rm NP}) }
=
\frac{C_1^q (\mu =  \mu_{\rm had}) m_q (\mu =  \mu_{\rm had}) }{ C_1^q (\mu = M_{\rm NP}) m_q (\mu = M_{\rm NP})}
=
0.79
,
\label{eq:qedm_running}
\end{equation}
for $ \mu_{\rm had} = 1$ GeV.
The running of the quark mass is
\begin{equation}
m_q (\mu =  \mu_{\rm had}) / m_q (\mu = M_{\rm NP})
=
2.0
.
\end{equation}
We have used the quark masses $m_t (\mu = m_t ) = 160 \, {\rm GeV}$, $m_b (\mu = m_b) = 4.18$ GeV, and $m_c (\mu = m_c) = 1.27$ GeV as input \cite{pdg}.

If the quark chromo-EDM is dominant at $\mu = M_{\rm NP} $, the Wilson coefficients at the hadronic scale mixes with the quark EDM:
\begin{eqnarray}
\frac{d_q (\mu =  \mu_{\rm had}) }{ d^c_q (\mu = M_{\rm NP}) }
&=&
\frac{C_1^q (\mu =  \mu_{\rm had}) m_q (\mu =  \mu_{\rm had}) }{ C_2^q (\mu = M_{\rm NP}) m_q (\mu = M_{\rm NP})}
=
-0.80
, \ \ \ \ \ 
\\
\frac{d^c_q (\mu =  \mu_{\rm had}) }{ d^c_q (\mu = M_{\rm NP}) }
&=&
\frac{C_2^q (\mu =  \mu_{\rm had}) m_q (\mu =  \mu_{\rm had}) }{ C_2^q (\mu = M_{\rm NP}) m_q (\mu = M_{\rm NP})}
=
0.89
.
\label{eq:chromo-edm_running}
\end{eqnarray}
Note that the flavor of the quark $q$ is conserved during the running in the leading logarithmic order. It is also to be noted that the 
running of the quark EDM in Eq. (\ref{eq:qedm_running}) and the chromo-EDM in Eq. (\ref{eq:chromo-edm_running}) are additionally affected by the 
integration of the Higgs boson, heavy electroweak gauge bosons and the top quark, if those particles have CP violating interactions in
the BSM physics \cite{dekens}.

In the case where only the Weinberg operator is present at $\mu = M_{\rm NP}$, we have
\begin{eqnarray}
C_1^q (\mu =  \mu_{\rm had}) / C_3 (\mu = M_{\rm NP})
&=&
7.7 \times 10^{-2}
,
\\
C_2^q (\mu =  \mu_{\rm had}) / C_3 (\mu = M_{\rm NP})
&=&
-0.14
,
\\
C_3 (\mu =  \mu_{\rm had}) / C_3 (\mu = M_{\rm NP})
&=&
0.16
.
\label{eq:weinberg_running}
\end{eqnarray}
Here the Wilson coefficients $C_1^q$ and $C_2^q$ are generated for all relevant quark flavors ($q=u,d,s$).
It is also important to note that $C_3$ is sizably suppressed after the running.
By comparing Eqs. (\ref{eq:chromo-edm_running}) and (\ref{eq:weinberg_running}), we see that the chromo-EDM becomes 
large at the hadronic scale, even if the Wilson coefficients of the Weinberg operator and the chromo-EDM are of the same order of magnitude.
This is the case for 2HDM, where the contribution from Barr-Zee type diagrams are the most important.

We also show the evolution of the four-quark operator of the Left-right symmetric model [see Sec. \ref{sec:LRSM}].
The CP-odd four-quark coupling of Eq. (\ref{eq:4-q_lagrangian}), renormalized at the electroweak scale $\mu = m_W$,
is evolved down to the hadronic scale as \cite{dekens1}
\begin{eqnarray}
&&
\frac{C_4^{u} (\mu =  \mu_{\rm had}) }{ {\rm Im} (\Xi) (\mu = m_W ) }
=
\frac{C_4^{d} (\mu =  \mu_{\rm had}) }{ {\rm Im} (\Xi) (\mu = m_W ) }
\nonumber\\
&=&
-\frac{C_1^{ud} (\mu =  \mu_{\rm had}) }{ 2\, {\rm Im} (\Xi) (\mu = m_W ) }
=
-\frac{C_1^{du} (\mu =  \mu_{\rm had}) }{ 2\, {\rm Im} (\Xi) (\mu = m_W ) }
\nonumber\\
&=&
4.8
\ \ \ \ \ (\mu_{\rm had} = 1\, {\rm GeV})
.
\label{eq:LR4q_rge}
\end{eqnarray}
Although we obtain several other Wilson coefficients at the hadronic scale, here we focus on $C_4^{u}$, $C_4^{d}$, $C_1^{du}$ and $C_1^{ud}$, since their corresponding operators are the components of the operator $\bar q q \, \bar q i \gamma_5 \tau_z q$, which is suggested to be the leading contribution of the isovector pion-nucleon interaction (see Sec. \ref{sec:pion-nucleon}).
We also note again that the running of the Wilson coefficient ${\rm Im} (\Xi)$ begins at the electroweak scale $\mu = m_W$, since the $W$ boson has to be integrated out to generate the four-quark operator in Left-right symmetric model.
At the scale above $\mu = m_W$, the coupling of the right-handed $W_R$ boson with quarks does not run.
In running from $\mu = m_W$ to $\mu_{\rm had}$, the left-right four-quark operator mixes with several other four-quark operators, but it is interesting to note that it does not mix with the quark EDM, the quark chromo-EDM, and the Weinberg operator.

In the case where several CP-odd processes are simultaneously relevant at the TeV scale, the RGE of them down to the hadronic scale is just given by the linear combination of Wilson coefficients seen above.
This is because the RGE is calculated only in QCD and the effect of CP-odd interactions on the running is negligible.

Finally, let us also briefly present the running of SM contribution, although we do not discuss the detail.
The SM contribution at the electroweak scale is expressed by ten $|\Delta S|=1$ four-quark operators \cite{buras2}.
The next-to-next-to-leading logarithmic order evolution of the SM contribution enhances one of the penguin operator (see Fig. \ref{fig:tree_penguin}) by a factor of about 40 when the scale is varied from $\mu = m_W$ to $\mu = 1$ GeV \cite{buras2,buras1,smnuclearedm}.
This effect is nontrivial and enhances the SM contribution to the nucleon level CP-odd processes from the naive estimation.

Note that the RGE of this subsection is calculated in the perturbative framework, and systematics due to nonperturabative effects may be important at the hadronic scale $\mu =1$ GeV.

\subsubsection{RGE of CP-odd e-q interaction}

We now present the QCD RGE of the CP-odd e-q interactions.
The change of the Wilson coefficients of the CP-odd e-q interactions depends on the Lorentz structure 
of the quark bilinears. For the S-PS and PS-S type ones [terms with $C_{\rm SP}$ and $C_{\rm PS}$ of Eq. (\ref{eq:p,cp-odd_e-q_interaction}), respectively], the renormalization is the same as that of the quark mass.
We therefore have
\begin{eqnarray}
\frac{C_{eq}^{\rm SP} (\mu =  \mu_{\rm had}) \ }{ C_{eq}^{\rm SP} (\mu = M_{\rm NP}) }
&=&
\frac{C_{eq}^{\rm PS} (\mu =  \mu_{\rm had}) }{ C_{eq}^{\rm PS} (\mu = M_{\rm NP}) }
=
\frac{m_q (\mu =  \mu_{\rm had}) }{  m_q (\mu = M_{\rm NP})}
\nonumber\\
&=&
\left\{
\begin{array}{rr}
2.0 & (\mu_{\rm had} = 1\, {\rm GeV}) \cr
1.8 & (\mu_{\rm had} = 2\, {\rm GeV}) \cr
\end{array}
\right.
,
\end{eqnarray}
with $M_{\rm NP} = 1$ TeV.
Here we also show the ratio for $\mu_{\rm had} = 2\, {\rm GeV}$, for which we have less theoretical uncertainty due to the nonperturbative effect of QCD.
This renormalization point is often used in the lattice QCD calculations of nucleon matrix elements.

For the T-PT CP-odd e-q interaction [the term with $C_{\rm T}$ of Eq. (\ref{eq:p,cp-odd_e-q_interaction})], the renormalization is the same as that of the quark EDM.
The renormalization group evolution is then
\begin{eqnarray}
\frac{C_{eq}^{\rm T} (\mu =  \mu_{\rm had}) \ }{ C_{eq}^{\rm T} (\mu = M_{\rm NP}) }
&=&
\frac{d_q (\mu =  \mu_{\rm had}) }{ d_q (\mu = M_{\rm NP}) } \nonumber \\
&=& 
\frac{C_1^q (\mu =  \mu_{\rm had}) m_q (\mu =  \mu_{\rm had}) }{ C_1^q (\mu = M_{\rm NP}) m_q (\mu = M_{\rm NP})}
\nonumber\\
&=&
\left\{
\begin{array}{rr}
0.79 & (\mu_{\rm had} = 1\, {\rm GeV}) \cr
0.83 & (\mu_{\rm had} = 2\, {\rm GeV}) \cr
\end{array}
\right.
.
\end{eqnarray}

The S-PS and S-PS type P,CP-odd e-q interactions with heavy quarks are integrated out at scale below the quark masses, but their effects remain relevant 
through the P,CP-odd electron-gluon (e-g) interaction. The P,CP-odd e-g 
interaction is defined as
\begin{equation}
{\cal L}_{eg} = -\frac{G_F}{\sqrt{2}} 
\left[
C_{eg}^{\rm SP}  G_{\mu \nu}^a G^{\mu \nu}_a \, \bar e i \gamma_5 e
+C_{eg}^{\rm PS} \tilde G_{\mu \nu}^a G^{\mu \nu}_a \, \bar e e
\right] \, 
.
\label{eq:p,cp-odd_e-g_interaction}
\end{equation}
The matching of the couplings at each quark mass threshold works as
\begin{eqnarray}
C_{eg}^{\rm SP} (\mu = m_Q - \epsilon ) 
&=& 
C_{eg}^{\rm SP} (\mu = m_Q + \epsilon ) 
\nonumber\\
&&
+ \frac{ \alpha_s (\mu = m_Q ) }{12 \pi m_Q} C_{eQ} ^{\rm SP} (\mu = m_Q + \epsilon)
, \ \ \ \ \ 
\\
C_{eg}^{\rm PS} (\mu = m_Q - \epsilon ) 
&=& 
C_{eg}^{\rm PS} (\mu = m_Q + \epsilon ) 
\nonumber\\
&&
+ \frac{ \alpha_s (\mu = m_Q ) }{8 \pi m_Q} C_{eQ} ^{\rm PS} (\mu = m_Q + \epsilon)
,
\end{eqnarray}
where $\epsilon$ is the infinitesimal shift of energy scale.
As $\alpha_s G_{\mu \nu}^a G^{\mu \nu}_a$ is invariant under the RGE, the couplings 
$C_{eg}^{\rm SP}$ and $C_{eg}^{\rm PS}$ run in the same way as the strong coupling $\alpha_s (\mu)$.

If there is only one type of CP-odd e-q interaction $C_{eQ}^{\rm SP}$ ($Q=t,b$) at the scale $\mu = 1$ TeV, the running of its effect down to the hadronic scale is given by
\begin{eqnarray}
12 \pi m_t \frac{C_{eg}^{\rm SP} (\mu =  \mu_{\rm had}) }{ C_{et}^{\rm SP} (\mu = M_{\rm NP}) }
&=&
\frac{ \alpha_s (\mu =  \mu_{\rm had}) m_q (\mu = m_t) }{ \alpha_s (\mu = m_t) m_q (\mu = M_{\rm NP})}
\nonumber\\
&=&
\left\{
\begin{array}{rr}
3.7 & (\mu_{\rm had} = 1\, {\rm GeV}) \cr
2.8 & (\mu_{\rm had} = 2\, {\rm GeV}) \cr
\end{array}
\right.
, 
\label{eq:eg_running_top}
\\
12 \pi m_b \frac{C_{eg}^{\rm SP} (\mu =  \mu_{\rm had}) }{ C_{eb}^{\rm SP} (\mu = M_{\rm NP}) }
&=&
\frac{ \alpha_s (\mu =  \mu_{\rm had}) m_q (\mu = m_b) }{ \alpha_s (\mu = m_b) m_q (\mu = M_{\rm NP})}
\nonumber\\
&=&
\left\{
\begin{array}{rr}
2.7 & (\mu_{\rm had} = 1\, {\rm GeV}) \cr
2.0 & (\mu_{\rm had} = 2\, {\rm GeV}) \cr
\end{array}
\right.
.
\label{eq:eg_running_bottom}
\end{eqnarray}

If we consider a hadronic scale lower than the charm quark mass, the charm quark is also integrated out.
The CP-odd e-g coupling generated by $C_{ec}^{\rm SP}$ is then
\begin{eqnarray}
12 \pi m_c \frac{C_{eg}^{\rm SP} (\mu =  \mu_{\rm had}) }{ C_{ec}^{\rm SP} (\mu = M_{\rm NP}) }
&=&
\frac{ \alpha_s (\mu =  \mu_{\rm had}) m_q (\mu = m_c) }{ \alpha_s (\mu = m_c) m_q (\mu = M_{\rm NP})}
\nonumber\\
&=&
2.2 \ \  (\mu_{\rm had} = 1\, {\rm GeV}) 
.
\label{eq:eg_running_charm}
\end{eqnarray}
Obviously, the contributions of the CP-odd electron-heavy quark interactions are suppressed as the quark mass increases.
This additional damping is because the CP-odd e-g operator has one mass dimension higher than that of the CP-odd electron-quark interaction.
We reiterate that the same running of the Wilson coefficients of Eqs. (\ref{eq:eg_running_top}), (\ref{eq:eg_running_bottom}) and (\ref{eq:eg_running_charm}) also applies for $C_{eg}^{\rm PS}$ (we must replace $12 \pi m_Q$ by $8 \pi m_Q$ in the right-hand side of the equalities).

\subsection{$\theta$-term and Peccei-Quinn mechanism\label{sec:theta}}

The QCD $\theta$-term is a dimension-4, P and CP violating interaction [see Eq. (\ref{eq:theta-term})], which is not constrained by symmetries 
in the SM.
In the point of view of the naturalness, $\bar{\theta} \sim O(1)$, but it is known to generate a too large EDM of neutron.
The contribution of $\bar \theta$ to the neutron EDM was extensively studied  
\cite{pospelovreview,crewther,pich,borasoy,dib,kuckei,narison,pospelovtheta1,pospelovtheta2,nedmholography1,nedmholography2,nedmlattice1,shintani1,nedmlattice3,shintani2,nedmlattice5,nedmlattice6,nedmetm,shintani3}, 
and the most recent analysis based on the chiral effective field theory (EFT) is giving \cite{ottnad,mereghetti2,mereghetti1,guo1,devriessplitting,devriesreview}
\begin{eqnarray}
d_n
&= &
-(2.7 \pm 1.2) \times 10^{-16} \bar \theta \,
e\, {\rm cm}
,
\label{eq:dntheta}
\\
d_p 
&=&
(2.1 \pm 1.2) \times 10^{-16} \bar \theta \,
e\, {\rm cm}
.
\label{eq:dptheta}
\end{eqnarray}
From the experimental data \cite{baker,Pendlebury}
\begin{equation}
d_n < 3.0 \times 10^{-26}e \, {\rm cm} 
,
\label{eq:neutronedmexp}
\end{equation}
we therefore have
\begin{equation}
\bar \theta < 10^{-10}
,
\label{eq:nedmconstraint}
\end{equation}
which is a too strong constraint to the $\theta$-term, which should naturally be of the same order of magnitude as the CP-even QCD 
Lagrangian.
This problem is known as the {\it Strong CP Problem}. This problem is also accentuated in the context of new sources of CP violation 
of BSM. A large $\theta$-term is also generated in many models of new physics such as 
SUSY models \cite{pospelovreview,ellismssmedm,buchmullermssmedm,duganmssmedm}, and this gives rise to a serious fine-tuning problem, as their effects must cancel to fulfill the constraint (\ref{eq:nedmconstraint}).
If we want to extend the SM to a theory with large source of CP violation, a mechanism which makes the $\theta$-term irrelevant to observables are at least mandatory.

A natural solution to the strong CP problem was proposed by Peccei and Quinn \cite{peccei}.
Their mechanism forces the $\theta$-term to have a zero expectation value by adding a new scalar field, the {\it axion}.
The newly introduced lagrangian of the axion is
\begin{equation}
{\cal L}_a = \frac{1}{2} \partial_\mu a \partial^\mu a + \frac{a(x)}{f_a} \frac{\alpha_s}{8\pi} G_{\mu \nu}^a \tilde G^{\mu \nu , a} \ ,
\label{eq:axionlagrangian}
\end{equation}
where the axion field $a$ has replaced the parameter $\bar \theta$ of the strong CP lagrangian.

The effective potential of the axion will then become
\begin{equation}
{\cal L}^{ eff}_a 
= 
\frac{1}{2} \partial_\mu a \partial^\mu a 
-K_1 \left( \frac{a}{f_a} + \bar \theta \right) 
-\frac{1}{2} K \left( \frac{a}{f_a} + \bar \theta \right)^2 
+ \cdots 
\ ,
\end{equation}
where $K= - m_* \langle 0 | \bar q q |0 \rangle + O( m_*^2) $ is the topological susceptibility with
\begin{equation}
m_* \equiv 
\frac{m_u m_d m_s}{m_u m_d + m_u m_s + m_d m_s} \approx \frac{m_u m_d }{m_u + m_d}
,
\label{eq:mstar}
\end{equation}
and $K_1$ is the correlation between the topological charge and the isoscalar CP-odd operators with high mass dimensions.
The decay constant $f_a$ is given by the spontaneous breaking of a chiral $U(1)_{\rm PQ}$ symmetry 
of BSM. Here the vacuum expectation value of the axion becomes the $\theta$-term $\langle a \rangle / f_a = -\bar \theta$.
If there are no other CP-odd operators than the $\theta$-term, this value is zero, which means that the $\theta$-term is dynamically canceled.
This mechanism of Peccei and Quinn is the most attracting scenario to naturally resolve the strong CP problem.

In the presence of flavor $SU(3)$ singlet CP-odd operators other than the $\theta$-term, the vacuum expectation value of the axion is not canceled, and becomes $\bar \theta = \bar \theta_{\rm ind} \equiv - K_1 (O_{CP}) / K$ \cite{shifmantheta}.
It is controlled by the coefficient $K_1$, which is expressed as \cite{bigi}
\begin{eqnarray}
K_1 (O_{CP})
&=& -i \lim_{k \to 0} \int d^4 x e^{ikx} \nonumber \\ & \times & \left< 0 \left| \, T \left\{ \, \frac{\alpha_s }{8\pi}G_{\mu \nu}^a \tilde G^{\mu \nu , a} (x) O_{CP} (0) \right\} \right| 0 \right>
.
\label{eq:k1theta}
\end{eqnarray}
In this review, 
the relevant ones are the quark chromo-EDM ${\cal L}_{\rm cedm} = -\frac{i}{2} d^c_q \bar q g_s \sigma^{\mu \nu} G_{\mu \nu}^a t_a \gamma_5 q $ and the left-right four-quark operator ${\cal L}_{LR} = C_{LR} \sum_{q,q' = u,d} \bar q q \, \bar q' i \gamma_5 \tau_z q'$.
The coefficient of the CP-odd four-quark operator is given by $C_{LR} = C_1^u = C_1^d = -C_4^{ud} = -C_4^{du}$ (the corresponding operators are defined in Eqs. (\ref{eq:o4}) and (\ref{eq:o1'})). 
In the case of the quark chromo-EDM, the evaluation of the correlator gives \cite{bigi}
\begin{equation}
K_1 ({\cal L}_{\rm cedm})
= \frac{m_*}{2} \sum_{q=u,d,s} \frac{d_q^c}{m_q} \langle 0 | \bar q g_s \sigma^{\mu \nu} G_{\mu \nu}^a t_a q |0 \rangle \ 
,
\end{equation}
where $m_0^2 \equiv -\frac{\langle 0 | g_s \bar q \sigma_{\mu \nu} t_a G_a^{\mu \nu} q | 0 \rangle}{\langle 0 | \bar q q | 0 \rangle} =(0.8 \pm 0.1)\, {\rm GeV}^2$ \cite{belayev1,belayev2}.
The induced $\theta$-term is then
\begin{equation}
\bar \theta_{\rm ind} ({\cal L}_{\rm cedm})
= 
- \frac{m_0^2}{2} \sum_{q=u,d,s} \frac{d^c_q}{m_q}
\ .
\label{eq:inducedthetacedm}
\end{equation}

For the left-right four-quark operator, the induced $\theta$-term is given by \cite{LRLHC2,4-quark5}
\begin{equation}
\bar \theta_{\rm ind} ({\cal L}_{\rm LR})
\approx
-C_{LR}
\frac{f_\pi^2 m_\pi^2}{m_u m_d}
,
\end{equation}
where the vacuum saturation approximation $\langle 0 | \bar q q \, \bar q' i \gamma_5 \tau_z q' | \pi^0 \rangle \approx \langle 0 | \bar q q |0 \rangle \langle 0| \bar q' i \gamma_5 \tau_z q' | \pi^0 \rangle$ was used.
From the large $N_c$ analysis, the correction to this formula is $O(1/N_c )$.

The Weinberg operator is also a flavor $SU(3)$ singlet, but the induced $\theta$-term is suppressed by a factor of light quark mass, so it becomes negligible for the case of interest.

\section{Hadron level effective P,CP-odd interactions\label{sec:hadron_physics}}

The atomic EDM receives contribution from the hadron level CP violation. 
The effective hadronic CP-odd interaction is generated by quark and gluon level CP-odd processes, but the calculation of their relations is a highly nontrivial 
task due to the nonperturbative nature of QCD. Here we summarize the current situation of the derivation of the hadron level CP violation from the QCD level physics.

\subsection{Hadron level effective interaction at the hadronic scale}

After obtaining the QCD level operators and their Wilson coefficients at the hadronic scale, we must now match them to the hadron level effective interactions.
The P,CP-odd hadronic interaction we consider is
\begin{equation}
{\cal L}_{\rm hadron} 
=
{\cal L}_{e N } + {\cal L}_{\rm Nedm} 
+ {\cal L}_{ \pi NN} \, ,
\label{eq:hadron_level_lagrangian}
\end{equation}
with 
\begin{itemize}

\item
The P,CP-odd e-N interaction
\begin{eqnarray}
{\cal L}_{eN} 
&=& 
-\frac{G_F}{\sqrt{2}} \sum_{N=p,n} \Biggl[
C_N^{\rm SP} \bar NN \, \bar e i \gamma_5 e
+C_N^{\rm PS} \bar Ni\gamma_5 N \, \bar e e 
\nonumber \\ 
&& \hspace{8em}
+\frac{1}{2} C_N^{\rm T} \epsilon^{\mu \nu \rho \sigma} \bar N \sigma_{\mu \nu} N \, \bar e \sigma_{\rho \sigma} e
\Biggr]
. \ \ \ \ \ \ 
\label{eq:pcpvenint}
\end{eqnarray}

\item
The nucleon EDM
\begin{equation}
{\cal L}_{\rm Nedm} = -\frac{i}{2} \sum_{N=p,n} d_N \bar N \sigma_{\mu \nu} \gamma_5 N F^{\mu \nu},
\end {equation}

\item
The P,CP-odd pion-nucleon ($\pi$-N-N) interaction \cite{Barton,haxton}
\begin{eqnarray}
{\cal L}_{\pi NN} &=& 
\sum _{N=p,n} \Biggl[ \sum_{a=1}^3 \bar g_{\pi NN}^{(0)} \bar N \tau^a N \pi^a 
+ \bar g_{\pi NN}^{(1)} \bar N N \pi^0 
\nonumber \\ 
&& \hspace{3em}
+\sum_{a=1}^3 \bar g_{\pi NN}^{(2)} ( \bar N \tau^a N \pi^a - 3\bar N \tau^3 N \pi^0) 
\Biggr]
, \ \ \ \ \ \ \ 
\label{eq:pcpvpinnint}
\end{eqnarray}
where $a$ denotes the isospin index.

\end{itemize}
The schematic dependences of the hadronic scale operators on the quark level operators are shown in Fig. \ref{fig:flow_diagram}.
In this subsection, we review the currently available results of the calculation of the hadronic effective CP violation.

We can also extend Eq. (\ref{eq:hadron_level_lagrangian}) by adding several interactions with low chiral indices.
For example we have the three-pion interaction \cite{mereghetti2}
\begin{eqnarray}
{\cal L}_{3\pi}
&=& 
m_N \Delta_{3\pi} \, \pi^z  \sum_{a=1}^3 \pi_a^2
,
\label{eq:three-pion}
\end{eqnarray}
the CP-odd $\eta$-nucleon interaction \cite{gorchtein,blin}
\begin{equation}
{\cal L}_{\eta NN} 
=
\sum _{N=p,n} \Big[ \bar g_{\eta NN}^{(0)} \bar N N \eta
+ \bar g_{\eta NN}^{(1)} \bar N \tau^z N \eta
\Big] \, .
\label{eq:pcpvetannint}
\end{equation}
and the isoscalar CP-odd contact interaction
\begin{eqnarray}
{\cal L}_{\rm C}
&=& 
\bar C_1 \bar N N \partial_\mu (\bar N S^\mu N )
+ \sum_{a=1}^3 \bar C_2 \bar N \tau_a N \cdot \partial_\mu (\bar N S^\mu \tau_a N )
. \ \ \ \ 
\label{eq:cpv_contact}
\end{eqnarray}
It is important to note that the above interactions give the leading contribution in the power counting of the chiral EFT for several dimension-six CP-odd interactions \cite{chiraleft3nucleon}.
For example, the three-pion interaction in Eq. (\ref{eq:three-pion}) receives the leading contribution from the isovector quark chromo-EDM or the left-right four-quark interaction in Eq. (\ref{eq:4-q_lagrangian}).
It is known to give a sizable next-to-leading order correction to the isovector CP-odd pion-nucleon interaction coupling $\bar g_{\pi NN}^{(1)}$, so can be used to estimate the theoretical uncertainty.
Regarding the CP-odd contact interaction in Eq. (\ref{eq:cpv_contact}), it receives the leading contribution from the Weinberg operator or some CP-odd four-quark interactions.
Its effect is however of short range, and we can expect it to be subleading at the nuclear level, since the nuclear EDM is a low energy phenomena dominated by the pion exchange physics.
This fact is suggested by the {\it ab initio} analyses of the EDM of light nuclei, which have smaller sensitivities on the heavy meson exchange CP-odd nuclear forces than those with the pion exchanges \cite{liu,stetcu,song,bsaisou,bsaisou2,yamanakanuclearedm}.

\subsection{CP-odd e-N interaction\label{sec:cpve-n}}

The CP-odd e-N interaction (\ref{eq:pcpvenint}) is a CP violating effect which can 
specifically be probed with the atomic EDM. It is related to the CP-odd e-q interaction 
(\ref{eq:p,cp-odd_e-q_interaction}) and the CP-odd e-g interaction 
(\ref{eq:p,cp-odd_e-g_interaction}) by several nucleon matrix elements, as
\begin{eqnarray}
- \frac{G_F}{\sqrt{2}} C_N^{\rm SP} \bar N N \, \bar e i \gamma_5 e
&=&
C_q^{\rm SP} \langle N| \bar q q | N \rangle \, \bar e i \gamma_5 e \nonumber \\
 + C_{eg}^{\rm SP} 
&& \hspace{-1.5em}
 \langle N | G_{\mu \nu}^a G^{\mu \nu}_a | N \rangle \, \bar e i\gamma_5 e
, 
\label{eq:ce-nsp} 
\\
- \frac{G_F}{\sqrt{2}} C_N^{\rm PS} \bar N i\gamma_5 N \, \bar e e
&=&
C_q^{\rm PS} \langle N| \bar qi\gamma_5 q | N \rangle \, \bar e e \nonumber \\
 + C_{eg}^{\rm PS} 
&& \hspace{-1.5em}
\langle N | G_{\mu \nu}^a \tilde G^{\mu \nu}_a | N \rangle \, \bar e e
,
\label{eq:ce-nps} 
\\
- \frac{G_F}{\sqrt{2}} C_N^{\rm T} \frac{1}{2} \epsilon^{\mu \nu \rho \sigma} 
\bar N \sigma_{\mu \nu} N \, \bar e \sigma_{\rho \sigma} e
&=&
 \frac{1}{2} C_q^{\rm T} \epsilon^{\mu \nu \rho \sigma} \nonumber \\ & \times & \langle N| \bar q \sigma_{\mu \nu} q | N \rangle \, \bar e \sigma_{\rho \sigma} e
.
\label{eq:ce-nt}
\end{eqnarray}
To calculate the nucleon matrix elements, evaluations of nonperturbative effects of QCD are required.
Here it is an excellent opportunity to use the results of the lattice QCD, which has recently made significant progress \cite{brambilla,collinslattice2016}.

\begin{figure}[tbp]
\begin{center}
\resizebox{8.0cm}{!}{\includegraphics{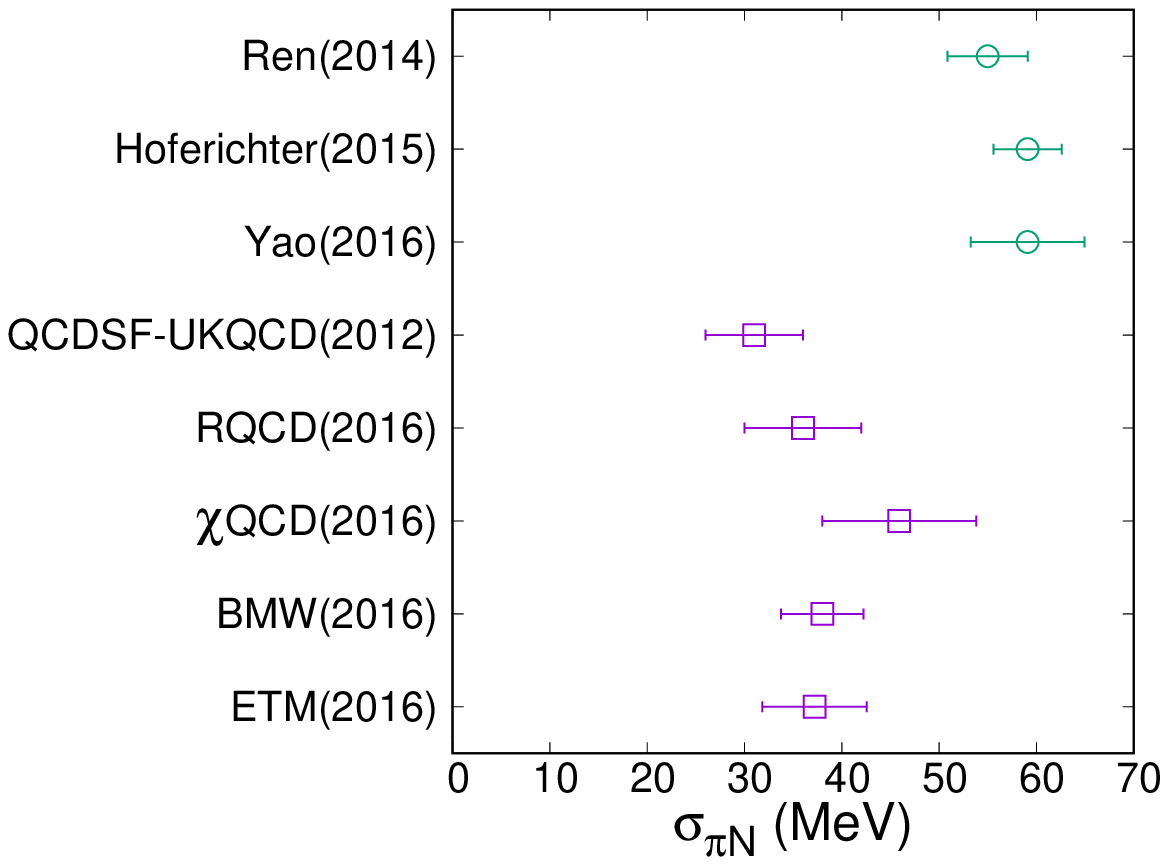}}
\resizebox{8.0cm}{!}{\includegraphics{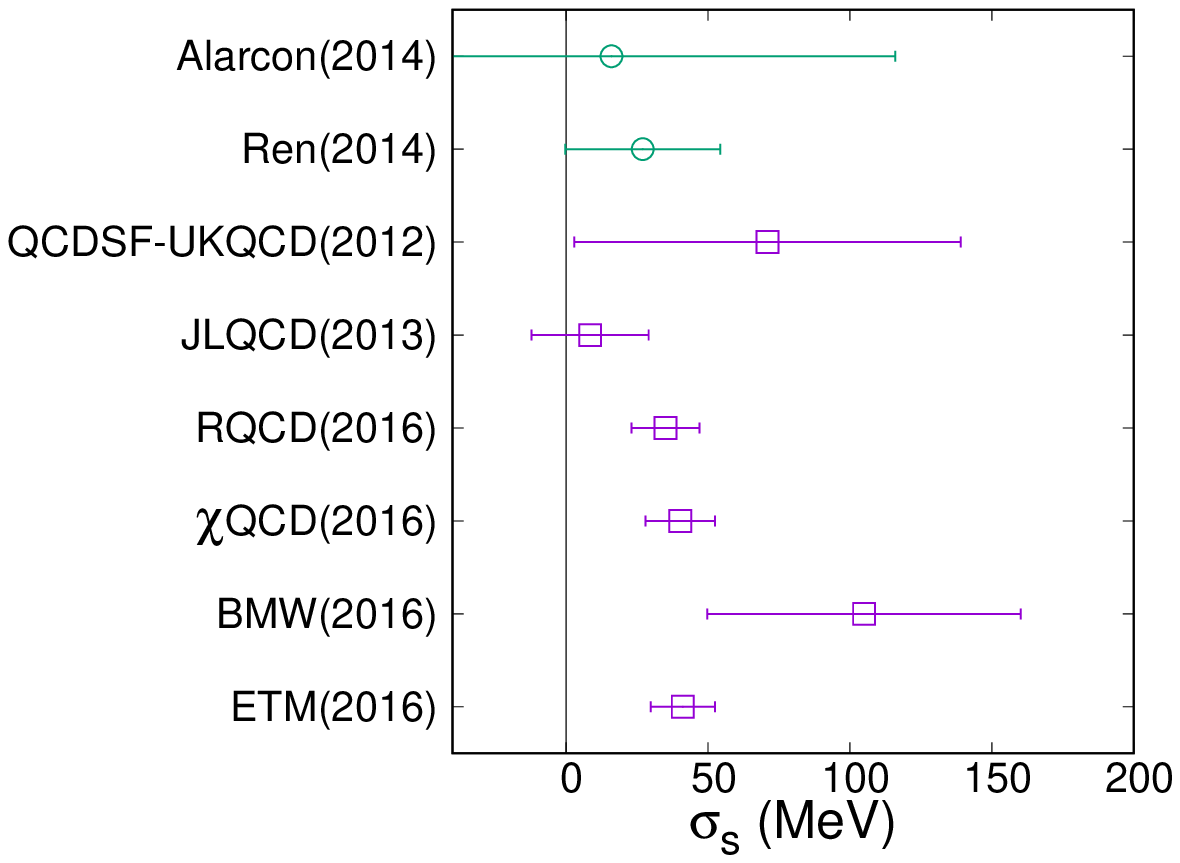}}
\resizebox{8.0cm}{!}{\includegraphics{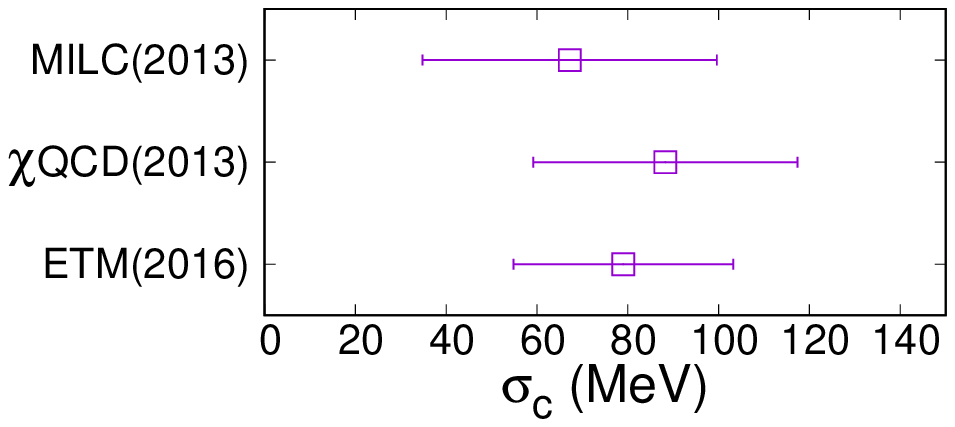}}
\caption{\label{fig:sigma-term-lattice}
Comparison of the results of several calculations of the nucleon sigma term ($\sigma_{\pi N}$) \cite{ren,Hoferichter,yao,qcdsf-ukqcdsigmaterm,durr2,rqcdsigmaterm,etmsigmaterm,chiqcdsigmaterm}, the strange content of nucleon ($\sigma_s$) \cite{ren,qcdsf-ukqcdsigmaterm,durr2,rqcdsigmaterm,etmsigmaterm,chiqcdsigmaterm,ohki2,alarcon3} and the charm content of nucleon ($\sigma_c$) \cite{etmsigmaterm,milccharmcontent,chiQCDcharmcontent}.
} 
\end{center}
\end{figure}

To determine the S-PS CP-odd e-N interaction (\ref{eq:ce-nsp}), the nucleon scalar density matrix elements $\langle N| \bar q q | N \rangle$ and $\langle N | G_{\mu \nu}^a G^{\mu \nu}_a | N \rangle$ are required.
To obtain the light quark contribution, we combine the isoscalar and isovector nucleon scalar densities.
The isoscalar one can be derived from the nucleon sigma term $\sigma_{\pi N} \equiv \frac{m_u + m_d}{2}\langle N | \bar uu + \bar dd | N \rangle$, 
which has extensively been discussed in phenomenology \cite{chengdashen,gasser,gasserleutwyler1,gasserleutwyler2,gasserleutwyler3,alarcon1,alarcon2,ren,Hoferichter,yao} and in lattice QCD \cite{ohki,young1,durr1,dinter,qcdsf1,qcdsf2,qcdsf-ukqcdsigmaterm,durr2,rqcdsigmaterm,etmsigmaterm,chiqcdsigmaterm} (see Fig. \ref{fig:sigma-term-lattice}).
The result is giving
\begin{equation}
\sigma_{\pi N} 
=
(30 -60) \, {\rm MeV}
.
\label{eq:nucleonsigmaterm}
\end{equation}
The phenomenological extractions are centered to 60 MeV, whereas the results of lattice QCD calculations are showing values around 40 MeV.
We consider this deviation as a systematic error.
By using the Particle data group value of up and down quark masses renormalized at $\mu =2$ GeV $m_u = 2.2^{+0.6}_{-0.4}$ MeV and $m_d = 4.7^{+0.5}_{-0.4}$ MeV \cite{pdg}, 
the isoscalar nucleon scalar density is then
\begin{equation}
\langle N | \bar uu + \bar dd | N \rangle
\sim
15
 \ \ \ \ \ (\mu =2 \, {\rm GeV})
,
\label{eq:isoscalarnucleonscalardensity}
\end{equation}
with a theoretical uncertainty of about 30\%.

The isovector nucleon scalar density can be derived in the leading order of the current quark masses in terms of the proton-neutron mass 
splitting $\Delta m_N^{(0)}$ as 
\cite{rpvedm1,yamanakabook,adler,rpvbetadecay1,rpvbetadecay2,gonzales-alonso}
\begin{equation}
\langle p | \bar uu - \bar dd | p \rangle
=
\frac{\Delta m_N^{(0)} }{m_d -m_u}
=
0.9 \ \ \ (\mu = 2\, {\rm GeV})
,
\label{eq:isovectornucleonscalardensity}
\end{equation}
with a theoretical uncertainty of about 30\%.
Here $\Delta m_N^{(0)}= 2.33 \pm 0.11$ MeV is the nucleon mass splitting without electromagnetic effects \cite{pdg,mohr,thomas}.
The isovector nucleon scalar density has also been studied on lattice, and consistent results with the above phenomenological value are given \cite{linisovectoraxial,rbcukqcdisovectortensor,green,pndmeconnected,rqcdisovector,chiqcdisovector,pndmeisovector,etmisovector}.

By combining Eqs. (\ref{eq:isoscalarnucleonscalardensity}) and (\ref{eq:isovectornucleonscalardensity}), we obtain
\begin{eqnarray}
\langle p | \bar uu  | p \rangle
&=&
\langle n | \bar dd | n \rangle
\sim
8
,
\\
\langle p | \bar dd | p \rangle
&=&
\langle n | \bar uu  | n \rangle
\sim
7
,
\end{eqnarray}
at $\mu = 2$ GeV, with a theoretical uncertainty of 30\%.
It is important to note that this error bar is mainly due to the uncertainty of the current quark mass and to that of the determination of the nucleon sigma term.
The nucleon scalar densities due to light quarks are substantially enhanced compared with the prediction of the nonrelativistic quark model ($\langle p | \bar uu  | p \rangle = 2$, $\langle p | \bar dd  | p \rangle = 1$).
This enhancement is understood by the dynamical gluon dressing effect \cite{hatsudakunihiro1,hatsudakunihiro2,hatsudareview,yamanakasde2}.

For the strange and charm contents of the nucleon $\sigma_s \equiv m_s \langle N | \bar ss | N \rangle$ and $\sigma_c \equiv m_c \langle N | \bar cc | N \rangle$, there are also available data from lattice QCD calculations \cite{young1,durr1,dinter,qcdsf1,qcdsf2,qcdsf-ukqcdsigmaterm,durr2,rqcdsigmaterm,etmsigmaterm,chiqcdsigmaterm,takeda,ohki2,etmdisconnected,etm1,milc1,engelhardt,junnarkar,milccharmcontent,chiQCDcharmcontent} (see Fig. \ref{fig:sigma-term-lattice}).
Their averaged values at the renormalization point $\mu = 2$ GeV are
\begin{eqnarray}
\langle N | \bar ss | N \rangle
&\sim &
0.4 
,
\label{eq:strangecontent}
\\
\langle N | \bar cc | N \rangle
&\sim &
0.07
.
\label{eq:charmcontent}
\end{eqnarray}
Here we have used $m_s = 96^{+8}_{-4}$ MeV for the current strange quark mass \cite{pdg}.
For the charm quark mass, we have adopted $m_c = 1.17$ GeV, which is obtained by running $m_c = 1.27$ GeV from the renormalization point $\mu = m_c = 1.27$ GeV \cite{pdg} to $\mu =2$ GeV.
The theoretical uncertainty is not less than 100\% for the strange content.
The results of phenomenological analyses have a large error bar and cannot be used in the determination of the strange content of nucleon \cite{ren,alarcon3}.
Lattice QCD results also seem to have a systematic error, as some values are not consistent (see Fig. \ref{fig:sigma-term-lattice}).
This situation may be improved in the future by refining lattice QCD analyses.
We may also expect improvement from a new experimental approach which can constrain the strange content of nucleon through the measurement of $\phi$ mesons in nuclear medium \cite{gubler1,gubler2,gubler3}.

For the charm content the uncertainty of lattice QCD data is about 30\%. 
It agrees with the heavy quark expansion formula \cite{hatsudakunihiro2,hatsudareview,wittenheavyquark,shifmanheavyquark,zhitnitskyheavyquark,franzheavyquark}
\begin{eqnarray}
\langle N | \bar cc | N \rangle
&= &
-\frac{\alpha_s (\mu = 2\, {\rm GeV})}{12 \pi m_c} \langle N | G_{\mu \nu}^a G^{\mu \nu}_a | N \rangle
+ O (1/m_c^2)
\nonumber\\
&\approx &
0.054
.
\end{eqnarray}

To obtain the gluonic content of the nucleon, we use the two-loop level trace anomaly formula of the nucleon mass:
\begin{eqnarray}
m_N
&=&
\frac{\beta_{\rm QCD}}{2 g_s} \langle N | G_{\mu \nu}^a G^{\mu \nu}_a | N \rangle
+ \sum_{q} m_q \langle N | \bar qq | N \rangle
\nonumber\\
&\approx &
\frac{ \alpha_s (\mu ) }{8 \pi} 
\Biggl[ \beta_0 + \beta_1 \frac{\alpha_s (\mu ) }{4 \pi }\Biggr] 
\langle N | G_{\mu \nu}^a G^{\mu \nu}_a | N \rangle
\nonumber\\
&&
+2\sigma_{\pi N} + \sigma_s + \sigma_c
,
\label{eq:trace_anomaly}
\end{eqnarray}
with $\beta_0 =\frac{25}{3}$ and $\beta_1 =\frac{154}{3}$ ($n_f = 4$ at $\mu = 2$ GeV).
By subtracting the quark contents of nucleon in Eqs. (\ref{eq:nucleonsigmaterm}), (\ref{eq:strangecontent}), and (\ref{eq:charmcontent}), 
it yields
\begin{eqnarray}
\langle N | G_{\mu \nu}^a G^{\mu \nu}_a | N \rangle
&\approx &
(-6000 \pm 450) {\rm MeV}
,
\end{eqnarray}
where we have used $\alpha_s (\mu = 2\, {\rm GeV}) = $ 0.30 (from two-loop level renormalization group equation).
Here the error bar is only due to the nucleon scalar densities, and amounts to about 8\%.
The error of the perturbative expansion of Eq. (\ref{eq:trace_anomaly}) can be estimated by evaluating the three-loop level correction.
The relative error is given by the ratio between the one-loop level and three-loop level term $\frac{\alpha_s^2 \beta_2}{(4 \pi )^2 \beta_0} = 3 \%$, where $\beta_2 = \frac{1}{2} \Bigl[ 2857 - \frac{20132}{9} + \frac{5200}{27}\Bigr]$ [$n_f = 4$, $\alpha_s (\mu = 2 \, {\rm GeV}) = 0.3$].
Here the most important source of theoretical uncertainty of $\langle N | G_{\mu \nu}^a G^{\mu \nu}_a | N \rangle$ is the error bars of the strange and charm contents of nucleon.
To improve the accuracy of the S-PS type CP-odd e-N interaction, continuous efforts in lattice QCD calculations are required.
We also have to remark that the contributions from the light and charm quarks are determined within 30\%, 
and quantitative discussions are becoming possible.

To know the coupling of the PS-S type CP-odd e-N interaction (\ref{eq:ce-nps}), values of pseudoscalar nucleon matrix elements are required.
Let us first evaluate the gluonic pseudoscalar nucleon matrix element $\langle N | G_{\mu \nu}^a \tilde G^{\mu \nu}_a | N \rangle$.
It can phenomenologically be calculated as \cite{hycheng,cheng,dienes}
\begin{eqnarray}
- \frac{\alpha_s }{8 \pi} \langle N | G_{\mu \nu}^a \tilde G^{\mu \nu}_a | N \rangle
&=&
m_N 
\Biggl[ \sum_q \frac{m_*}{m_q} \Delta q \Biggr]
+m_* \eta
\nonumber\\
&\approx &
- 400 \, {\rm GeV} 
,
\label{eq:gluon_spin}
\end{eqnarray}
where the nucleon axial charges are given by $\Delta u = [0.82 ,0.85]$, $\Delta d = [-0.45,-0.42]$ and $\Delta s = [-0.11,-0.08]$ (renormalized at $\sqrt{3}$ GeV) \cite{compass}.
The results of lattice QCD analyses for $\Delta s$ is smaller than the experimental value \cite{etmdisconnected,qcdsfprotonspin}.
Here $\eta$ is the deviation of the relation $\sum_q \langle p | \bar q i\gamma_5 q | p \rangle =0$ which is valid in the large $N_c$ limit, and it should be considered as an $O(N_c^{-1})$ 
systematic error, which is negligible in this derivation \cite{hycheng,dienes}. The largest theoretical uncertainty to $\frac{\alpha_s }{8 \pi} \langle N | G_{\mu \nu}^a \tilde G^{\mu \nu}_a | N \rangle$ 
comes from the current quark mass of $u$ and $d$ quarks and the total error bar may not be less than 30\%.

By using the anomalous Ward identity
\begin{equation}
m_N \Delta q 
=
m_q \langle p | \bar q i\gamma_5 q | p \rangle 
- \frac{\alpha_s (\mu )}{8 \pi} 
\langle N | G_{\mu \nu}^a \tilde G^{\mu \nu}_a | N \rangle
,
\end{equation}
the quark pseudoscalar contents of nucleon can be calculated phenomenologically as
\cite{rpv4f1,yamanakabook,gonzales-alonso,hycheng,cheng,dienes,hycheng2,scopel}
\begin{eqnarray}
\langle p| \bar ui\gamma_5 u |p\rangle 
&=& 
180 
,
\label{eq:u5u}
\\
\langle p| \bar di\gamma_5 d |p\rangle 
&=& 
-170 
,
\label{eq:d5d}
\\
\langle p| \bar si\gamma_5 s |p\rangle 
&=& 
-5.1 
, 
\label{eq:s5s}
\\
\langle p| \bar ci\gamma_5 c |p\rangle 
&=& 
-0.34.
\label{eq:c5c}
\end{eqnarray}
For the charm quark contribution, we have neglected $\Delta c$ and used $m_c (\mu = 2 \, {\rm GeV}) = 1170$ MeV.
We remark that the pseudoscalar nucleon matrix elements for light quarks are large.
This is due to the pion pole effect \cite{yamanakasde2}.
This enhancement has an important impact in the evaluation of the atomic EDM, because it can counterbalance the nonrelativisic
suppression of the effect of PS-S CP-odd e-N interaction in Eq. (\ref{eq:ce-nps}).
The nucleon pseudoscalar density $\langle p| \bar qi\gamma_5 q |p\rangle$ receives the dominant contribution from the gluonic matrix 
element $\frac{\alpha_s }{8 \pi} \langle N | G_{\mu \nu}^a \tilde G^{\mu \nu}_a | N \rangle$. The main source of its theoretical
uncertainty, therefore, comes from the current quark masses. Since we do not know the higher order correction in the chiral expansion
reliably, we can conservatively estimate the total uncertainty to $\langle p| \bar qi\gamma_5 q |p\rangle$ is about 50\%.

The T-PT CP-odd e-N couplings are given in terms of the nucleon tensor charge $\delta q$ as
\begin{eqnarray}
- C_p^{\rm T} 
&=&
\delta u \, C_{eu}^{\rm T} 
+\delta d \, C_{ed}^{\rm T} 
\sum_{q=s (,c)}
\delta q \, C_{eq}^{\rm T} 
,
\\
- C_n^{\rm T} 
&=&
\delta d \, C_{eu}^{\rm T} 
+\delta u \, C_{ed}^{\rm T} 
+\sum_{q=s (,c)}
\delta q \, C_{eq}^{\rm T} 
,
\end{eqnarray}
where we have assumed the isospin symmetry and  the proton tensor charge $\delta q$ are defined by
\begin{equation}
\langle p (k,s) | \bar q i \sigma_{\mu \nu} \gamma_5 q | p (k,s) \rangle
= 
2 (s_\mu k_\nu - s_\nu k_\mu ) \delta q
,
\label{eq:nucleon_tensor_charge}
\end{equation}
with $s$ and $k$ the 4-vector polarization and momentum of the proton, respectively.
The nucleon tensor charge is the transversely polarized quark contribution to the nucleon polarization, where the nucleon is transversely polarized against its momentum $k$.
Currently, lattice QCD is giving the most accurate data, and the results are giving 
\cite{linisovectoraxial,rbcukqcdisovectortensor,green,pndmeconnected,rqcdisovector,chiqcdisovector,pndmeisovector,etmisovector,etmdisconnected,aokitensorcharge,pndmetensor1,pndmetensor2,jlqcd4}
\begin{eqnarray}
\delta u & \approx & 0.8
,
\label{eq:deltau}
\\
\delta d & \approx & -0.2
,
\label{eq:deltad}
\\
| \delta s | & < & 0.02
,
\end{eqnarray}
at the renormalization point $\mu = 2$ GeV, 
with theoretical uncertainties of roughly 10\% for the up and down quark contributions.
For $\delta s$, these are currently the only results consistent with zero at the physical point \cite{pndmetensor1,pndmetensor2}.
In the literature, the nonrelativistic quark model predictions $\delta u = \frac{4}{3}$ and $\delta d = -\frac{1}{3}$ were often quoted.
The suppression of the nucleon tensor charges in Eqs. (\ref{eq:deltau}) and (\ref{eq:deltad}) from the nonrelativistic quark model prediction is 
partially understood by the gluon dressing effect, which superposes spin flipped states of spin 1/2 quarks due to the emission and 
absorption of spin 1 gluons, as suggested by the Schwinger-Dyson analyses \cite{yamanakasde1,pitschmann}.
The nucleon tensor charges can also be extracted from the experimental data, but the analysis may have sizable systematics due to the functional form used in the analysis \cite{bacchetta,anselmino,courtoy,radici,kang,yez}.

\begin{figure}
\begin{center}
\resizebox{6.0cm}{!}{\includegraphics{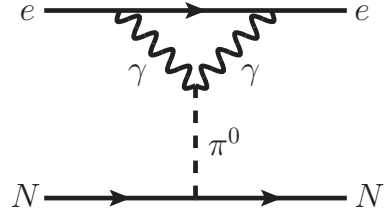}}
\caption{\label{fig:sm_e-n}
SM contribution to the CP-odd e-N interaction.
} 
\end{center}
\end{figure}

Let us also discuss the SM contribution to the CP-odd e-N interaction, generated by the CKM matrix elements.
The leading contribution is given by the S-PS type interaction $C_N^{\rm SP}$ (see Fig. \ref{fig:sm_e-n}) \cite{pospelovsmatomicedm,hee-Nedm}.
We estimate this effect as
\begin{equation}
C_N^{\rm SP} 
\sim
\bar g_{\pi NN , {\rm SM}}^{(1)} \frac{3 \alpha_{\rm em}^2 m_e}{2 \pi^2 f_\pi } 
\frac{\sqrt{2}}{m_\pi^2 G_F}
\ln \frac{m_\pi}{m_e} 
=
O(10^{-17})
,
\end{equation}
where we have estimated the effective $\pi^0 \bar e i \gamma_5 e$ vertex by solving the one-loop level renormalization group equation with the nonrenormalizable Wess-Zumino-Witten term $\frac{\pi^0}{4 \pi f_\pi} F_{\mu \nu} \tilde F^{\mu \nu}$ \cite{choithetaedm}.
The CP-odd $\pi$-N-N coupling in SM $\bar g_{\pi NN , {\rm SM}}^{(1)} = O(10^{-17})$ which was given 
in the factorization approach with the $|\Delta S|=1$ four-quark interaction calculated obtained from the two-loop level renormailzation group evolution (see also the end of Sec. \ref{sec:pion-nucleon}) \cite{smnuclearedm}.
The CP-odd e-N interaction generated by the exchange of the Higgs boson in the SM is negligibly small \cite{Chubukov}. 
In the SM, the PS-S and T-PT CP-odd e-N interactions are generated at higher order than for the S-PS one, so we neglect them.

\subsection{\label{sec:pion-nucleon}The CP-odd $\pi$-N-N interaction}

Now let us present the calculation of the hadron level CP violation. The most important CP-odd interaction is the CP-odd $\pi$-N-N interaction, which is the base of hadronic effective CP-odd interaction.

Let us first see the $\theta$-term contribution to the CP-odd $\pi$-N-N  interaction in Eq. (\ref{eq:pcpvpinnint}).
Using the partially conserved axial current relation (PCAC), the isoscalar CP-odd $\pi$-N-N coupling is given as
\begin{equation}
\bar g_{\pi NN}^{(0)}
\approx
\frac{\bar \theta m_* }{f_\pi}
\langle N | 
\bar u u - \bar d d 
| N \rangle
,
\end{equation}
where $m_*$ is defined in Eq. (\ref{eq:mstar}).
A more refined calculation in chiral EFT yields \cite{devriessplitting}
\begin{equation}
\bar g_{\pi NN}^{(0)}
=
(15.5 \pm 2.5) \times 10^{-3} \bar \theta
.
\label{eq:g0theta}
\end{equation}
The $\theta$-term contribution to the isovector coupling was also estimated in Ref. \cite{bsaisou1}, as
\begin{equation}
\bar g_{\pi NN}^{(1)}
=
-(3.4 \pm 2.4) \times 10^{-3} \bar \theta
.
\end{equation}
The large error bar is due to the uncertainty of the low energy constants as well as higher order corrections.
It can be noted that the $\bar \theta$ contribution to the CP-odd $\pi$-N-N couplings is important even in the case when the 
Peccei-Quinn mechanism is active, since the quark chromo-EDM induces $\bar \theta_{\rm ind}$ [see Eq. (\ref{eq:inducedthetacedm})].

Let us now see the contribution of the quark chromo-EDM to the CP-odd $\pi$-N-N couplings.
The chromo-EDM contributes to the CP-odd $\pi$-N-N interaction through two leading processes.
The first one is the short distance contribution, which can be obtained by applying the PCAC relation to the 
$\pi$-N-N matrix element
\begin{equation}
\langle B_a \pi^c | {\cal L}_{\rm cEDM} | B_b \rangle
\approx
\frac{d_q^c}{f_\pi}
\langle B_a | \bar q g_s \sigma_{\mu \nu} G^{\mu \nu}_a t_a T_c  q | B_b \rangle
.
\end{equation}
The nucleon matrix element in the right-hand side of the above equation cannot be reduced further, and we have to quote the result of 
calculations using phenomenological models. Here we use the result of QCD sum rules \cite{pospelovreview,pospelov1,fuyuto}:
\begin{eqnarray}
D_u 
\equiv
\langle p | \bar u g_s \sigma_{\mu \nu} G^{\mu \nu}_a t_a  u | p \rangle
&=&
-0.26 \, {\rm GeV}^2
,
\\
D_d 
\equiv
\langle p | \bar d g_s \sigma_{\mu \nu} G^{\mu \nu}_a t_a  d | p \rangle
&=&
-0.17 \, {\rm GeV}^2
,
\end{eqnarray}
at the renormalization point $\mu =  1$ GeV.
These matrix elements have an uncontrolled systematic uncertainty which is certainly not less than $O(100\%)$.

The quark chromo-EDM also contributes to the CP-odd $\pi$-N-N interaction through the pion pole \cite{dashencp}.
Combining the second order term of the chiral Lagrangian with the pion tadpole generated by the chromo-EDM, 
we have
\begin{eqnarray}
\bar g_{\pi NN}^{(0)}
&\approx &
-\frac{D_u -D_d }{4 f_\pi} (d_u^c + d_d^c)
\nonumber\\
&&
-\frac{m_* }{2 f_\pi}
\langle p| \bar u u - \bar d d  | p \rangle
\Bigl[
2 \bar \theta + \frac{m_0^2}{2} \Bigl( \frac{m_u -m_d}{m_u m_d} (d_u^c - d_d^c) 
\nonumber \\ 
&& \hspace{10em}
+ \frac{d_u^c + d_d^c - 2 d_s^c}{m_s} \Bigr)
\Bigr]
, \ \ \ \ \ 
\\
\bar g_{\pi NN}^{(1)}
&\approx &
-\frac{d_u^c - d_d^c}{4 f_\pi} \Bigl[ 
D_u +D_d +m_* \langle p| \bar u u + \bar d d  | p \rangle m_0^2 
\Bigr]
,
\end{eqnarray}
where $m_* \equiv \frac{m_u m_d m_s}{m_u m_d + m_d m_s + m_u m_s}$.
Here we have also written the $\theta$-term contribution to take into account the induced $\theta$-term for the case where the Peccei-Quinn mechanism is active \cite{peccei}.
By substituting the induced $\theta$-term in Eq. (\ref{eq:inducedthetacedm}), the isoscalar CP-odd $\pi$-N-N  coupling becomes
\begin{equation}
\bar g_{\pi NN, {\rm PQ}}^{(0)}
\approx 
-\frac{d_u^c + d_d^c}{4 f_\pi} \Bigl[ 
D_u -D_d +m_* \langle p| \bar u u - \bar d d  | p \rangle m_0^2 
\Bigr]
.
\end{equation}

The total quark chromo-EDM contribution to $\bar g_{\pi NN, {\rm PQ}}^{(0)} $ and $\bar g_{\pi NN}^{(1)} $ is then \cite{pospelovreview,pospelov1}
\begin{eqnarray}
\bar g_{\pi NN, {\rm PQ}}^{(0)} (d^c_q)
&=&
\tilde \omega^{(0)}_{\rm PQ} \frac{d^c_u + d_d^c}{10^{-26}{\rm cm}}
,
\\
\bar g_{\pi NN}^{(1)} (d^c_q)
&=&
\tilde \omega^{(1)} \frac{d^c_u - d_d^c}{10^{-26}{\rm cm}}
,
\label{eq:g1cedm}
\end{eqnarray}
with
\begin{eqnarray}
\tilde \omega^{(0)}_{\rm PQ}
&=&
-6.9\times 10^{-13} \times \frac{|\langle 0 |\bar qq|0\rangle | }{(265\, {\rm MeV})^3}
\frac{|m_0^2|}{0.8 \, {\rm GeV}^2}
,
\label{eq:cEDM_gpinn0_coef}
\\
\tilde \omega^{(1)} 
&=&
-1.0 \times 10^{-11} \times \frac{|\langle 0 |\bar qq|0\rangle | }{(265\, {\rm MeV})^3}
\frac{|m_0^2|}{0.8 \, {\rm GeV}^2}
.
\label{eq:cEDM_gpinn1_coef}
\end{eqnarray}
Here the coefficient of the isovector component of the quark chromo-EDM is enhanced, due to the large value of the isoscalar nucleon scalar density $\langle p| \bar u u + \bar d d  | p \rangle$ ($ \sim O(10)$ at $\mu = 2$ GeV) [see Eq. (\ref{eq:isoscalarnucleonscalardensity})].
It is also important to note that the dominant contribution comes from the pion pole effect (or the vacuum alignment).
The short distance effect, due to the matrix elements $\langle N | \bar q g_s \sigma_{\mu \nu} G^{\mu \nu}_a t_a  q | N \rangle$, is less than 20\% for $\bar g_{\pi NN, {\rm PQ}}^{(0)}$, and less than 10\% for $\bar g_{\pi NN}^{(1)}$.
From this fact, we can estimate the theoretical uncertainty of the quark chromo-EDM contribution to the CP-odd $\pi$-N-N interaction.
The pion pole effect depends on the light quark mass, scalar densities and the mixed condensate $m_0^2$, which have all sizable error bars. 
The largest should be the light quark masses, which is about $O(30\%)$.
We also have keep in mind that the above analysis was performed at the leading order of the 
chiral expansion. There are also substantial uncertainties due to the unknown higher order contributions, which are expected to be quantifiable in the next-to-leading order analysis \cite{devriessplitting,senghigherorder}.
Being optimistic, the error bars of the coefficients given by Eqs. (\ref{eq:cEDM_gpinn0_coef}) 
and (\ref{eq:cEDM_gpinn1_coef}) are expected to be about 50\%.

The last important contribution to the CP-odd $\pi$-N-N interaction to be investigated is the
P,CP-odd 4-quark interactions. Here let us see the effect of the left-right four-quark interaction
\begin{eqnarray}
{\cal L}'_{LR}
&=&
C_{LR1}' \bar d\gamma^\mu P_L u \, \bar u\gamma^\mu P_R d
+C_{LR2}' \bar d_\alpha \gamma^\mu P_L u_\beta \, \bar u_\beta \gamma^\mu P_R d_\alpha 
\nonumber\\
&&
+ {\rm h.c.}
,
\label{eq:lr2}
\end{eqnarray}
which gives the leading contribution to the isovector $\pi$-N-N interaction ($\bar g_{\pi NN}^{(1)}$).
Here we use the operator basis of Ref. \cite{cirigliano}, where it is pointed that the isovector CP-odd pion-nucleon coupling is generated by the vacuum alignment effect (pion pole effect).
We have
\begin{eqnarray}
\bar g_{\pi NN}^{(1)}
&\approx &
-{\rm Im} (C'_{LR1}) \times 0.63 \, {\rm GeV}^2
\nonumber\\
&&
-{\rm Im} (C'_{LR2})\times 2.9 \, {\rm GeV}^2
,
\label{eq:g1lrpionpole}
\end{eqnarray}
where the renormalization scale is set to $\mu = 3$ GeV.
Here we note that the direct contribution to the CP-odd pion-nucleon vertex, which is given by the short distance physics, is not considered.

The direct contribution is not easy to evaluate, and we have to use model calculations which will give us the theoretical uncertainty.
In the estimation of the direct contribution of the CP-odd four-quark interaction, the vacuum saturation approximation is often used \cite{4-quark5,4-quark1,4-quark2,4-quark3,4-quark4}. 
The contribution of the left-right four-quark interaction [see Eq. (\ref{eq:4-q_lagrangian})] to the isovector CP-odd $\pi$-N-N coupling can be calculated as
\begin{eqnarray}
\bar g_{\pi NN}^{(1)} 
&=& 
\langle \pi^0 N | {\cal L}'_{LR} | N \rangle
\nonumber\\
&\approx& 
\Bigl[
\frac{1}{3}{\rm Im} (C'_{LR1}) 
+\frac{1}{2}{\rm Im} (C'_{LR2}) 
\Bigr]
\nonumber\\
&&
\times
\langle \pi^0 N | \bar q i\gamma_5 \tau_z q \, (\bar u u +\bar d d) | N \rangle \nonumber\\
&\approx& 
\Bigl[
\frac{1}{3}{\rm Im} (C'_{LR1}) 
+\frac{1}{2}{\rm Im} (C'_{LR2}) 
\Bigr]
\nonumber\\
&&
\times
\langle \pi^0 | \bar q i\gamma_5 \tau_z q | 0 \rangle \langle N | \bar u u + \bar dd | N \rangle \nonumber\\
&=& 
-
\Bigl[
\frac{2}{3}{\rm Im} (C'_{LR1}) 
+{\rm Im} (C'_{LR2}) 
\Bigr] 
\nonumber\\
&&
\times
\frac{\langle 0 | \bar qq | 0 \rangle}{f_\pi} \langle N | \bar u u +\bar d d | N \rangle
\nonumber\\
&=& 
{\rm Im} (C'_{LR1}) \times 4 \, {\rm GeV}^2
\nonumber\\
&&
+{\rm Im} (C'_{LR2}) \times 5 \, {\rm GeV}^2
, \ \ \ \ \ 
\end{eqnarray}
where we have used the PCAC assertion in the fourth equality, and
$\langle 0 | \bar qq | 0 \rangle \approx -\frac{m_\pi^2 f_\pi^2}{m_u + m_d}$.
Here it is important to remark that 
the direct contribution 
is enhanced by the scalar density of the light quarks $\langle N | \bar u u +\bar d d | N \rangle \sim O(10)$, and is as large as that of the pion pole (\ref{eq:g1lrpionpole}).
The renormalization scale of $\langle 0 | \bar qq | 0 \rangle$ and $\langle N | \bar qq | N \rangle$ is taken as $\mu = 3$ GeV.
This feature is suggested by the model analysis where the scalar density is enhanced by the constructive interference due to dynamical gluon emissions and absorptions \cite{yamanakasde2}.
In the large $N_c$ analysis, the error bar of the direct contribution is estimated as $O(100\%)$ due to the presence of baryons.
In addition to the model dependence, higher order correction due to the three-pion interaction (\ref{eq:three-pion}), which is known to contribute to $\bar g_{\pi NN}^{(1)}$, may also be sizable \cite{dekens,devriessplitting,senghigherorder}.
We therefore conclude that the theoretical uncertainty of the left-right four-quark interaction to $\bar g_{\pi NN}^{(1)}$ is $O(100\%)$.
For the strange quark contribution, the systematics due to the direct contribution is smaller, due to the small value of $\langle N | \bar ss | N \rangle \sim O(0.1)$ [see Eq. (\ref{eq:strangecontent}) and discussion below it], and the pion pole contribution is the dominant process.

We have to note that the contribution of the Weinberg operator to the CP-odd $\pi$-N-N interaction (\ref{eq:pcpvpinnint}) is suppressed by at least a factor of light quark, due to its chiral symmetry breaking nature.

The SM contribution generated by the CP phase of the CKM matrix can be estimated by using the factorization of $|\Delta S|=1$ four-quark operators.
By combining the hyperon-nucleon transition and the $|\Delta S|=1$ meson-baryon interaction, we can calculate $\bar g_{\pi NN}^{(0)}$ and $\bar g_{\pi NN}^{(1)}$.
The result is of the order of $\bar g_{\pi NN, {\rm SM}}^{(0)} \sim \bar g_{\pi NN, {\rm SM}}^{(1)} \sim 10^{-17}$ \cite{khriplovich,smnuclearedm,sushkov,Flambaum86,hesmedm,smdeuteronedm}.
This estimation also involves a theoretical uncertainty of $O(100\%)$, due to the gluonic correction in the $1/N_c$ expansion.

\subsection{\label{sec:nedm}The nucleon EDM}

\begin{figure}
\begin{center}
\resizebox{6.0cm}{!}{\includegraphics{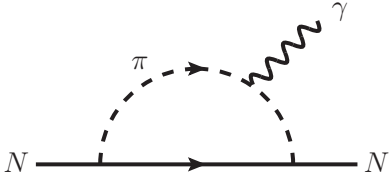}}
\caption{\label{fig:chiral_loop} 
Diagrammatic representation of the meson-loop contribution to the nucleon EDM.
The solid line represents a baryon, the dashed line a light pseudoscalar meson, and the wavy line a photon.
}
\end{center}
\end{figure}

The nucleon EDM receives the leading order contribution from the quark EDM, the quark chromo-EDM, and the Weinberg
operator, without suppression by the light quark mass:\footnote{It is to be noted that the Left-right four-quark interaction also sizably contributes to the nucleon EDM through the induced $\theta$-term when the PQ mechanism is active \cite{LRLHC2} (see Section \ref{sec:theta}).
}
\begin{equation}
d_N
=
d_N (d_q) + d_N (d_q^c) + d_N (w)
,
\end{equation}
where $N=p,n$.
The quark EDM contribution to the nucleon EDM $d_N (d_q)$ is simply given by the proton tensor charges (\ref{eq:nucleon_tensor_charge}), as
\begin{eqnarray}
d_p (d_q)
&=& 
\delta u \, d_u +\delta d \, d_d  
+ \delta s \, d_s 
,
\label{eq:qedm_proton}
\\
d_n (d_q) 
&=& 
\delta d \, d_u +\delta u \, d_d  
+ \delta s \, d_s 
,
\label{eq:qedm_neutron}
\end{eqnarray}
where we have assumed the isospin symmetry.
Note that the quark EDM can only be probed with the nucleon EDM.
It therefore plays an important role in probing the CP violation of several scenarios such as the split SUSY (see Sub-sec. \ref{sec:susy}).

The quark chromo-EDM contribution to the nucleon EDM $d_N (d_q^c)$ can be estimated in the chiral approach.
As we have seen in the previous section, the quark chromo-EDM generates the CP-odd $\pi$-N-N interaction.
We can therefore infer that the most important part of $d_N (d_q^c)$ is given by the long distance effect, the meson cloud diagram of Fig. \ref{fig:chiral_loop}.
The isoscalar and isovector nucleon EDMs defined by $d_0 \equiv d_p +d_n $ and $d_1 \equiv d_p - d_n$, respectively, are given in the leading order chiral perturbation by the following formulae \cite{hisanoshimizu2,faessler2,yamanakabook,linearprogramming,eft6dim,sengisovector,crewther,borasoy,ottnad,mereghetti2,mereghetti1,guo1,devriessplitting,devriesreview,fuyuto,Khatsymovsky1,devries1}
\begin{equation}
d_0
=
\bar d_0
- \frac{e g_A \bar g^{(0)}_{\pi NN} }{4 \pi f_\pi} \Biggl(
\frac{3 m_\pi}{4 m_N} 
\Biggr)
- \frac{e g_A \bar g^{(1)}_{\pi NN} }{16 \pi f_\pi} \frac{m_\pi}{m_N}
,
\label{eq:d_0}
\end{equation}
and 
\begin{eqnarray}
d_1
&=&
\bar d_1 
- \frac{e g_A \bar g^{(0)}_{\pi NN} }{4 \pi^2 f_\pi} \Biggl(
\frac{2}{4-d} - \gamma_E+ \ln \frac{4 \pi \mu^2}{m_\pi^2}
+\frac{5 \pi m_\pi}{4 m_N} 
\Biggr)
\nonumber\\
&&
- \frac{e g_A \bar g^{(1)}_{\pi NN} }{16 \pi f_\pi} \frac{m_\pi}{m_N}
,
\label{eq:d_1}
\end{eqnarray}
where the isovector axial coupling is given by $g_A = 1.27$ \cite{ucna}.
The low energy constants $\bar d_0$ and $\bar d_1$ are the counter terms of the one-loop level diagram (Fig. \ref{fig:chiral_loop}), and 
include the short distance effect which does not come from the meson cloud.
Roughly, they originate from short distance effect (shorter than the renormalization scale, $\mu =1$ GeV in our case). 
Let us show the quark chromo-EDM contribution to the nucleon EDM by neglecting $\bar d_n$ and $\bar d_p$.
The leading order chiral analysis of the nucleon EDM generated by the quark chromo-EDM, taking into account the effect of hadrons with strange quark, was done in Ref. \cite{fuyuto}.
These results are given by
\begin{eqnarray}
d_n ( d_q^c) 
&=&
e\tilde \rho_n^u d_u^c
+e\tilde \rho_n^d d_d^c
+e\tilde \rho_n^s d_s^c
,
\\
d_p ( d_q^c) 
&=&
e\tilde \rho_p^u d_u^c
+e\tilde \rho_p^d d_d^c
+e\tilde \rho_p^s d_s^c
,
\end{eqnarray}
so that we have $\tilde \rho_n^u \approx -0.76$, $\tilde \rho_n^d \approx -0.17$, $\tilde \rho_n^s \approx 0.55$, $\tilde \rho_p^u \approx -0.026$, $\tilde \rho_p^d \approx -1.1$, and $\tilde \rho_p^s \approx 1.3$ when there is no Peccei-Quinn mechanism, and
$\tilde \rho_n^u \approx 1.5$, $\tilde \rho_n^d \approx 0.93$, $\tilde \rho_n^s \approx 0.60$, $\tilde \rho_p^u \approx 0.37$, $\tilde \rho_p^d \approx -0.93$, and $\tilde \rho_p^s \approx 1.3$ when the Peccei-Quinn mechanism is active. 
The neutron EDM was also evaluated using QCD sum rules \cite{pospelovchromoedm,hisanochromoedm}, giving smaller results.
It is possible that the QCD sum rules approach could not take into account the long distance physics due to the pion loop which enhances the nucleon EDM.
On the contrary, the QCD sum rules can quantify the short distance physics which is in principle impossible to treat in the calculation using the effective CP-odd meson-nucleon interaction without the knowledge of the low energy constants.
It is to be noted that the above result may be affected by a sizable theoretical uncertainty due to the effect of higher order corrections \cite{devriessplitting,senghigherorder}.
The ideal way to obtain $d_N (d_q^c)$ is to evaluate it on lattice.
There are currently continuous efforts to achieve this goal \cite{bhattacharya4,bhattacharya5,bhattacharya6,Abramczyk}.

The final important process contributing to the nucleon EDM to be discussed is the Weinberg operator.
The derivation is based on the CP violating rotation of the nucleon state evaluated using the QCD sum rules \cite{bigi,pospelovweinbergop}.
By quoting the calculation using the QCD sum rules, the nucleon EDM generated by the Weinberg operator constant $w$ is 
\begin{eqnarray}
d_N (w) 
&\sim &
\frac{a_N}{2 m_N}  w \frac{3g_s m_0^2}{32 \pi^2 } \ln \frac{M_b^2}{\mu_{\rm IR}^2}
\nonumber\\
&\approx&
\left\{
\begin{array}{rl}
- w \times 20 \, e \, {\rm MeV} & (N = n ) \cr
w \times 5 \, e \, {\rm MeV} & (N = p ) \cr
\end{array}
\right.
,
\label{eq:weinbergop_gpinn}
\end{eqnarray}
where $\frac{M_b^2}{\mu_{\rm IR}^2} = 2$ and $g_s = 2.1$.
The anomalous magnetic moment ($g-2$) of the nucleon $N$ is given by $a_N$ ($a_n =$ -3.91, $a_p =$ 0.79).
As we have seen previously, the Weinberg operator also induces a $\theta$-term when the Peccei-Quinn mechanism is relevant, but its contribution is suppressed by a factor of light quark mass.
The theoretical uncertainty of Eq. (\ref{eq:weinbergop_gpinn}) is large due to the model dependence, and certainly exceeds $O(100\%)$.

As mentioned in Sec. \ref{sec:smcpv}, the nucleon EDM in the SM is of order $O(10^{-(31 - 32)})e$ cm.
It is estimated by the long distance effect generated by the chiral loop diagram (see Fig. \ref{fig:chiral_loop}) with $|\Delta S|=1$ 
interactions \cite{ellissmedm,nanopoulossmedm,Deshpande,gavelasmedm,smneutronedmkhriplovich,eeg,hamzaouismedm,smneutronedmmckellar,seng}.
A loopless process with high mass dimension operator was also pointed out to contribute to the nucleon EDM with the same order of magnitude \cite{mannel}.
Like the other CP violating processes, the nucleon EDM in the SM is much smaller than that generated by typical models of new physics
with TeV scale parameters.

\subsection{CP-odd nuclear force}

The CP-odd nuclear force is the leading CP violating process in generating CP-odd nuclear moments.
The most general CP-odd nuclear force is given by 
\begin{eqnarray}
H_{P\hspace{-.5em}/\, T\hspace{-.5em}/\, }
& = &
\big \{  
\vec{\sigma}_{-} V_1 (r)
+{\bf \tau}_{1} \cdot {\bf \tau}_{2}\, \vec{\sigma}_{-} V_2 (r)
\nonumber \\ 
&& \hspace{1em}
+\frac{1}{2} ( \tau_{+}^{z}\, \vec{\sigma}_{-} +\tau_{-}^{z}\,\vec{\sigma}_{+} ) V_3 (r)
+\frac{1}{2} ( \tau_{+}^{z}\, \vec{\sigma}_{-} -\tau_{-}^{z}\,\vec{\sigma}_{+} ) V_4 (r)
\nonumber \\ 
&& \hspace{1em}
+(3\tau_{1}^{z}\tau_{2}^{z}- {\bf \tau}_{1}\cdot {\bf \tau}_{2})\,\vec{\sigma}_{-} V_5 (r)
\big \}
\cdot
\hat{\vec{r}} \,
,
\label{eq:CP-odd_nuclear_force}
\end{eqnarray}
where $\hat{\vec{r}} \equiv \frac{\vec{r}_1 - \vec{r}_2}{|\vec{r}_1 - \vec{r}_2|}$ with $\vec{r}_1$ and $\vec{r}_2$ are the coordinates of the interacting two nucleons.
The spin and isospin matrices are given by ${\vec{\sigma}}_{-} \equiv \vec{ \sigma}_1 -\vec{\sigma}_2$, $\vec{\sigma}_{+} \equiv \vec{\sigma}_1 + \vec{\sigma}_2$, 
${\bf \tau}_{-} \equiv {\bf \tau}_1 -{\bf \tau}_2$, and ${\bf \tau}_{+} \equiv {\bf \tau}_1 + {\bf \tau}_2$.
As we can see, the CP-odd nuclear force is a spin dependent interaction, so the CP-odd nuclear polarization arises only for systems with nonzero angular momenta.

\begin{figure}
\begin{center}
\resizebox{5.0cm}{!}{\includegraphics{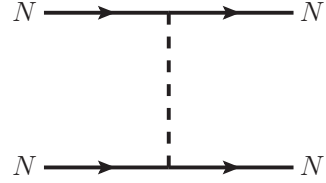}}
\caption{\label{fig:CPVNN} 
Diagrammatic representation of one-meson exchange CP-odd nuclear force.
The solid line represents the nucleon and the dashed line a light pseudoscalar meson ($\pi ,\eta$).
}
\end{center}
\end{figure}

At the scale of nuclear physics, with the cutoff scale $\mu = 500$ MeV, the pion exchange CP-odd nuclear force provides the leading contribution to the CP-odd nuclear moment.
At the leading order, the CP-odd nuclear force is a one-pion exchange process made by combining the CP-even and CP-odd pion-nucleon interactions (see Fig. \ref{fig:CPVNN}).
Its nonrelativistic potential is given by \cite{haxton,pvcpvhamiltonian2,pvcpvhamiltonian3}
\begin{eqnarray}
V_2^\pi (r) \hat{\vec{r}}
& = &
-\frac{g_{\pi NN} \bar g^{(0)}_{\pi NN} }{2 m_N}
\nabla {\cal Y} (m_\pi , r)
,
\label{eq:g0NN}
\\
V_3^\pi (r) \hat{\vec{r}}
& = &
-\frac{g_{\pi NN} \bar g^{(1)}_{\pi NN} }{2 m_N}
\nabla {\cal Y} (m_\pi , r)
,
\label{eq:g1NN}
\\ 
V_5^\pi (r) \hat{\vec{r}}
& = &
\frac{g_{\pi NN} \bar g^{(2)}_{\pi NN} }{2 m_N}
\nabla {\cal Y} (m_\pi , r)
,
\label{eq:g2NN}
\end{eqnarray}
where ${\cal Y} (m_\pi , r) \equiv \frac{e^{-m_\pi  r}}{4 \pi r}$, and $g_{\pi NN} \equiv \frac{g_A m_N}{f_\pi}$.
Note the sign change for the isoscalar and isovector couplings which is due to the difference of conventions \cite{pospelovreview,korkin}.

We should also present some subleading processes.
The first one is the contact interaction [terms with $\bar C_1$ and $\bar C_2$ of Eq. (\ref{eq:cpv_contact})], which can be written as
\begin{eqnarray}
V_1^C (r) \hat{\vec{r}}
& = &
-m_N \bar C_1 \nabla \delta ( r)
,
\\
V_2^C (r) \hat{\vec{r}}
& = &
-m_N \bar C_2 \nabla \delta ( r)
,
\end{eqnarray}
where the delta function is valid up to the cutoff (renormalization) scale.
In practice, it is smeared with a Gaussian or a Yukawa function with the cutoff scale as their range.
An example of the effect contributing to the contact interaction is the CP-odd nuclear force with $\eta$ meson exchange (see Fig. \ref{fig:CPVNN}).
It can be matched with the isoscalar CP-odd N-N interaction as \cite{yamanakareview}
\begin{eqnarray}
g_{\eta NN} \bar g_{\eta NN}^{(0)}
&\approx &
-m_N m_\eta^2 \bar C_1
,
\label{eq:contact1eta}
\end{eqnarray}
where $g_{\eta NN} = 2.24$ \cite{tiator} is the CP-even $\eta$-nucleon coupling.
Quantifying the effect of the CP-odd contact interaction is potentially important since it receives contributions from the Weinberg operator.
Unfortunately, there are currently no hadron level evaluations available.

Another possible subleading contribution is the CP-odd three-nucleon interaction \cite{eft6dim}.
This interaction is generated by the three-pion interaction [terms with $\Delta$ of Eq. (\ref{eq:three-pion})].
This process however interacts with the spin and isospin of all the three relevant nucleons.
This kind of configurations is suppressed in nuclei due to the pairing of nucleons, so the effect of the CP-odd three-nucleon interaction is not important.
Neglecting this interaction should therefore be a good approximation.

Before going to the nuclear physics, we have to point out that the renormalization scale where the CP-odd nuclear force Eq. (\ref{eq:CP-odd_nuclear_force}) was defined is not the same as that used in the calculational approach adopted for heavy nuclei.
If we respect the ``bare'' CP-odd nuclear force, the model space of many-body nuclear systems becomes too large, with exponentially growing calculational cost.
To treat the CP-odd nuclear force in some many-body approaches, we actually have to construct an effective theory for heavy nuclei, respecting the model space.
We can expect that the long range pion exchange is not much affected by this change of model space, since the pion exchange is the most relevant interaction in low energy nuclear physics.
We must however note that the contact interaction which contains the short range physics suffers from the renormalization in the reduction of the model space, required in the construction of the effective interaction in heavy nuclei.

In the context of the change of model space, a notable CP-odd contribution is the isotensor CP-odd nuclear force.
At the scale of the hadron level effective theory [renormalization scale of Eq. (\ref{eq:CP-odd_nuclear_force}), e.g. $\mu = 500$ MeV], this interaction is suppressed by the isospin splitting of the quark mass.
The situation may however change in heavy nuclei, where the CP-odd nuclear force must be renormalized into an effective theory with a reduced model space.
As heavy nuclei have a medium with large isospin breaking, they certainly induce an isotensor CP-odd nuclear force through the renormalization of isovector CP-odd interaction.
There are currently no evaluations of the effective CP-odd nuclear forces for heavy nuclei, but this is an important subject to be discussed in the future.
At the same time, the evaluation of the effect of the isotensor CP-odd nuclear force to the CP-odd nuclear moments are almost mandatory in quantifying the EDM of heavy atoms.

The leading SM contribution to the CP-odd nuclear force is generated by the pion-exchange process, which is just the combination of the 
CP-even and CP-odd pion-nucleon interactions (see Sec. \ref{sec:pion-nucleon}). In SM, we can also consider additional effects due to the exchange of $K$ and $\eta$ mesons, which contribute to all terms of Eq. (\ref{eq:CP-odd_nuclear_force}) \cite{smnuclearedm}.

\section{Nuclear structure calculation}
\label{Sec:Nuclear structure calculation}

In this section we first give the definition of the NSM and then review 
how it is evaluated in framework of the nuclear shell model 
and other calculations based on the mean 
field theories.

\subsection{Definition of NSM}

The effective electric field $\Phi (\vec {r})$ which an electron at position $\vec {r}$ feels 
can be expressed as \cite{Senkov2008}
\begin{eqnarray}
\left\langle \Psi  \right|e \Phi (\vec {r})-\frac{1}{Z}\left\langle {{\rm 
{\bf \vec {d}}}_N } \right\rangle \cdot \vec {\nabla }\Phi (\vec {r})\left| 
\Psi \right\rangle  \nonumber \\
= -\frac{Ze^2}{\vert r\vert }+4\pi e{\rm {\bf \vec {\rm S}}}
\cdot \vec {\nabla }\delta ({\rm {\bf \vec {r}}})+\cdot \cdot \cdot ,
\label{Yeq1}
\end{eqnarray}
where $\left| \Psi  \right\rangle $ indicates the nuclear ground state, 
$\left\langle {{\rm {\bf \vec {d}}}_N } \right\rangle $ is the nuclear 
electric dipole moment of the nuclear ground state and $Z$ is the number of protons. 
Here the $k$th component $\left( {k=x,y,z} \right)$ of the NSM 
$S_k$ may be given as 
\begin{equation}
S_k =S_{\rm{ch},k} + S_{\rm{int},k} .
\end{equation}
The $ S_{\rm{ch},k} $  is caused 
by the charge asymmetry of a nucleus and is given as
\label{Yeq2}
\begin{equation}
{S}_{\rm{ch},k} =\frac{1}{10}\int {\left( {r^2r_k -\frac{5}{3}\left\langle 
{r^2} \right\rangle _{\rm{ch}} r_k -\frac{2}{3}\left\langle {Q_{kk'} } 
\right\rangle r_{k'} } \right)} \rho (\vec {r})d\vec {r},
\label{Yeq3}
\end{equation}
where $r_k$ represents position of a nucleon, $\rho 
(\vec {r})$ nuclear charge density, $\left\langle {Q_{kk'} } \right\rangle $ 
nuclear quadrupole moment of the nuclear ground state, 
which can be dropped for the spin $I=1/2$ nuclear state.
$\left\langle {r^2} \right\rangle _{\rm{ch}} $ is the
charge mean square radius. 

Then the NSM is
defined as the maximum projection
of the NSM operator on the nuclear axis, $S= < \hat {S}_z >$
and is calculated as 
\begin{equation}
S_{\rm{ch}} =\left\langle {\Psi \left| {\hat {S}_{\rm{ch},z} } \right|\Psi } , 
\label{Yeq11}
\right\rangle ,
\end{equation}
where $\left| \Psi \right\rangle $ is the $PT$-violating ground state,
which is usually evaluated by the nuclear mean field theories.
The operator $\hat{\bf{S}}_{\rm ch}$ is expressed in terms
of nucleon's degree of freedom as 
\begin{equation}
\hat{\bf{S}}_{\rm ch} =\frac{1}{10}\sum\limits_{i=1}^A e_i 
\left( r_i^2 -\frac{5}{3}\left\langle r^2 
\right\rangle _{\rm ch} \right) \boldsymbol{r}_i .
\end{equation}
Here $A$ is the mass number of a specific nucleus, and 
$e_i$ is the charge for the $i$th nucleon. 
We take $e_i =0$ for a neutron and $e_i =e$ for a proton. 
As inclusion of the relativistic effects, Flambaum et al 
had pointed out the contribution to 
the NSM operator as \cite{Flambaum12}
\begin{eqnarray}
{\bf S}' = \frac{Ze}{10}\frac{1}{1-\frac{5}{14}Z^2\alpha^2}
\left\{ \left[ \langle {\bf r}r^2 \rangle - \frac{5}{3} \langle {\bf r} 
\rangle \langle r^2 \rangle - \frac{2}{3} \langle r_i \rangle \langle q_{ij} \rangle \right] \right.\nonumber\\
\left. -\frac{5}{28}\frac{Z^2 \alpha^2}{R^2_N} \left[ \langle {\bf r}r^4 \rangle - \frac{7}{3}\langle {\bf r} \rangle \langle r^4 \rangle - \frac{2}{3}\langle r_i \rangle \langle q_{ij} r^2\rangle \right] \right\}, \nonumber\\
\end{eqnarray}
where $R_N$ is the nuclear radius, $q_{ij}$ is the nuclear quadrupole moment operator
and $Z$ is the atomic number of the nucleus.

If P, T-odd, which is equivalent to P, CP-odd,
interaction $ \hat {V}_{\pi (T)}^{PT} $ exists 
in the total Hamiltonian,
we have
\begin{eqnarray}
H = H_0 + \hat {V}_{\pi (T)}^{PT}, 
\label{Yeq5}
\end{eqnarray}
where $H_0$ does not break P and T. 
Here $\hat {V}^{\pi}_{PT} $ is the pion-exchange CP-odd nuclear force, given by the sum of the isoscalar [$T=0$ , defined in Eq. (\ref{eq:g0NN})], isovector [$T=1$ , defined in Eq. (\ref{eq:g1NN})] and isotensor [$T=2$ , defined in Eq. (\ref{eq:g2NN})] terms.
The coupling constants can be rewritten as
\begin{eqnarray}
\bar{g}^{(0)} g 
&=&
-g_{\pi NN} \bar{g}_{\pi NN}^{(0)} , \\
\bar{g}^{(1)} g 
&=&
-g_{\pi NN} \bar{g}_{\pi NN}^{(1)} , \\
\bar{g}^{(2)} g 
&=&
g_{\pi NN} \bar{g}_{\pi NN}^{(2)}  ,
\end{eqnarray}
to respect the convention often adopted in the nuclear structure calculations.
We note again that the bare CP-odd nuclear force obtained at the hadron scale and that used in the calculations of the CP-odd moments of heavy nuclei (that of this section) are not the same, 
since the model spaces where they are defined are different.
The nuclear forces in the media are usually calculated in terms of 
the Brueckner-Bethe-Goldstone (BBG) many-body theory~\cite{Baldo99}. 
Another complexity comes in from the shell-model calculation. 
In the shell-model calculation
we divide the model space into the core and valence spaces. 
The effective interaction in the valence shell
should be modified by taking into account the core excitations.
Since the treatment of both of them is rather involved, 
we do not take into account these effects in this paper.

Since $\hat {V}_{\pi (T)}^{PT}$ is very weak,
the NSM which is P,T-odd is calculated perturbatively as
\begin{equation}
\label{Yeq6}
S_{\mbox{ch}} =\sum\limits_{k=1} {\frac{\left\langle {I_1^+ \left| {\hat 
{S}_{ch,z} } \right|} \right.\left. {I_k^- } \right\rangle \left\langle 
{I_k^- \left| {\hat {V}_{\pi (T)}^{PT} } \right|} \right.\left. {I_1^+ } 
\right\rangle }{E_1^+ -E_k^- }} +c.c.
\end{equation}
 Here $\left| 
{I_1^+ } \right\rangle $ represents the lowest state with spin $I$ and 
positive parity (parity of the nuclear ground state is assumed to positive here)  
and $\left| {I_k^- } \right\rangle  $, the $k$th state with spin 
$I$ and negative parity. The energy $E_k^\pi $ of the $k$th state  with parity 
$\pi $, is obtained by diagonalizing the original shell model 
Hamiltonian $H_0$. i.e. 
\begin{eqnarray}
H_0 \left| {\left. {I_k^\pi } \right\rangle =} \right.E_k^\pi 
\left| {\left. {I_k^\pi } \right\rangle } \right. . \nonumber
\end{eqnarray}

Then the NSM is expressed in terms of  $\bar {g}^{(T)}$ as
\begin{equation}
\label{Yeq9}
S_{ch} =a_0 \mbox{ }\bar {g}^{(0)}g\,+a_1 \mbox{ }\bar {g}^{(1)}g+a_2 \mbox{ 
}\bar {g}^{(2)}g,
\end{equation}
where coefficients $a_T$ with $T=1,2,3$ in front of $\mbox{ }\bar 
{g}^{(T)}g\,$'s are given in units of $e$fm$^{3}$. 
Coefficients $a_T $ are tabulated in the following.

Another contribution to ${\rm {\bf \hat {S}}}_{\rm int} $ coming from the nucleon 
intrinsic EDM is given by as \cite{Dmitriev03}
\begin{equation}
{\rm {\bf \hat {S}}}_{\rm int} =\frac{1}{6}\sum\limits_{i=1}^A {{\rm {\bf d}}_i 
\left( {r_i^2 -\left\langle {r^2} \right\rangle _{ch} } \right)} 
+\frac{1}{5}\sum\limits_{i=1}^A {\left[ {{\rm {\bf r}}_i \left( {{\rm {\bf 
r}}_i \cdot {\rm {\bf d}}_i } \right)-{\rm {\bf d}}_i r_i^2 /3} \right]} ,
\label{Yeq4}
\end{equation}
where ${\rm {\bf d}_i}$ is the $i$th nucleon intrinsic dipole moment (either proton or neutron).
Then the intrinsic component of NSM is expressed as
\begin{equation}
\label{Yeq14}
S_{\rm int}=s_p d_p +s_n d_n, 
\end{equation}
where $d_p $ and $d_n $ are EDMs of the proton and the 
neutron, respectively. Here $s_p$ and $s_n$ are the unknown coefficients 
that have to be calculated using nuclear many-body methods.

\subsection{Evaluation of NSMs}
\subsubsection{Simple Shell model approaches}

As a simple shell model estimate, an odd-mass nucleus is expressed as
a one-particle (either neutron or proton) plus the core (even-even part
of the nucleus). 
In Ref. \cite{Dmitriev05} NSMs
were calculated for a set 
of nuclei ($^{199}$Hg, $^{129}$Xe, $^{211}$Rn, $^{213}$Ra, $^{225}$Ra, $^{133}$Cs, and $^{223}$Fr) 
with full account of core polarization effects (namely from the even-even part of the nucleus). 
Their results are given in Table~\ref{NTab01} without core polarization
and in Table~\ref{YTab01} with core polarization effects.
The effects of core polarization
are found to have in general a large effect on the reduction of the Schiff moments (ten to hundred times reduction).
It is also found that
the dominant contribution comes from the isovector ($T = 1$) term for $^{199}$Hg.

\begin{table}
\caption{Coefficients $a_T$ in Ref.~\cite{Dmitriev05} in units of $e \rm{fm}^3$.
The bare values of the Schiff moment in Eq.~(\ref{Yeq9}), without core polarization, 
are calculated. Note that the sign of tensor type interaction is changed
from the original paper in accordance with the definition in the present paper.
}
\begin{center}
\begin{tabular}{cccc}\hline
                & $a_0$        & $a_1$    & $a_2$ \\\hline
$^{199}$Hg & $-$0.09    & $-$0.09   & $-$0.18 \\
$^{129}$Xe & 0.06         & 0.06        & 0.12 \\
$^{211}$Rn & $-$0.12    & $-$0.12   & $-$0.24  \\
$^{213}$Ra & $-$0.012   & $-$0.021 & $-$0.016 \\
$^{225}$Ra & 0.08         & 0.08        & 0.16\\
$^{133}$Cs & 0.08         & $-$0.02   & 0.21  \\
$^{223}$Fr  & $-$0.122  & $-$0.052  & $-$ 0.300      \\\hline
\end{tabular}
\end{center}  
\label{NTab01}
\end{table}  

\begin{table}
\caption{Coefficients $a_T$ in Ref.~\cite{Dmitriev05} with core polarization  in units of $e \rm{fm}^3$.}
\begin{center}
\begin{tabular}{cccc}\hline
                & $a_0$        & $a_1$      & $a_2$ \\\hline
$^{199}$Hg & $-$0.00004 & $-$0.055  & $-$0.009 \\
$^{129}$Xe & 0.008         & 0.006      & 0.009 \\
$^{211}$Rn & $-$0.019    & 0.061       & $-$0.053  \\
$^{213}$Ra & $-$0.012    & $-$0.021  & $-$0.016 \\
$^{225}$Ra & 0.033         & $-$0.037  & 0.046\\
$^{133}$Cs & 0.006         & $-$0.02   & 0.04  \\
$^{223}$Fr  & $-$0.009    & $-$0.016 & $-$ 0.030      \\\hline
\end{tabular}
\end{center}  
\label{YTab01}
\end{table}

In Ref. \cite{Engel03}, the Skyrme-Hartree-Fock method is used to 
calculate the  NSM for the octupole 
deformed nucleus $^{225}$Ra. $^{225}$Ra is known as a possible candidate 
which has a large Schiff moment. The first $I^\pi =\textstyle{1 \over 2}^-$ 
state is located at 55~keV above the ground state with spin $I_{g.s.}^\pi 
=\textstyle{1 \over 2}^+$ and the energy denominator in Eq. (\ref{Yeq6}) becomes 
large. 
As intermediate states in perturbation theory, they took only the first 
$I^\pi =\textstyle{1 \over 2}^-$ state with excitation energy of $\Delta 
E=55\mbox{ keV}$.

Then, to a very good approximation, we have
\begin{equation}
\label{Yeq16}
S =  
- \frac{{\left\langle {1/{2^ + }\left| {{{\hat S}_z}} 
\right|1/{2^ - }} \right\rangle \left\langle {1/{2^ - }
\left| {{{\hat V}^{PT}}} \right|1/{2^ + }} \right\rangle }} {{\Delta E}} + c.c.,
\end{equation}
This is further simplified as
\begin{equation}
S =  - 2\frac{J}{{J + 1}}\frac{{\left\langle {{{\hat S}_z}} \right\rangle 
\left\langle {{{\hat V}^{PT}}} \right\rangle }}{{\Delta E}},
\end{equation}
where $J=1/2$ and $ \left\langle {\hat S}_z \right\rangle $ and 
$ \left\langle {\hat V}^{PT} \right\rangle $ are expectation values
in terms of mean fields (intrinsic-states).
Their result is summarized in Table \ref{KTab01}.

\begin{table}
\caption{Coefficients $a_T$ in Ref.~\cite{Engel03} for $^{225}$Ra in units of $e \rm{fm}^3$.}
\begin{center}
\begin{tabular}{cccc}\hline
                & $a_0$       & $a_1$     & $a_2$ \\\hline
$^{225}$Ra & $-$ 5.06         & 10.4       & $-$10.1 \\\hline
\end{tabular}
\end{center}  
\label{KTab01}
\end{table}  

The effect of the intrinsic nucleon EDM to the Schiff moment can also be estimated in the simple shell model.
It is given by \cite{khriplovich,ginges,yamanakabook}
\begin{eqnarray}
S_{\rm int} 
=
\left\{
\begin{array}{ll}
d_N
\Bigl[
\frac{1}{10} \frac{2+j}{1+j} \langle r^2 \rangle_{\rm val}
-\frac{1}{6} \langle r^2 \rangle_{\rm ch}
\Bigr]
 &  (j= l+1/2)
\cr
d_N
\Bigl[
\frac{1}{10} \frac{1-j}{1+j} \langle r^2 \rangle_{\rm val}
+\frac{1}{6} \frac{j}{1+j} \langle r^2 \rangle_{\rm ch}
\Bigr]
& (j=l-1/2)
\cr
\end{array}
\right.
,
\ \ \ \ \ 
\label{eq:intrinsic_nsm}
\end{eqnarray}
where $\langle r^2 \rangle_{\rm val}$ is the mean square radius of the valence nucleon $N$.
The nuclear angular momentum and the orbital angular momentum of the single valence nucleon are denoted by $j$ and $l$, respectively.
In ordinary nuclei, $\langle r^2 \rangle_{\rm val} \approx \langle r^2 \rangle_{\rm ch} \approx A^{\frac{2}{3}} (1.1\, {\rm fm})^2$ with $A$ the nucleon number.
In the simple shell model, $\frac{1}{2}^+$ nuclei have an s-wave valence nucleon (e.g. $^{129}$Xe, $^{225}$Ra). From Eq. (\ref{eq:intrinsic_nsm}), $S_{\rm int}$ of those nuclei vanishes.
In reality, the single valence nucleon approximation does not hold due to the configuration mixing, and the intrinsic nucleon EDM contribution does not cancel.
We note that the effect of intrinsic nucleon EDM is not enhanced, since the relativistic effect is weak in nuclei, in contrast to that for the electrons in atoms \cite{sandars1,sandars2,flambaum2}.
For $\frac{1}{2}^+$ nuclei, we should consider that $|S_{\rm int}| \sim \frac{d_N}{6} \langle r^2 \rangle_{\rm ch}$ is an upper limit.

\subsubsection{Mean field framework}

In Ref.~\cite{Jesus05}, the NSM of the nucleus $^{199}$Hg is 
calculated by $\pi$-N-N interaction vertices that are  P,T-odd. Their approach, formulated in diagrammatic perturbation 
theory with important core-polarization diagrams summed to all orders, gives 
a close approximation to the expectation value of the Schiff operator in the 
odd-A Hartree-Fock-Bogoliubov ground state generated by a Skyrme interaction 
and a weak P,T-odd pion-exchange  potential.  
In the following their method is reviewed in short.

The NSM is approximately expressed as the expectation value of the Schiff operator $S_z$
in the completely self-consistent one-quasiparticle ground state of $^{199}$Hg, constructed 
from a two-body interaction that includes both a Skyrme potential and the P,T-odd 
potential $V^{PT}$. It is an approximation because   $V^{PT}$ is not treated
in a completely self-consistent way.
The mean-field calculation in $^{199}$Hg itself is not carried out.
Instead, the HF+BCS ground-state of the even-even nucleus $^{198}$Hg 
is first calculated
and add a neutron in the $2p_{1/2}$ level. 
The core-polarizing effects of this neutron are treated in the QRPA framework. 

Following a spherical HF+BCS calculation in $^{198}$Hg, 
the Hamiltonian is divided into unperturbed 
and residual parts. 
The unperturbed part, expressed in the quasiparticle basis, is
\begin{eqnarray}
H_0 = T + V_{00} + V_{11} ,
\end{eqnarray}
where $T$ is the kinetic energy and $V$ the Skyrme interaction, with subscripts referring
to the numbers of quasiparticles which the interaction creates and destroys. 
The perturbed part is
\begin{eqnarray}
H_{res} = V^{PT} + V_{22} + V_{13} + V_{31} + V_{04} + V_{40}.
\end{eqnarray}
The interaction $V^{PT}$ can also be expanded in terms of quasiparticle creation 
and annihilation operators.
The model space of effective operator theory is one-dimensional: a 
quasiparticle in the $a \equiv \left( {2{p_{1/2}},m = 1/2} \right)$ level. The unperturbed ground state
$\left| {\Phi _a} \right\rangle $  is simply this one-quasiparticle state. Then the expectation value of $S^z$,
in the full correlated ground state is given by
\begin{eqnarray}
\left\langle {{\Psi _a}\left| {{S_z}} \right|{\Psi _a}} \right\rangle  
&=& {N^{ - 1}}\left\langle {{\Phi _a}} 
\right|\left[ {1 + {H_{res}}\left( {\frac{Q}{{{\varepsilon _a} - {H_0}}}} \right) +  \cdots } \right]{S_z} 
\nonumber \\
&\times & \left[ {1 + \left( {\frac{Q}{{{\varepsilon _a} - {H_0}}}} \right){H_{res}} +  \cdots } \right]
\left| {{\Phi _a}} \right\rangle 
\end{eqnarray}
Here $\varepsilon _a$ is the single-quasiparticle energy of the valence nucleon, the operator $Q$ projects 
onto all other single quasiparticle states, $N$ is the normalization factor. 

The terms that are first order in $H_{res}$ do not include the strong interaction 
$V$ because it has a different parity from the Schiff operator. Thus the lowest order contribution 
to the NSM is
\begin{eqnarray}
{\left\langle {{\Psi _a}\left| {{S^z}} \right|{\Psi _a}}  \right\rangle^{LO}}  
= \left\langle  -  \right|{c_a}\left[ {{V^{PT}}\left( {\frac{Q}{{{\varepsilon _a} - {H_0}}}} \right){S^z}} \right]
c_a^\dag \left|  -  \right\rangle  + c.c.,  \nonumber \\ 
\end{eqnarray}
where $c_a^\dag $ is the creation operator for a quasiparticle in the valence level $a$ and 
$\left|  -  \right\rangle$ is the no-quasiparticle BCS vacuum describing the even-even core, 
so that $\left| {{\Phi _a}} \right\rangle $  is just $c_a^\dag \left|  -  \right\rangle $.
The core polarization is also considered, implemented through a certain QRPA method.
To assess the uncertainty in the results, they carried out the calculation with several Skyrme 
interactions, the quality of which is tested by checking predictions for the 
isoscalar-$E1$ strength distribution in $^{208}$Pb. 
Their final results are summarized in Table~\ref{YTab02}.

\begin{table}
\caption{Coefficients $a_T$ in Ref.~\cite{Jesus05} in $^{199}$Hg 
for the five different Skyrme interactions in units of $e \rm{fm}^3$.}
\begin{center}
\begin{tabular}{cccc}\hline
        & $a_0$   & $a_1$  & $a_2$ \\\hline
SkM   & 0.009   & 0.070  & 0.022  \\
SkP   & 0.002   & 0.065  & 0.011  \\
SIII    & 0.010   & 0.057  & 0.025  \\
SLy4 & 0.003   & 0.090  & 0.013  \\
SkO'  & 0.010   & 0.074  & 0.018  \\\hline
\end{tabular}
\end{center}  
\label{YTab02}
\end{table}

In Ref.~\cite{Dobaczewski05} they present a comprehensive mean-field 
calculation of the  NSM of the nucleus $^{225}$Ra, the quantity 
that determines the static electric-dipole moment of the corresponding atom 
if T is violated in the nucleus. The calculation 
breaks all possible intrinsic symmetries of the nuclear mean-field and 
includes both exchange and direct terms from the full 
finite-range T-violating N-N interaction, and the effects of 
short-range correlations. The resulting NSM, which depends on 
three unknown T-violating  $\pi$-N-N coupling constants, is much larger 
than in $^{199}$Hg, the isotope with the best current experimental limit on 
its atomic  EDM. In the following their work
is reviewed briefly.

The asymmetric shape of $^{225}$Ra implies parity doubling, namely, 
the existence of a very low energy   
$\left| {1/{2^ - }} \right\rangle $ state, in this case 55 keV above the ground state  
$\left| {{\Psi _0}} \right\rangle  \equiv \left| {1/{2^ + }} \right\rangle $ 
that dominates the sum in Eq.~(\ref{Yeq6}) because 
of the corresponding small denominator.
With the approximation that the shape deformation is rigid, the ground state 
and its negative-parity partner in the octupole-deformed nucleus are projections onto good parity 
and angular momentum of the same ``intrinsic state", 
which represents the wave function of the nucleus in its own body-fixed frame 
with the total angular momentum aligned along the symmetry axis. Equation~(\ref{Yeq6})
then reduces to
\begin{equation}
S \approx  - \frac{J}{J+1}\left\langle {{{\hat S}_z}} \right\rangle \frac{{\left\langle {{V^{PT}}} 
\right\rangle }}{{\left( {{\rm{55 keV}}} \right)}}
\label{Yeq17}
\end{equation}
where $J=1/2$ and the brackets indicate expectation values in the intrinsic state. 
The octupole deformation enhances $\left\langle {{{\hat S}_z}} \right\rangle $,  
making it collective, robust, 
and straightforward to calculate with an error of a factor of 2 or less. 
To evaluate  $\left\langle {{V^{PT}}} \right\rangle $  
they constructed a new version of the code Hartree-Fock code (HFODD). 
HFODD works with any Skyrme energy functional. 
Their results for various types of Skyrme interactions are given in Table~\ref{YTab03}.

\begin{table}
\caption{Coefficients $a_T$ in Ref.~\cite{Dobaczewski05} in $^{225}$Ra, 
calculated with the different types of Skyrme interactions in units of $e \rm{fm}^3$.
}
\begin{center}
\begin{tabular}{cccc}\hline
           & $a_0$    & $a_1$  & $a_2$ \\\hline
SkO'    & $-$1.5   & 6.0    & $-$4.0  \\
SIII      & $-$1.0   & 7.3     & $-$3.9  \\
SkM*   & $-$4.7   & 21.5   & $-$11.0  \\
SLy4   & $-$3.0   & 16.9   & $-$8.8 \\\hline
\end{tabular}
\end{center}  
\label{YTab03}
\end{table}  

In Ref.~\cite{Ban10}, they calculate the  NSMs of the 
nuclei $^{199}$Hg and $^{211}$Rn in completely self-consistent odd-nucleus 
mean field theory by modifying the Hartree-Fock-Bogoliubov code HFODD. They 
allow for arbitrary shape deformation and include the effects of nucleon 
dipole moments alongside those of a $\pi$-N-N
interaction that violates charge-parity (CP) symmetry. The results for 
$^{199}$Hg differ significantly from those of previous calculations when the 
CP-violating interaction is of isovector character.

Here they do not use perturbation theory, but instead  the 
NSM is directly calculated, 
by including the $PT$ violating interaction. Namely 
they have
\begin{equation}
\label{eq11}
S_{\mbox{ch}} =\left\langle {\Psi \left| {\hat {S}_{ch,z} } \right|\Psi } 
\right\rangle ,
\end{equation}
where $\left| \Psi \right\rangle $ is the P,T-odd ground state. 
Their results are given in Tables~\ref{YTab04} and \ref{YTab05}.

They also calculate the Schiff moment coming from the Schiff moment operator due to 
nucleon intrinsic electric dipole moment \cite{Dmitriev05}
as in Eq.~(\ref{Yeq4})
where the nucleon EDM operator in the leading chiral approximation [see also Eqs. (\ref{eq:d_0}) and (\ref{eq:d_1})] can be 
written as
\begin{eqnarray}
{\rm {\bf \hat {d}}}_i 
&=&
\frac{eg}{4\pi ^2m_N }\ln \frac{m_N }{m_\pi }\left( 
{\bar {g}^{\left( 0 \right)}-\bar {g}^{\left( 2 \right)}} \right){\rm {\bf 
\hat {\sigma }}}_i ( -\tau _i^z ) 
\nonumber\\
&\approx &
5.2 \times 10^{-2} {\rm GeV}^{-1}
\,
e g \left( {\bar {g}^{\left( 0 \right)}-\bar {g}^{\left( 2 \right)}} \right)
{\rm {\bf \hat {\sigma }}}_i (- \tau _i^z )
.
\label{Yeq13}
\end{eqnarray}
where $i$ represents $i$th nucleon.
Here the minus sign of the isospin matrix is due to the difference of convention.
In terms of the coefficient $b$ of Tables \ref{YTab04} and \ref{YTab05}, Eq. (\ref{Yeq14}) is written as
\begin{equation}
S_{\rm int}
=
\frac{b \, d_n}{1.0 \times 10^{-2} e\, {\rm fm}}
,
\end{equation}
where it is assumed that only the intrinsic EDM of the neutron contributes.

\begin{table}
\caption{Coefficients $a_T$ in Ref.~\cite{Ban10} in $^{211}$Rn in units of $e \rm{fm}^3$.}
\begin{center}
\begin{tabular}{ccccc}\hline
            & $a_0$   & $a_1$         & $a_2$ & $b$ \\\hline
SLy4     & 0.042   & $-$0.018     & 0.071 & 0.016 \\
SkM*     & 0.042   & $-$0.028    & 0.078  & 0.015 \\
SIII        & 0.034   & $-$0.0004  & 0.064  & 0.015 \\\hline
\end{tabular}
\end{center}  
\label{YTab04}
\end{table}

\begin{table}
\caption{Coefficients $a_T$ in Ref.~\cite{Ban10} in $^{199}$Hg  in units of $e \rm{fm}^3$.
The first three lines are in the HF approximation, and the next two are in the HFB approximation.
}
\begin{center}
\begin{tabular}{ccccc}\hline
            & $a_0$   & $a_1$         & $a_2$   & $b$ \\\hline
SLy4     & 0.013   & $-$0.006     & 0.022   & 0.003 \\
SIII        & 0.012   & 0.005         & 0.016   & 0.004 \\
SV        & 0.009   & $-$0.0001   & 0.016   & 0.002 \\ \hline
SLy4     & 0.013   & $-$0.006     & 0.024   & 0.007 \\
SkM*     & 0.041   & $-$0.027    & 0.069   & 0.013 \\\hline
\end{tabular}
\end{center}  
\label{YTab05}
\end{table}

In Ref.~\cite{Dmitriev03}, 
they calculated the contribution of internal  $d_N$
to the  NSM of
$^{199}$Hg. The contribution of the  $d_p$ was 
obtained via core polarization effects
that were treated in the framework of random phase approximation (RPA) 
with effective residual forces.
Their results are given in Table~\ref{YTab09}.

\begin{table}
\caption{Values of $s_p$ and $s_n$ for different $g_s$ and $g'_s$
in $^{199}$Hg
where $g_s$ and $g'_s$ are strengths for the Landau-Migdal interaction
in Ref.~\cite{Dmitriev03} .}
\begin{center}
\begin{tabular}{ccccc}\hline
            & $s_p$   & $s_n$  & $g_s$   & $g'_s$ \\\hline
SIII        & 0.18    & 1.89    & 0.25    & 0.9 \\
SV        & 0.19    & 1.86    & 0.25    & 1.0 \\
SLy4     & 0.20    &1.93     & 0.19    & 0.9 \\
SkM*     &0.22    & 1.90    & 0.19    & 1.0 \\\hline
\end{tabular}
\end{center}  
\label{YTab09}
\end{table}

The  NSM is predicted to be enhanced 
in nuclei with static quadrupole and octupole deformation.
The analogous suggestion of the enhanced contribution 
to the  NSM from the soft collective quadrupole
and octupole vibrations in spherical nuclei is tested 
in this article in the framework of the quasi RPA (QRPA) with separable quadrupole and octupole forces 
applied to the odd $^{217-221}$Ra and $^{217-221}$Rn
isotopes. In this framework, we confirm the existence 
of the enhancement effect due to the soft modes, but only in
the limit when the frequencies of quadrupole 
and octupole vibrations are close to zero. 
According to the QRPA,
in realistic cases the enhancement in spherical nuclei 
is strongly reduced by a small weight of the corresponding
``particle+phonon'' component in a complicated wave function 
of a soft nucleus. 
They considered the following weak P,T-odd interaction
\begin{equation}
W = \frac{G_F}{ {\sqrt 2 } } 
\frac{1}{{2m}} \eta (\boldsymbol{\sigma} \boldsymbol{n}) 
\frac{1}{{4\pi }} \frac{{d\rho (r)}}{{dr}}
\end{equation}
where $G_F$ is the Fermi constant of the
weak interaction, and $\eta$ is the strength of the P,T-odd
interaction and $\rho (r)$ is the nuclear charge distribution.

\begin{table}  
\caption{Schiff moments in units of $\eta~10^{-8} e \rm{fm}^3$ in Ref.~\cite{Auerbach06} }
\begin{center}
\begin{tabular}{ccccccc}\hline
    & $^{217}$Ra  & $^{217}$Rn  & $^{219}$Ra  & $^{219}$Rn & $^{221}$Ra & $^{221}$Rn \\\hline
    & $-$0.03      & $-$0.01       & 0.30           & $-$0.03      & $-$0.07      & 0.06 \\ \hline
\end{tabular}
\end{center}
\label{YTab13}
\end{table}

\subsubsection{Configuration mixing shell model approaches}

In Ref.~\cite{Yoshinaga13}, the NSMs for the lowest $1/2^+$ states of 
Xe isotopes are calculated. 
The nuclear wave functions beyond mean-field theories are 
calculated in terms of the nuclear 
shell model, which contains P, T-odd two-body interactions. 
In the following their approach is reviewed in detail.

For a description of the first $1/2^+$ states (the 
$1/2_1^+$ states) of odd-mass nuclei, the 
pair-truncated shell model (PTSM)~\cite{Higashi03,Yoshi04,Higashi11} is adopted, 
where the gigantic 
shell model space is restricted to an efficient and 
dominant model space in terms of collective pairs. 
In the low-lying states angular momenta zero ($S$) and two 
($D$) collective pairs are most important. 
The $S$ and $D$ pairs are defined as
\begin{eqnarray}
\hat{S}^{\dag}   &=& \sum_{j}     \alpha_j        \hat{A}^\dag{}_{0}^{(0)}(jj)      , \label{eq:spair} \\
\hat{D}^{\dag}_M &=& \sum_{j_1j_2}\beta_{j_1j_2}  \hat{A}^\dag{}_{M}^{(2)}(j_1j_2)  , \label{eq:dpair}
\end{eqnarray}
where the structure coefficients $\alpha$ and $\beta$ are 
determined by variation. 
Here the creation operator of a pair of nucleons in 
orbitals $j_1$ and $j_2$ with total angular momentum $J$, 
and its projection $M$ is written by
\begin{equation}
\hat{A}^\dag{}_{M}^{(J)}(j_1j_2)
=\sum_{m_1m_2} \left( \left. j_1 m_1 j_2 m_2 \right| JM \right)
\hat{c}^{\dag}_{j_1m_1} \hat{c}^{\dag}_{j_2m_2}
\label{eq:paircr},
\end{equation}
where $\hat{c}_j^\dag$ is the nucleon creation operator 
for the $j$ orbital. 

The many-body wave functions of even-nucleon systems for neutrons or protons 
are created by applying the pair creation operators $\hat{S}^\dag$ and 
$\hat{D}^\dag$ to the inert core $\left| - \right\rangle$: 
\begin{equation}
\left| S^{n_s}D^{n_d}~\gamma I \right>
=
\big( \hat{S}^{\dag} \big)^{n_s} \big( \hat{D}^{\dag} \big)^{n_d}
\left | - \right >
\label{eq:evsys}   ,
\end{equation}
where $I$ is a total angular momentum of the many-body system, and $\gamma$ 
an additional quantum number required to completely specify 
the states. The $n_s$ and $n_d$ are numbers of $S$ and $D$ pairs, respectively.
The total number of $S$ and $D$ pairs ($n_s +n_d$) is 
restricted to half the number of valence nucleons in the even-nucleon system. 
For the description of odd-nucleon systems, an unpaired 
nucleon in the $j$ orbital is added to the even-nucleon system. The state is now written as 
\begin{equation}
\left| j S^{n_s}D^{n_d}~\gamma I \right>
=
\hat{c}_j^\dag \left| S^{n_s}D^{n_d}~\gamma I' \right>
\label{eq:odsys}   .
\end{equation}

As for single-particle levels, all the relevant five 
orbitals, $0g_{7/2}$, $1d_{5/2}$, $1d_{3/2}$, $0h_{11/2}$, 
and $2s_{1/2}$, in the major shell between the magic 
numbers 50 and 82 are taken into account for both neutrons 
and protons. 
In addition, four orbitals with negative parity, 
$1f_{7/2}$, $1f_{5/2}$, $2p_{3/2}$ and $2p_{1/2}$, are 
considered above the closed shell $Z=82$ for protons. 
This is because 
the shell model space is necessary to be expanded including the 
negative-parity states connected to the $1/2_1^+$ states
in order to calculate the NSMs coming from the P,T-odd two-body interactions.
For a description of those negative-parity states, 
introduce proton negative-parity ($N_k ,k=1,2,3,4,5)$ pairs 
that are necessary in addition to the $S$ and $D$ pairs. i.e.
\begin{eqnarray}
\hat{N}_{1}^\dag{}^{(K_1)}_M & = & \hat{A}_{M}^{\dag(K_1)}(g_{7/2},f_{7/2}) ,\\
\hat{N}_{2}^\dag{}^{(K_2)}_M & = & \hat{A}_{M}^{\dag(K_2)}(d_{5/2},f_{5/2}) ,\\
\hat{N}_{3}^\dag{}^{(K_3)}_M & = & \hat{A}_{M}^{\dag(K_3)}(s_{1/2},p_{1/2}) ,\\
\hat{N}_{4}^\dag{}^{(K_4)}_M & = & \hat{A}_{M}^{\dag(K_4)}(g_{7/2},f_{5/2}) ,\\
\hat{N}_{5}^\dag{}^{(K_5)}_M & = & \hat{A}_{M}^{\dag(K_5)}(d_{5/2},f_{7/2}) ,
\end{eqnarray}
where the coupled angular momenta take values of 
$K_{1,2}=0, 1, 2, 3, 4$, $K_3=0, 1$ and $K_{4,5}=1, 2, 3, 4$. 
Then the wave function of the even-nucleon system with 
negative parity is constructed as
\begin{equation}
\left| S^{n_s}D^{n_d}N_k~\gamma I \right>
=
\big( \hat{S}^{\dag} \big)^{n_s} \big( \hat{D}^{\dag} \big)^{n_d}
\hat{N}_k^{\dag}
\left | - \right >
\label{eq:evnsys}   ,
\end{equation}
where $n_s+n_d+1$ gives half the number of valence protons. 

The odd-mass (neutron odd and proton even) nuclear state with a total spin 
$I$ and its projection $M$ is written as a product of the above state in 
neutron space and that in proton space as
\begin{eqnarray}
\lefteqn{
\left | \Phi (IM\eta ) \right > 
}
\nonumber \\
&=&
\left [ 
\left| j_n S^{\bar{n}_s}D^{\bar{n}_d} I_n \eta_n \right>
\otimes 
\left| S^{n_s}D^{n_d}N_i^{n_n} I_p \eta_p \right>
\right ]^{(I)}_M
\label{eq:oenwf}   ,
\end{eqnarray}
where $2(\bar{n}_s + \bar{n}_d )+1$ and $2(n_s+n_d+n_n)$ 
are numbers of valence neutron holes and proton particles, 
respectively. 
In this mass region, valence neutrons are treated as holes, 
and valence protons are treated as particles. 
The number of the proton 
negative-parity pairs, $n_n$, is limited to at most one (i.e., 
$n_n =0$ or 1).

As an effective two-body interaction, the 
monopole pairing (MP) and quadrupole pairing (QP) plus quadrupole-quadrupole (Q-Q)
interaction is employed. The effective shell-model Hamiltonian is written as 
\begin{equation}
\hat{H} = \hat{H}_n+\hat{H}_p+\hat{H}_{np}
\label{eq:ham}   ,
\end{equation}
where $\hat{H}_n$, $\hat{H}_p$, and $\hat{H}_{np}$ 
represent the interaction among neutrons, the interaction 
among protons, and the interaction between neutrons and 
protons, respectively. 
The interaction among like nucleons $\hat{H}_t$ 
($t=n$ or $p$) consists of spherical single-particle 
energies,  MP, QP and QQ interactions. i.e.,
\begin{eqnarray}
\lefteqn{
\hat{H}_t = \sum_{jm} \varepsilon_{jt}\hat{c}^{\dag}_{jmt}\hat{c}_{jmt} 
}
\nonumber \\
&&
- G_{0t}\hat{P}_t^{\dag(0)}\hat{P}_t^{(0)}
- G_{2t}\hat{P}_t^{\dag(2)} \cdot \tilde{\hat{P}}_t^{(2)}
- \kappa_t:\hat{Q}_t \cdot \hat{Q}_t:   ,\nonumber \\
&&
\end{eqnarray}
where $: :$ denotes normal ordering.
The interaction between neutrons and protons $\hat{H}_{np}$ 
is given by the $QQ$ interaction, 
\begin{equation}
\hat{H}_{np} =-\kappa _{np} \hat{Q}_n \cdot \hat{Q}_p .
\end{equation}
As for the single-particle basis states,
the harmonic oscillator basis states with
the oscillator parameter $b=\sqrt{\hbar / M\omega}$ are employed.
Further details of the effective interaction are presented
in Ref.~\cite{Higashi03,Yoshi04,Higashi11}.

The Hamiltonian given in Eq.~(\ref{eq:ham}) is diagonalized 
in terms of the many-body basis wave functions in Eq.~(\ref{eq:oenwf}) as
\begin{equation}
\label{eqse}
\hat{H}\left| I^{\pi} ;k \right\rangle =E(I^{\pi} ;k)\left| I^{\pi} ;k 
\right\rangle ,
\end{equation}
where $\left| I^{\pi} ;k \right\rangle $ is the 
normalized eigenvector for the $k$th state with spin $I$ 
and parity $\pi$, and $E(I^{\pi} ;k)$ is the eigenenergy 
for the state $\left| I^{\pi} ;k \right\rangle $. 

\begin{figure}
\begin{center}
\includegraphics[scale=0.50]{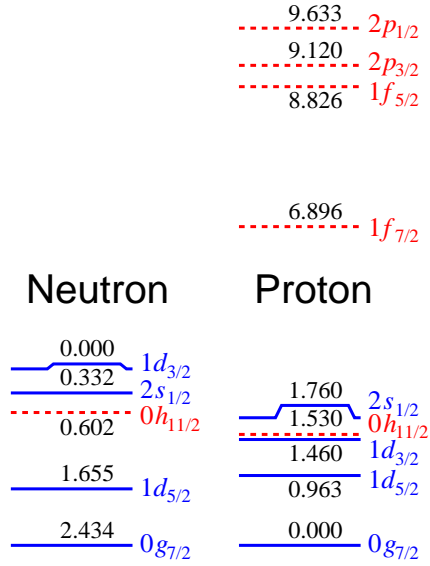}
\caption{\label{fig:spe}
(Color online) 
Model space for neutrons and protons adopted in the calculations. 
The single particle energy is listed for each single particle 
orbital in unit of MeV in the case of $^{129}$Xe from Ref.~\cite{Yoshinaga13}.
}
\end{center}
\end{figure}

The single particle energies are 
determined by the following procedure. 
Since the small change of the single particle energies 
hardly influences the energy levels of even-even nuclei, 
the single particle energies are determined primarily to 
reproduce the energy levels of low-lying states for 
odd-mass nuclei. 
Using the same set of two-body interactions adopted in 
the previous studies~\cite{Yoshi04}, 
the single 
particle energies are adjusted so as to approximately reproduce the 
energy levels of low-lying states for odd-mass nuclei. 
Next, the strengths of the two-body interactions are 
determined to reproduce the energy spectra of even-even 
nuclei. As shown later, the strengths of the two-body 
interactions are changed linearly with the number of the 
valence particles. 
Finally, the single particle energies are again modified 
to get an improved fitting to the low-energy levels of 
odd-mass nuclei. 
The single particle energies are thus obtained by 
repeating the above procedure, iteratively. 
Fig.~\ref{fig:spe} shows model space for neutrons and protons adopted. 
The single particle energy is listed for each single particle 
orbital in $^{129}$Xe. In order to investigate the systematics of low-lying 
states in the mass $A\sim 130$ region, it is assumed that 
the strengths of the two-body interactions change linearly 
with the number of the valence neutron holes $\bar{N}_n$ 
and the valence proton particles $N_p$~\cite{Higashi11}.

In Table~\ref{tab:DIM} total dimensions of 
the $1/2^+$ states and the $1/2^-$ states are shown for Xe isotopes.
\begin{table}
\begin{center}
\caption{
Dimensions of the $1/2^+$ states of $^{129}$Xe
built on the neutron space with parity + and 
the proton space with parity -,
and those of the $1/2^-$ states arising from
the proton negative-parity pairs
from Ref.~\cite{Yoshinaga13}.
}
\begin{tabular}{ccccc}\hline
Nucleus            & $^{129}$Xe & $^{131}$Xe & $^{133}$Xe& $^{135}$Xe \\
\hline
$1/2^+$  & 168      & 84      & 32        & 7 \\
$1/2^-$  & 4077     & 1968    & 718       & 419 \\\hline
\end{tabular} 
\label{tab:DIM}
\end{center}
\end{table}

The partial contribution of the $k$th state 
$\big| {\textstyle \frac{1}{2}}_{k}^{-} \big\rangle$
to the Schiff moment is defined by
\begin{equation}
\label{eqsmcp}
S_{(I)}(k)= 
\frac{
\big\langle {\textstyle \frac{1}{2}}_{1}^{+} \big| \hat{S}_{{\rm ch},z} 
\big| {\textstyle \frac{1}{2}}_{k}^{-} \big\rangle 
\big\langle {\textstyle \frac{1}{2}}_{k}^{-} \big| V^{PT}_{{\pi}(I)}
\big| {\textstyle \frac{1}{2}}_{1}^{+} \big\rangle 
}
{E^{(+)}_1-E^{(-)}_k} +c.c..
\end{equation}

\begin{figure}
\includegraphics[scale=0.35]{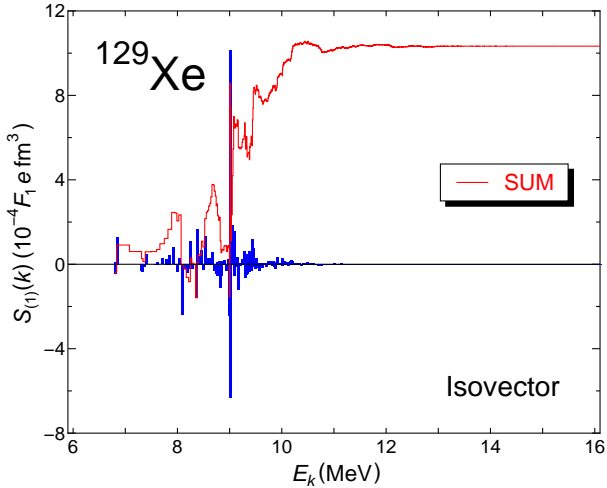}
\caption{\label{fig:schiv}
(Color online) 
Partial contribution to the Schiff moment
for the isovector type two-body interaction
in $^{129}$Xe as a function of excitation energies of $1/2^-$ states
from Ref.~\cite{Yoshinaga13}.
}
\end{figure}
In Fig.~\ref{fig:schiv}, the partial contribution $S_{(I)}(k)$ to the Schiff moment
for the isovector ($I=1$) two-body interaction 
in $^{129}$Xe is shown as a function of the excitation 
energy $E_k = E^{(-)}_k - E^{(+)}_1$.
The ``SUM" indicates the sum of each NSM contribution
defined by
\begin{equation}
\label{eqsum}
S_{(I)}^{\rm SUM} (k) = \sum\limits_{i=1}^{k} S_{(I)}(i), 
\end{equation}
where the summation takes over 
contributions from the first state to
the $k$th state with spin 1/2 and parity $-$.
There are four large contributions around $E_k =9.0$~MeV,
one positive and others negative.
Almost no
contributions are seen above 12.0~MeV.

\begin{figure}
\includegraphics[scale=0.35]{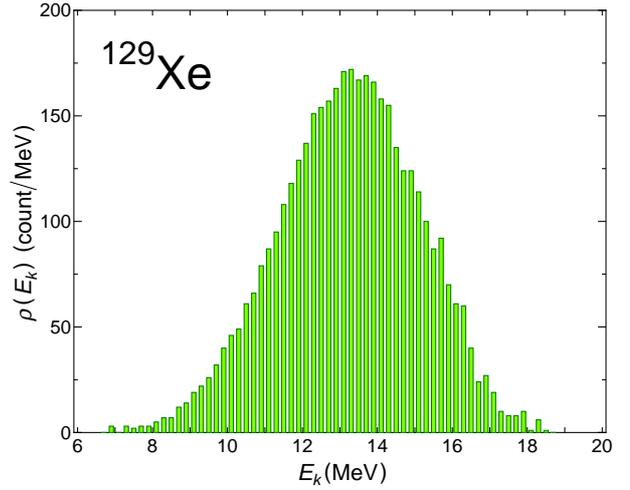}
\caption{\label{fig:dens}
(Color online) 
Density of the $1/2^-$ states
$\rho (E_k)$ for $^{129}$Xe from Ref.~\cite{Yoshinaga13}.
}
\end{figure}

In Fig.~\ref{fig:dens} the density of the $1/2^-$ states is shown
\begin{equation}
\rho (E_k) = \frac{d N}{d E}
\end{equation}
for $^{129}$Xe, where $d E = 0.2$~MeV is taken
and $d N$ is the number of the $1/2^-$states in the range $d E$.
It is seen from the figure that the $\rho$ has a  Gaussian shape and
increases 
exponentially between 8 and 12~MeV. Around 13~MeV
it becomes maximum, but the contribution of each state to the NSM
is marginal above 12~MeV as seen from Fig.~\ref{fig:schiv}.
The density of the $1/2^-$ states
presented in Fig.~\ref{fig:dens} is actually large enough to accommodate the most
contribution to the  NSMs.

\begin{figure}
\includegraphics[scale=0.35]{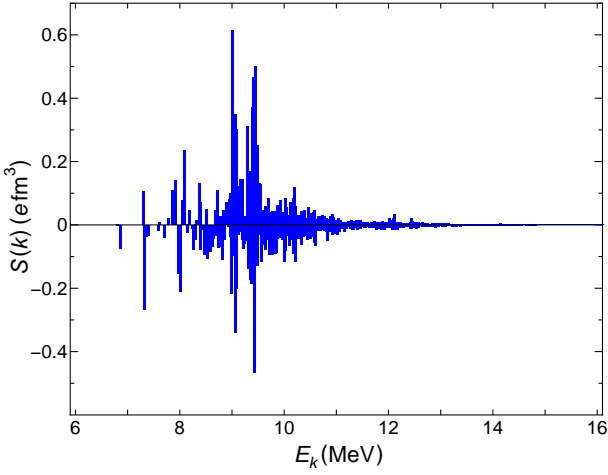}
\caption{\label{fig:sope}
(Color online) 
Strength function for the Schiff moment operator 
in $^{129}$Xe from Ref.~\cite{Yoshinaga13}.
}
\end{figure}

To investigate the components of the Schiff moment,
the strength function is evaluated for the NSM operator defined by
\begin{equation}
\label{eqsi}
S \left( k \right) = 
\big\langle {\textstyle \frac{1}{2}}_{1}^{+} \big|
\hat{S}_{{\rm ch}, z}
\big| {\textstyle \frac{1}{2}}_{k}^{-}
\big\rangle,
\end{equation}
which is shown for $^{129}$Xe in Fig.~\ref{fig:sope}.
There are several strengths in the range between 9 MeV 
and 10 MeV, but four large strengths around 9.0 MeV contribute
to the Schiff moment.

\begin{figure}
\includegraphics[scale=0.35]{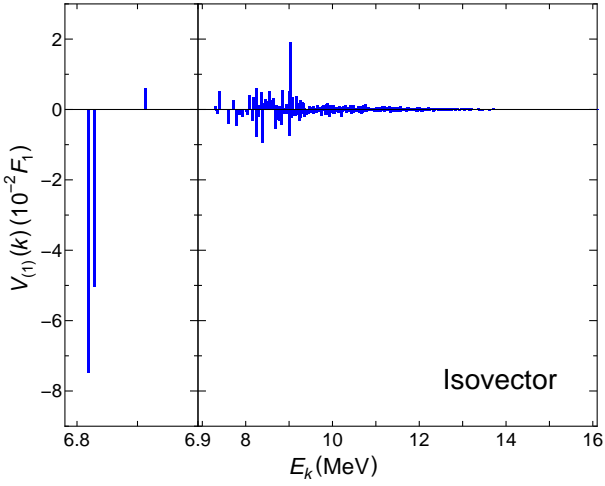}
\caption{\label{fig:vptiv}
(Color online) 
Off-diagonal potential matrix elements between $1/2^+_1$ state and 
$1/2^-_k$ state within the energy ranges below 6.9 MeV 
(left panel) and above 6.9 MeV (right panel) from Ref.~\cite{Yoshinaga13}.
}
\end{figure}

\begin{figure}
\includegraphics[scale=0.35]{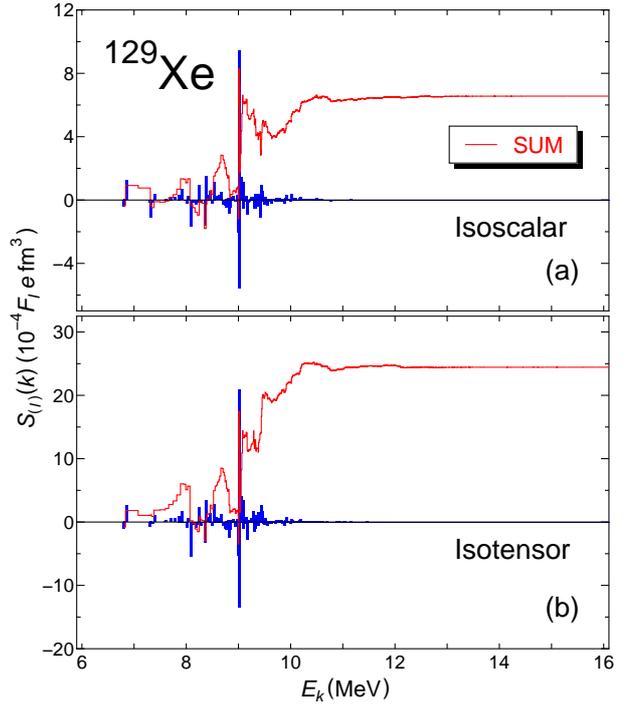}
\caption{\label{fig:schisit}
(Color online) 
Same as in Fig.~\ref{fig:schiv}, but for the (a) isoscalar 
($I=0$) and (b) isotensor ($I=2$) type two-body interactions from Ref.~\cite{Yoshinaga13}.
}
\end{figure}

In Fig.~\ref{fig:vptiv}, the off-diagonal potential matrix elements
is shown for the isovector ($I=1$) part
\begin{equation}
\label{eqvi}
V_{(I)} \left( k \right) =
\big\langle {\textstyle \frac{1}{2}}_{k}^{-} \big|
V^{PT}_{{\pi}(I)}
\big| {\textstyle \frac{1}{2}}_{1}^{+}
\big\rangle .
\end{equation}
In contrast to the strength function for the NSM, 
there are now two large contributions just above 6.8 MeV
in enlarged scale.
However, they do not contribute to the total
NSM at all since there are no strong NSM
strength functions in that corresponding regime.
In Fig.~\ref{fig:schisit} 
the partial contributions to the  NSMs are shown
and their total sums, respectively for isoscalar ($I=0$) 
and isotensor ($I=2$) two-body interactions.
All the three isospin NSMs
resemble to one another, but especially isovector and isotensor moments 
look quite similar to each other besides absolute values.

\begin{figure}
\includegraphics[scale=0.35]{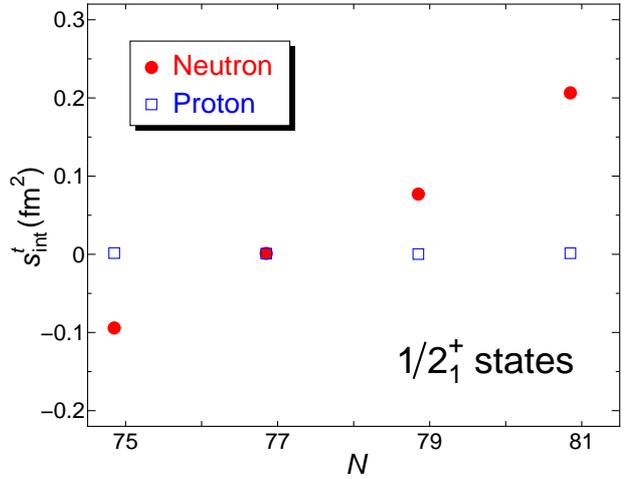}
\caption{\label{fig:smint} 
(Color online) 
The factors $s_{\rm int}^t$ ($t = p$ or $n$) for the Xe isotopes. 
The circles and squares represent 
the $s_{\rm int}^n$ and $s_{\rm int}^p$ values, respectively from Ref.~\cite{Yoshinaga13}. 
}
\end{figure}

The contribution to the  NSM also comes from
the intrinsic $d_N$.
By assuming the intrinsic nucleon EDM, the
NSMs are evaluated for the
$1/2_1^+ $ states in Xe isotopes, which are shown in 
Fig.~\ref{fig:smint}. 
These factors for neutrons are positively large for $^{135}$Xe, 
and negative for $^{129}$Xe. 
For all the nuclei, the factors for protons are almost 
zero.  $s_{\rm int}^p =+0.00156$~fm$^2$ and $s_{\rm int}^n =-0.09420$~fm$^2$
are obtained for $^{129}$Xe.

\begin{table}
\caption{
Coefficients $a_T$ 
in units of $10^{-3}~e \mbox{fm}^3$ ($I=0,1,2$)
from Ref.~\cite{Yoshinaga13}.
}
\begin{center}
\begin{tabular}{ccccc}\hline
Nucleus           & $^{129}$Xe & $^{131}$Xe & $^{133}$Xe & $^{135}$Xe \\
\hline
isoscalar ($I=0$) & 0.507      & 0.514      & 0.464      & 0.630 \\
isovector ($I=1$) & 0.399      & 0.352      & 0.285      & 0.323 \\
isotensor ($I=2$) & 1.89       & 1.60       & 1.24       & 1.31 \\ \hline
\end{tabular} 
\end{center}
\label{tab:SM3}
\end{table}

\begin{table}
\caption{
Coefficients $a_T$ 
in units of $10^{-3}~e \mbox{fm}^3$ ($I=0,1,2$)
obtained by the closure approximation
from Ref.~\cite{Yoshinaga13}.
}
\begin{center}
\begin{tabular}{ccccc}\hline
Nucleus           & $^{129}$Xe & $^{131}$Xe & $^{133}$Xe & $^{135}$Xe \\
\hline
isoscalar ($I=0$) &   0.701    &   0.691    &   0.666    &   0.733    \\
isovector ($I=1$) &   0.501    &   0.448    &   0.394    &   0.379    \\
isotensor ($I=2$) &   2.30     &   2.00     &   1.70     &   1.54 \\ \hline
\end{tabular} 
\end{center}
\label{tab:SMclo}
\end{table}

There is one limitation which requires
better discussion, namely, limitations of the model space used
in the shell model. 
The density of the $1/2^-$
states after 12 MeV, presented in Fig.~\ref{fig:dens}, 
is a result of the model space cutoff.
It is a question whether the model space is rich enough for the
converged determination of the  NSMs. 

Only one negative parity pair with different kinds are considered
as in Eq.~(\ref{eq:evnsys}). It should be examined how the bell-shape 
density of the $1/2^-$ states,
presented in Fig.~\ref{fig:dens} is large enough
to accommodate the most contribution
to the NSMs.

In order to consider all the intermediate states in 
Eq.~(\ref{eqsmcp}), the summation is carried out
using the closure approximation.
Here the denominator $ E^{(+)}_1-E^{(-)}_k$ is set constant as 
a representative value
$\langle E\rangle = \big\langle E^{(+)}_1-E^{(-)}_k \big\rangle$.
Using this approximation, the NSM is expressed as

\begin{eqnarray}
\label{eqsik}
S^{\rm clo}_{\rm{ch} (\it{I})} &=& 
\sum\limits_{k=1}
\frac{
\big\langle {\textstyle \frac{1}{2}}_{1}^{+} \big| \hat{S}_{{\rm ch},z} 
\big| {\textstyle \frac{1}{2}}_{k}^{-} \big\rangle 
\big\langle {\textstyle \frac{1}{2}}_{k}^{-} \big| V^{PT}_{{\pi}(I)}
\big| {\textstyle \frac{1}{2}}_{1}^{+} \big\rangle 
}
{\langle E\rangle} +c.c.  \nonumber \\
&=& \frac{
\big\langle {\textstyle \frac{1}{2}}_{1}^{+} \big| 
\hat{S}_{{\rm ch},z} V^{PT}_{{\pi}(I)}
\big| {\textstyle \frac{1}{2}}_{1}^{+} \big\rangle 
}
{\langle E\rangle}
+c.c.   ,
\end{eqnarray}
where the identity
$ \sum\limits_{k=1}
\big| {\textstyle \frac{1}{2}}_{k}^{-} \big\rangle 
\big\langle {\textstyle \frac{1}{2}}_{k}^{-} \big| =1 $
is used.
As shown in Figs.~\ref{fig:schiv} and \ref{fig:schisit}, 
the contribution to the  NSM $S_{(I)}(k)$ is 
dominant around the excitation energy
$E_k (= E^{(-)}_k - E^{(+)}_1) = 9.0$~MeV.
Thus  $\langle E\rangle = -9.0$~MeV is adopted.

The NSMs using the closure approximation
are shown in Table~\ref{tab:SMclo}.
Each value is consistently about 1.3 times larger than
the corresponding one shown in Table~\ref{tab:SM3},
which validates their discussion that the model space is rich enough
to accommodate the most contribution to the  NSMs.

\subsubsection{Recent shell model calculation with configuration mixing}

In a recent paper \cite{Teruya16}
any intermediate state given in Eq.~(\ref{Yeq6}) is represented 
as a one-particle and one-hole excited state ($1p1h$-state)
from the state  $\left| I^+_{1} \right>$.
Since the  NSM operator is a one-body operator working only on protons, 
it is enough to consider proton excited $1p1h$-states.
To evaluate the  NSM in Eq.~(\ref{Yeq6}), 
$k$th intermediate $1p1h$-state  is approximately given as 
\begin{eqnarray}
\big| I_k^- \big\rangle \sim  \left|(ij)K;I^-\right>
= N^{(K)}_{ij}\left[ \left[ c^{\dagger}_{i\pi} \tilde{c}_{j\pi} \right]^{(K)} \otimes \left| I^+_1 \right> \right]^{(I)} ,\nonumber\\ 
\end{eqnarray}
where $c^{\dagger}_{i\pi}$ ($c_{j\pi}$) represents the proton creation (annihilation) operator in the orbital $i$ ($j$),
with $\tilde{c}_{jm}=(-1)^{j-m} c_{j-m}$.
Namely, a $1p1h$-state with spin $K$, in which one proton excites 
from orbital $j$ to orbital $i$ by the  NSM operator,
is coupled with the nuclear ground state $\left| I^+_1 \right>$ to form an excited state $\left|(ij)K;I^-\right>$. 
$N^{(K)}_{ij}$ is the normalization constant determined as 
$$\left < (ij) K; I^- \right. \left | (ij) K; I^- \right > =1.$$
Here $K$ can take 1 or 0 for $I=1/2$.

Originally an intermediate state should satisfy 
$H \left|I_k^-\right> = E^{-}_k \left|I_k^-\right>$, where $H$ is the original shell model Hamiltonian. 
Here it is approximately assumed that $H \left| (ij)K;I^- \right>= (\varepsilon _{i}- \varepsilon _{j} +E^{+}_1) \left|(ij)K;I^-\right>$,
where $\varepsilon _{j}$ ($\varepsilon _{i}$) represents the single-particle energy in the orbital $j$ ($i$).
With this approximation in mind, Eq.~(\ref{Yeq6})  is written as
\begin{eqnarray}
\label{eq:SchiffMTD} 
S_{(T)} 
= \sum_{Kij}\frac{\langle I^+_1 | S^{(1)}_{0} | (ij)K;I^-\rangle 
\langle (ij)K;I^- |  V^{PT}_{{\pi}(T)}| I^+_1 \rangle  } 
{\Delta E_{ij} } +c.c.  \nonumber\\
\end{eqnarray}
Here the denominator $\Delta E_{ij}$ is explicitly written as $\Delta E_{ij} 
\equiv \varepsilon _{j}- \varepsilon _{i}$.

Here three types of $1p1h$-excitations are considered.
The first type  is a set of excitations from an orbital between 50 and 82 to an orbital over 82.
These excitations are called {\it type-I} excitations.
The second type is a set of excitations  from an orbital under 50 to an orbital between 50 and 82.
These excitations are called {\it type-II} excitations.
The third type is a set of excitations from an orbital under 50 to an orbital over 82.
These excitations are called {\it type-III} excitations.
Note that excitations among orbitals between 50 and 82 are vanished since these orbitals are not connected 
by the NSM operator.

For the type-I excitation, an intermediate state is explicitly written as
\begin{equation}
\left|(ph)K;I^- \right>_{\rm type-I} = N^{(K)}_{ph}\left[ \left[ a^{\dagger}_{p\pi} \tilde{c}_{h\pi} \right]^{(K)} \otimes \left| I^+_1 \right> \right]^{(I)} .
\end{equation}
Here $a^{\dagger}_{p\pi}$ represents the proton creation operator in the orbital $p$, 
where $p$ indicates an orbital over 82.
$\tilde{c}_{h\pi}$ represents the proton annihilation operator in the orbital $h$, 
where $h$ indicates an orbital between 50 and 82.
For the type-II excitation, an intermediate state is written as
\begin{equation}
\label{eq:core}
\left|(ph)K;I^- \right>_{\rm type-II} = N^{(K)}_{ph}\left[ \left[ c^{\dagger}_{p\pi} \tilde{b}_{h\pi} \right]^{(K)} \otimes \left| I^+_1 \right> \right]^{(I)} .
\end{equation}
Here $c^{\dagger}_{p\pi}$ represents the proton creation operator in an orbital $p$, 
where $p$ indicates an orbital between 50 and 82.
$\tilde{b}_{h\pi}$ represents the proton annihilation operator in the orbital $h$, 
where $h$ indicates an orbital below 50.
For the type-III excitation, an intermediate state is written as
\begin{equation}
\label{eq:core-over}
\left|(ph)K;I^- \right>_{\rm type-III} = N^{(K)}_{ph}\left[ \left[ a^{\dagger}_{p\pi} \tilde{b}_{h\pi} \right]^{(K)} \otimes \left| I^+_1 \right> \right]^{(I)} .
\end{equation}
Here $a^{\dagger}_{p\pi}$ represents the proton creation operator in an orbital $p$, 
where $p$ indicates an orbital over 82.
$\tilde{b}_{h\pi}$ represents the proton annihilation operator in the orbital $h$, 
where $h$ indicates an orbital below 50.  

In the study, all orbitals below the magic number 50 are considered for core orbitals.
However, $0d_{3/2}$, $1s_{1/2}$, and $0s_{1/2}$ orbitals are not connected 
by the  NSM operator.
For over-shell orbitals over the magic number 82, all orbitals up to 8~$\hbar \omega $ 
from the bottom are considered.
However, $2d_{5/2}$, $0j_{15/2}$, $0j_{13/2}$, $1h_{11/2}$, $0k_{15/2}$, $0k_{17/2}$, $2g_{7/2}$, $3d_{3/2}$, $3d_{5/2}$, and $4s_{1/2}$ orbitals are also not connected by the Schiff moment operator. 
Orbitals over 8~$\hbar \omega $ have less contributions to the NSM
because the  NSM operator proportionals to the $r$-square radius 
and these orbitals are not connected to low-lying orbitals by the NSM operator.   

The energy of each single particle orbital is taken from the Nillson energy as
\begin{equation}
\epsilon _{n\ell j}=\left( 2n + \ell + \frac{3}{2} \right) \hbar \omega -\kappa \Bigl( 2\boldsymbol{\ell}\cdot\boldsymbol{s} + \mu \left( \boldsymbol{\ell}^2 
-\left< \boldsymbol{\ell}^2 \right>_N \right) \Bigr) \hbar \omega , 
\end{equation}
with $\kappa=0.0637$ and $\mu=0.60$, where $\left< \boldsymbol{\ell}^2 \right>_N=\frac{1}{2}N(N+3)$ 
with the primary quantum number $N$ and $\hbar\omega = 41 A^{-1/3}$~MeV.

To analyze contributions to the  NSMs from each orbital,
a partial contribution of the  NSM from any orbital ($h$) between 50 and 82 
to a specific orbital ($p$) over 82 (type-I excitations) is defined in terms of $\bar{g}^{(T)}g$ as
\begin{eqnarray}
s^{\rm type-I}_{(T)}(p)=a^{\rm type-I}_{(T)}(p)~\bar{g}^{(T)}g, \label{at_over}
\end{eqnarray}
where
\begin{eqnarray}
&&s^{\rm type-I}_{(T)}(p) = \nonumber\\
&&\sum_{Kh}\frac{\langle I^+_1 | S^{(1)}_{0} | (ph)K;I^-\rangle 
\langle (ph)K;I^- |  V^{PT}_{{\pi}(T)}| I^+_1 \rangle  } 
{\Delta E_{ph} } +c.c. ,\nonumber\\
\end{eqnarray}
and $a^{\rm type-I}_{(T)}(p)$'s are coefficients so determined in evaluating 
the partial NSM $s^{\rm type-I}_{(T)}(p)$.

A partial contribution to any orbital ($p$) between 50 and 82 
from the specific orbital ($h$) below 50 (type-II excitations) is defined as 
\begin{eqnarray}
s^{\rm type-II}_{(T)}(h)=a^{\rm type-II}_{(T)}(h)~\bar{g}^{(T)}g, \label{at_core}
\end{eqnarray}
where
\begin{eqnarray}
&&s^{\rm type-II}_{(T)}(h) = \nonumber\\
&&\sum_{Kp}\frac{\langle I^+_1 | S^{(1)}_{0} | (ph)K;I^-\rangle 
\langle (ph)K;I^- |  V^{PT}_{{\pi}(T)}| I^+_1 \rangle  } 
{\Delta E_{ph}} +c.c.  \nonumber\\
\end{eqnarray}

A partial contribution from the specific orbital ($h$) below 50 
to any orbital ($p$) over 82 (type-III excitations) is also defined as 
\begin{eqnarray}
s^{\rm type-III}_{(T)}(h)=a^{\rm type-III}_{(T)}(h)~\bar{g}^{(T)}g, \label{at_c-o}
\end{eqnarray}
where
\begin{eqnarray}
&&s^{\rm type-III}_{(T)}(h) = \nonumber\\
&&\sum_{Kp}\frac{\langle I^+_1 | S^{(1)}_{0} | (ph)K;I^-\rangle 
\langle (ph)K;I^- |  V^{PT}_{{\pi}(T)}| I^+_1 \rangle  } 
{\Delta E_{ph}} +c.c.  \nonumber\\
\end{eqnarray}

Using these definitions, the  NSM is given as 
\begin{eqnarray}
S =\sum_{T} \left(s^{\rm type-I}_{(T)}  + s^{\rm type-II}_{(T)} + s^{\rm type-III}_{(T)} \right) ,
\end{eqnarray}
with
\begin{eqnarray}
&&s^{\rm type-I}_{(T)} = \sum _p s^{\rm type-I}_{(T)}(p) ,\\
&&s^{\rm type-II}_{(T)} = \sum _h s^{\rm type-II}_{(T)}(h) ,\\
&&s^{\rm type-III}_{(T)} = \sum _h s^{\rm type-III}_{(T)}(h) .
\end{eqnarray}

\begin{table}
\caption{\label{tab:Xe iso}
Calculated results of $a_{(T)}$ for the nuclear ground state $1/2^+$ state (in units of $ 10^{-3}  e \rm{fm}^3$) from Ref.  \cite{Teruya16}. 
Previous results ($a^{\rm prev}_{(T)}$) are taken from Ref.~\cite{Yoshinaga13}.}
\begin{tabular}{ccccccc}\hline
           &$T$& $a^{\rm type-I}_{(T)}$ & $a^{\rm type-II}_{(T)}$ & $a^{\rm type-III}_{(T)}$ & $a_{(T)}$ &$a^{\rm prev}_{(T)}$    \\\hline
           &$0$& 2.357 & 0.670 & $-$1.057 & 1.969 & 0.630\\
$^{135}$Xe &$1$& 1.297 & 1.693 & $-$0.602 & 2.389 & 0.323\\
           &$2$& 5.427 & 9.490 & $-$2.554 & 12.363& 1.31\\\hline
                
           &$0$& 1.812 & 1.716 & $-$1.047 & 2.481 & 0.464\\
$^{133}$Xe &$1$& 0.949 & 1.510 & $-$0.578 & 1.882 & 0.285 \\
           &$2$& 3.982 & 7.343 & $-$2.419 & 8.906 & 1.24\\\hline
                                               
           &$0$& 1.575 & 2.097 & $-$0.968 & 2.704 & 0.514\\
$^{131}$Xe &$1$& 0.787 & 1.282 & $-$0.530 & 1.539 & 0.352\\
           &$2$& 3.145 & 5.596 & $-$2.177 & 6.564 & 1.60\\\hline
             
           &$0$& 1.322 & 2.897 & $-$0.978 & 3.242 & 0.507\\
$^{129}$Xe &$1$& 0.586 & 1.140 & $-$0.522 & 1.204 & 0.399\\
           &$2$& 2.192 & 3.940 & $-$1.961 & 4.172 & 1.89 \\ \hline
\end{tabular} 
\end{table}

Table~\ref{tab:Xe iso} shows the calculated results of $a_{(T)}$ for the lowest $I=1/2$ states 
of Xe isotopes.
Here, using Eqs.~(\ref{at_over}), (\ref{at_core}), and (\ref{at_c-o}), $a_{(T)}$ is given as 
\begin{eqnarray}
a_{(T)} = a^{\rm type-I}_{(T)}  + a^{\rm type-II}_{(T)} + a^{\rm type-III}_{(T)} ,
\end{eqnarray}
with
\begin{eqnarray}
&&a^{\rm type-I}_{(T)} = \sum _p a^{\rm type-I}_{(T)}(p) ,\\
&&a^{\rm type-II}_{(T)} = \sum _h a^{\rm type-II}_{(T)}(h) ,\\
\end{eqnarray}
and
\begin{eqnarray}
a^{\rm type-III}_{(T)} = \sum _h a^{\rm type-III}_{(T)}(h) .
\end{eqnarray}
The contributions of the core excitations are a few times larger than those 
from the over-shell excitations for most of the components.
The isotensor ($T=2$) components are largest for all nuclei.

As a more elaborate configuration mixing framework of the shell model~\cite{Yoshinaga10},
NSMs for the ground $1/2^+$ states around the mass 130 are calculated 
in terms of the nuclear shell model. 
The intrinsic  NSM is evaluated as
\begin{equation}
\label{Neq14}
S=s_p d_p +s_n d_n, 
\end{equation}
where $d_p $ and $d_n $ are the electric dipole moments of the proton and the 
neutron, respectively. The factors $s_p $ and $s_n $ for the intrinsic  NSM of $^{129}$Xe 
are calculated as $s_p =+0.0061$ and $s_n =-0.3169$ (in $\mbox{fm}^2)$.

\subsubsection{Reliability, accuracy and uncertainty of the NSM calculations}

In this subsection we discuss the accuracy,  reliability and uncertainty 
of the theoretical calculations on the NSM. 

First of all, it is noted that the atoms considered in this article are closed-shell atoms,
whereas the corresponding nucleus is no more closed for both neutrons and protons.
Therefore the treatment in nuclear physics is expected to be different from atomic physics
even if many-body problems should be solved in both physics.

The accurate estimation of the NSM is very tough in nuclear physics.
There are various things we have to consider when arguing the accuracy and the reliability of
the calculations on the NSM.

We start from the following nuclear Hamiltonian (\ref{Yeq5}),
\begin{eqnarray}
H = H_0 + \hat {V}_{\pi (T)}^{PT} \nonumber, 
\end{eqnarray}
where $H_0$ is the nuclear strong interaction which conserves P and T.

(i) One major problem here is how to derive (calculate) the nuclear strong interaction $H_0$
in a nucleus, namely in the nuclear media. 
Firstly the difference between the atom 
and the nucleus is in that we do
not know precisely the interactions between nucleons. For atoms the interaction is the Coulomb force, 
but the nuclear force is considered to be a Van der Waals force derived 
from the fundamental QCD 
interaction among quarks and gluons.
Secondly we cannot use the bare interaction between free nucleons
because the nuclear interaction should be modified according to the medium effects.
This medium effect is taken into account by the Brueckner-Bethe-Goldstone 
(BBG) many-body theory~\cite{Baldo99}.
However, for nuclei in the medium and heavy region such as $^{129}$Xe and $^{199}$Hg, 
it is a hard task to obtain the effective interaction through BBG theory.
Thus we have to utilize some kinds of phenomenological interactions
such as the Skyrme interactions, surface delta interaction or pairing and quadrupole ($P+QQ$) interactions.
Usually for the mean field theories such as HF or HFB, the Skyrme interactions are often used,
but there are many kinds of Skyrme interactions.
For more elaborate calculations, the surface delta
interaction and the $P+QQ$ interaction are often used in the nuclear shell model.
The uncertainty on the NSM coming from these phenomenological interactions are supposed to be not so large
because the results calculated by using these interactions are tested by comparing them with the experimental data,
such as nuclear energy spectra, electro-magnetic moments and transitions.
On the other hand, in some simple mean field theories the binding energy is the only experimental source 
to be compared with theory
and the reliability is relatively small compared to the nuclear shell model.

(ii) Second problem is related to the CP-odd nuclear force  $\hat {V}_{\pi (T)}^{PT}$ appearing in Eq.~(\ref{Yeq5})
obtained at the hadron scale. The strength of the interaction used in the calculations of the CP-odd moments 
of heavy nuclei should not be the same with the bare CP-odd nuclear force. 
This is the same problem discussed above.
For the actual strength of the CP-odd nuclear force uncertainty is expected to be large and
we have no method to estimate its strength at present.

Since the PT-violating interaction is so weak compared to the nuclear force,
 the formula~(\ref{Yeq6}) by perturbation theory
 is good enough in the evaluation of the NSM,
\begin{equation}
S_{\mbox{ch}} =\sum\limits_{k=1} {\frac{\left\langle {I_1^+ \left| {\hat 
{S}_{ch,z} } \right|} \right.\left. {I_k^- } \right\rangle \left\langle 
{I_k^- \left| {\hat {V}_{\pi (T)}^{PT} } \right|} \right.\left. {I_1^+ } 
\right\rangle }{E_1^+ -E_k^- }} +c.c. \nonumber
\end{equation}

(iii)Third problem is related to the ground state wave functions  $ \left |  {I_1^+ } \right\rangle $
and the intermediate states  $ \left |  {I_k^- } \right\rangle $.
In order to obtain accurate eigenfunctions, Hamiltonian (\ref{Yeq5})  should be diagonalized
assuming a suitable set of basis states.
This cannot be done exactly due to the huge dimension of the configurations.
Therefore  the eigenfunctions  can be obtained only approximately.
There are basically two ways to calculate the ground state wave functions.
One is to use mean field theories such as HF or HFB theories.
Another is to use theories beyond the mean field (BMF) theories.
In mean field theories the wave functions are obtained with comparative ease
by breaking symmetries of rotation and the particle-number,
but the wave functions do not have specific spin and parity or particle-number, 
which are essential in a solitary system like a nucleus.
It is necessary to restore these symmetries.

In the following we discuss the accuracy of the results in the cases of $^{199}$Hg and  $^{129}$Xe.
For $^{199}$Hg, the results in
Table \ref{NTab01} and \ref{YTab01} are given by the same group~\cite{Dmitriev05}.
The results in Table \ref{NTab01} are calculated using a simple shell model,
where the whole nuclear state is determined 
by the last neutron in the orbital $p_{1/2}$ in the major shell of 50 and 82
without considering other nucleon's effects.
The results in Table \ref{YTab01} are calculated by considering
other nucleon's effects and as seen in Tables ten times difference is seen between them.
Certainly, the results in Table~\ref{YTab01} are more reliable in comparison with those in Table \ref{NTab01}.
The results in Table \ref{YTab02} are the most elaborate ones
up to date for the  $^{199}$Hg, 
but the method is based on the mean field theories and 
analyses using BMF theories are expected.
The results in Table \ref{YTab05} are given by mean field theories~\cite{Ban10}.
Even signs of the isovector components are different 
according to the various usage of the Skyrme interactions.

For $^{129}$Xe
the results in Table \ref{tab:SM3}, \ref{tab:SMclo} and \ref{tab:Xe iso} 
are calculated by the same group~\cite{Yoshinaga13,Teruya16}
based on the shell model.
The results in Table \ref{tab:SM3} are given in terms of the PTSM calculations~\cite{Yoshinaga13}
using the effective $P+QQ$ interaction.
The results in Table \ref{tab:SMclo} are given by the closure approximation
where intermediate energies $ E_k^ -$ are  set constant as $ \left\langle {E_k^-} \right\rangle $.
Then the summation over intermediate states are exactly evaluated as in Eq.~(\ref{eqsik}).
The results in Table \ref{tab:Xe iso} are  the most elaborate ones
using one-particle one-hole states for intermediate states 
$\left| {I_k^ - } \right\rangle $.
The results are most reliable up to now and only a factor of
two or three times difference compared to the results in Table \ref{YTab01}.

\subsection{Nuclear spin matrix elements}

To evaluate the effect of the nucleon spin dependent CP-odd $e-N$ interaction, namely $C^{\rm T}_N$ 
and $C^{\rm PS}_N$, on the atomic EDM, the values 
of the nuclear spin matrix elements $\langle \Psi | \sigma^p | \Psi \rangle$ 
and $\langle \Psi | \sigma^n | \Psi \rangle$ are required.
In the simple shell model, they are given as \cite{yamanakabook}
\begin{equation}
\left\langle {\left( {l{\textstyle{1 \over 2}}} \right)j,m 
= j\left| {{\sigma _z}} \right|\left( {l{\textstyle{1 \over 2}}} \right)j,m = j} \right\rangle
=
\left\{
\begin{array}{cl}
1 &  (j= l+1/2)
\cr
-\frac{j}{j+1}
& (j=l-1/2)
\cr
\end{array}
\right.
,
\end{equation}
where $\sigma_z$ is the single valence nucleon spin-operator.
The nuclear spin matrix element is also useful in evaluating the nucleon EDM effect to the nuclear EDM.
An extension of the above formula using the magnetic moment are also available \cite{Fujita12}.

In Ref.~\cite{Yoshinaga10,Yoshinaga14}, the 
nuclear spin matrix elements for the lowest $1/2^+$ states of Xe and Ba isotopes 
are calculated in terms of the nuclear shell model with configuration mixing. 
Values for several Xe and Ba isotopes (calculated in the context of the nuclear EDM) are given 
in Table~\ref{KTab06} of Ref.~\cite{Yoshinaga14}.
We see that the quenching of the nucleon spin becomes important in the nucleus as the nucleon number 
goes away from the magic number.
This is due to the superposition of configurations where the nucleon spin interferes destructively, 
due to the mixing with the orbital angular momentum.
This suppression means that the effect of $C^{\rm T}_N$ and $C^{\rm PS}_N$ are attenuated 
in nuclei far from magic numbers.
The nuclear spin matrix elements are still unknown for many nuclei, although their evaluation being easier than the nuclear Schiff moment.
It is therefore an important future subject to discuss for reducing the theoretical uncertainty of the atomic EDM.

\begin{table}
\caption{The neutron quenching factor $< \hat{\sigma}^n_z >$
for each nuclear $1/2^+_1$ state from Ref.~\cite{Yoshinaga14}.
}
\begin{center}
\begin{tabular}{cc}\hline
Nucleus      & $< \hat{\sigma}^n_z >$   \\\hline
$^{129}$Xe & 0.2306    \\
$^{131}$Xe & 0.4644     \\
$^{133}$Xe & 0.6546     \\
$^{135}$Xe & 0.9777     \\
$^{129}$Ba & 0.1090     \\
$^{131}$Ba & 0.3537     \\
$^{133}$Ba & 0.4360     \\
$^{135}$Ba & 0.9811     \\\hline
\end{tabular}
\end{center}  
\label{KTab06}
\end{table}  

\subsection{Enhancement due to octupole deformation} \label{ocmsec}

A charged particle, residing outside of the nucleus at distance $r$ like an electron, can see the potential due to electromagnetic 
interactions within the nucleus as \cite{ginges,auerbach}
\begin{eqnarray}
 \phi(r) = \int d^3r_N \frac{\rho(r_N)}{|r - r_N|} .
\end{eqnarray}
Carrying out the multipole expansion of $\frac{1}{|r-r_N|}$ can give rise to both P,T- odd and even potential terms. The first and 
dominant P,T- odd term can arise as
\begin{eqnarray}
 \phi(r) &=& - \int d^3r_N \rho(r_N) \left ( \vec{r}_N \cdot \vec{\nabla}_{r}\frac{1}{r} \right ) 
\end{eqnarray}
As per the Schiff theorem, this term will exactly cancel out with the NSM contribution for a point-like nucleus. In order to obtain the 
P,T- odd interaction term from here, it is therefore necessary to account the next leading order term, which yields
\begin{eqnarray}
 \phi_{oct}(r) & \simeq & - \frac{1}{6} \int d^3r_N \rho(r_N) r_i r_j r_k \nabla_i \nabla_j \nabla_k \frac{1}{r} ,
\end{eqnarray}
where subscript $oct$ implies that it corresponds to contribution from the electric octupole moment (EOM) for which the EOM tensor is given by
\begin{eqnarray}
 O_{ijk} &=& \int d^3r_N \rho(r_N) \nonumber \\ && \times \left [ r_i r_j r_k - \frac{1}{5} (\delta_{ij} r_k + \delta_{jk} r_i + \delta_{ki} r_j ) \right ] .
\end{eqnarray}
The EOM tensor $O_{ijk}$ has three units of angular momentum, hence it can only exist in nuclei with spin $I \ge 3/2$, whereas
the NSM can arise in nuclei with spin $I \ge 1/2$. Without the P,T-odd interactions the average value of the EOM 
for a rotational state in the laboratory system is zero. However, in the presence of such an interaction, the odd and even parity mixing 
of rotational doublet states gives rise to a finite value of the EOM. In particular for atoms with nuclei that have almost 
degenerate rotational doublets, there is a large enhancement of the EOM leading to an increase in the size of the observable 
EDMs of the atoms. This contribution needs to be extracted before estimating limits on various nuclear and particle physics
parameters from the observed atomic EDMs. From preliminary investigations it has been found that the EOM enhances the EDMs in $^{223}$Ra,
$^{225}$Ra and $^{223}$Rn atoms by 400, 300 and 1000 times more than due to the other P,T- odd interactions \cite{ginges,auerbach,Spevak}.

\section{Atomic structure calculations}

\subsection{P,T-odd sources in atoms}
\label{sec:1}

As has been discussed before, the dominant P,T-odd interactions in an atomic system can come from three important sources 
\cite{pospelov1,pospelovreview,ramsey,engel}. They are (i) EDMs of constituent particles such as the $d_e$, $d_n$ and $d_p$, 
(ii) P,T-odd e-N and N-N interactions, and (iii) P,T-odd pion exchange interactions.

Considering the dominant P,T-odd interactions in the diamagnetic atoms, the interaction Hamiltonian due to the NSM for the exchange of 
pions is given by \cite{ginges,dzuba1}
 \begin{eqnarray}
  H_{at}^{N}= \frac{3{\bf S.r}}{B_4} \rho(r),
  \label{nsmeq}
 \end{eqnarray}
where $B_4=\int_0^{\infty} dr r^4 \rho(r)$, and similarly by adding the coherent contributions from the individual nucleons the net 
electron-nucleus T-PT interaction Hamiltonian is given by
\begin{eqnarray}
H_{at}^{T} = i \sqrt{2} G_F C_{at}^T \sum_e \mbox{\boldmath $\sigma_N \cdot \gamma$} \rho(r),
\label{tpteq}
\end{eqnarray}
where $C_{at}^T$ is the T-PT electron-nucleus and {\boldmath$\sigma_N$}$=\langle \sigma_N \rangle {\bf I}/I$ is the Pauli spinor of the 
nucleus.

Nevertheless, in the diamagnetic atoms the $d_e$ and the P,T-odd or P,CP-odd type S-PS type e-N interaction corresponding to term with the 
coupling $C_N^{\rm SP}$ of Eq. (\ref{eq:pcpvenint}) can also contribute to the atomic EDM to some extent, mainly through the hyperfine 
induced interaction. Since these are not the dominant contributions in these atoms, we estimate their contributions using analytical
formulas known in the literature instead of performing rigorous numerical calculations.

The contribution of the electron EDM can analytically be related to the T-PT type P,T-odd e-N interaction [i.e. term with the coupling 
$C_N^{\rm T}$ of Eq. (\ref{eq:pcpvenint})] as \cite{khriplovich,ginges,flambaum1985}
\begin{equation}
d_e 
\leftrightarrow
\frac{3 m_N e }{7\pi \alpha_{\rm em} \mu } \frac{R}{R-1} \frac{G_F}{\sqrt{2}} 
\left( 
C_p^{\rm T} \sum_p \langle \Psi | \mathbf{\sigma}_p | \Psi \rangle
+C_n^{\rm T} \sum_n \langle \Psi | \mathbf{\sigma}_n | \Psi \rangle
\right)
,
\end{equation}
where ${\cal R}$ is the atomic enhancement factor to the atomic EDM due to T-PT e-N interaction and $\mu_I$ is the nuclear magnetic moment in unit of nuclear magneton $\mu_N$. 
The nuclear spin matrix elements $\langle \Psi | \mathbf{\sigma}_N | \Psi \rangle$ ($N=p,n$) is the expectation value of the nucleon 
spin polarized in the $z$-direction. 

Analogously, the contribution of $C_N^{\rm SP}$ is analytically related to $C_N^{\rm T}$ by
\begin{eqnarray}
\left( 
\frac{Z}{A} C_p^{\rm SP} 
+ \frac{A-Z}{A} C_n^{\rm SP}
\right)
\leftrightarrow 
\frac{1.9 \times 10^{3} }{ ( 1+ 0.3 Z^2 \alpha_{\rm em}^2 ) A^{-2/3} \mu } 
\nonumber\\
 \times
\left( 
C_p^{\rm T} \sum_p \langle \Psi | \mathbf{\sigma}_p | \Psi \rangle
+C_n^{\rm T} \sum_n \langle \Psi | \mathbf{\sigma}_n | \Psi \rangle
\right)
.
\end{eqnarray}
Assuming same number of protons and neutrons in the atom and their interaction strengths are of similar order, we can conveniently 
express \cite{dzuba1}
\begin{eqnarray}
C_{at}^P 
\leftrightarrow 
3.8 \times 10^3 \times \frac{A^{1/3}}{Z} C_{at}^T ,
\label{eqcp}
\end{eqnarray}
where $C_{at}^P$ is the corresponding P,T-odd S-PS coupling constant for the electron-nucleus interaction. Thus, with the knowledge 
of $C_{at}^{T}$ and its enhancement factor, we can estimate contributions due to $C_{at}^P$ and $d_e$ in the diamagnetic atoms. Hence, 
we only intend to estimate the $C_{at}^T$ coupling coefficient by accounting the interaction Hamiltonian given by Eq. (\ref{tpteq}).

Again, the magnetic quadrupole moment (MQM) of the nucleus can also contribute to the EDM of diamagnetic atoms through
hyperfine induced interaction, but that contribution will be extremely small and has been neglected here.

\subsection{Atomic many-body methods}
\label{sec:3}

The EDM of the ground state wave function ($|\Psi_0 \rangle$) in an atom is given by
\begin{eqnarray}
 d_{at} = \frac{\langle \Psi_0 | D | \Psi_0 \rangle}{\langle \Psi_0 | \Psi_0 \rangle },
 \label{edmeq}
\end{eqnarray}
where $D$ is the electric dipole moment operator. The evaluation of $|\Psi_0 \rangle$ should take into consideration the electromagnetic
and weak interactions in the atomic systems. In actual practice, the dominant one-photon electromagnetic interactions are included in the first 
step followed by if necessary higher order relativistic effects and the basic quantum electrodynamics (QED) corrections. The much weaker
P,T-odd interactions are added subsequently only to first order either in a perturbative or non-perturbative framework.
Such an approach is computationally simpler than including the P and T violating in the zeroth order Hamiltonian as it would would involve
atomic wave functions of a definite parity in the calculations as opposed to wave functions of mixed parity which would result from the latter
approach.

The starting point of the relativistic atomic many-body calculations is the Dirac-Coulomb (DC) Hamiltonian which is
\begin{eqnarray}
H^{DC} &=& \sum_i \left [ c\mbox{\boldmath$\alpha$}_i\cdot \textbf{p}_i+(\beta_i -1)c^2 + V_n(r_i) + \sum_{j>i} \frac{1}{r_{ij}} \right ], \ \ \ \ \ \
\end{eqnarray}
where $\mbox{\boldmath$\alpha$}$ and $\beta$ are the usual Dirac matrices and $V_n(r)$ represents for the nuclear potential. We 
evaluate the nuclear potential considering the Fermi-charge distribution defined by
\begin{equation}
\rho(r)=\frac{\rho_{0}}{1+e^{(r-b)/a}},
\end{equation}
for the normalization factor $\rho_0$, the half-charge radius $b$ and $a= 2.3/4(ln3)$ is related to the skin thickness. The half-charge radius 
is determined using the relation 
\begin{eqnarray}
b&=& \sqrt{\frac {5}{3} r_{rms}^2 - \frac {7}{3} a^2 \pi^2}
\end{eqnarray}
and the root mean square (rms) charge radius of the nucleus is evaluated by
\begin{eqnarray}
 r_{rms} =0.836 A^{1/3} + 0.570.
\end{eqnarray}
in $fm$.

The contribution from the frequency independent Breit interaction is estimated by adding the term 
\begin{eqnarray}
V_B =- \sum{i,j>i} \frac{1}{2r_{ij}}\{\mbox{\boldmath$\alpha$}_i\cdot \mbox{\boldmath$\alpha$}_j+
(\mbox{\boldmath$\alpha$}_i\cdot\bf{\hat{r}_{ij}})(\mbox{\boldmath$\alpha$}_j\cdot\bf{\hat{r}_{ij}}) \} ,
\end{eqnarray}
to the DC Hamiltonian; i.e. $H^{at} \equiv H^{DC} + V_B$

We have also estimated the lower order quantum electrodynamic corrections by considering the following QED potentials 
in the atomic Hamiltonian; i.e. $H^{at} \equiv H^{DC} + V_B+V_{QED}$ 
with $V_{QED}=\sum_i(V_U(r_i)+V_{WK}(r_i)+V_{SE}^{ef}(r_i)+
V_{SE}^{mg}(r_i))$ in a manner similar to that described in Ref. \cite{flambaum,bks-qed} but for the above nuclear Fermi-charge 
distribution. In this approximate approach, the lower order vacuum polarization (VP) effects are taken as the sum of
 the Uehling ($V_{U}(r)$) and 
the Wichmann-Kroll ($V_{WK}(r)$) potentials, which are given by
\begin{eqnarray}
V_{U}(r)&=&  - \frac{2 \alpha^2 }{3 r} \int_0^{\infty} dx \ x \ \rho(x) 
\int_1^{\infty}dt \sqrt{t^2-1} \nonumber \\ && \times 
\left(\frac{1}{t^3}+\frac{1}{2t^5}\right)  \left [ e^{-2ct|r-x|} - e^{-2ct(r+x)} \right ] \ \ \
\end{eqnarray}
and
\begin{eqnarray}
V_{WK}(r)&=&-\frac{8 Z^2 \alpha^4 }{9 r} (0.092) \int_0^{\infty} dx \ x \ \rho(x)  \nonumber \\ && \times \big ( 0.22 
\big \{ \arctan[1.15(-0.87+2c|r-x|)] \nonumber \\ && - \arctan[1.15(-0.87+2c(r+x))] \big \} \nonumber \\ && + 0.22 
\big \{ \arctan[1.15(0.87+2c|r-x|)] \nonumber \\ && - \arctan[1.15(0.87+2c(r+x))] \big \} \nonumber \\ 
&& - 0.11 \big \{ \ln[0.38 -0.87c|r-x|+c^2 (r-x)^2 ] \nonumber \\ 
&& - \ln[0.38 -0.87c (r+x) + c^2 (r+x)^2 ] \big \} \nonumber \\
&& + 0.11 \big \{ \ln[0.38 +0.87 c |r-x| + c^2 (r-x)^2 ] \nonumber \\ 
&& -  \ln[0.38 +0.87c (r+x) + c^2 (r+x)^2 ] \big \} \big ). \ \ \ \ \
\end{eqnarray}
The contributions from the self-energy (SE) interaction are evaluated by considering the contributions 
due to the electric form-factor given by
\begin{eqnarray}
V_{SE}^{ef}(r)&=& - A(Z) (Z \alpha )^4 e^{-Zr} + \frac{B(Z,r) \alpha^2 }{ r} \int_0^{\infty} dx  x  \rho(x) 
\nonumber \\ && \times \int^{\infty}_1 dt \frac{1}{\sqrt{t^2-1}} \big \{ \left( \frac{1}{t}-\frac{1}{2t^3} \right )\nonumber \\
&&\times \left [ \ln(t^2-1)+4 \ln \left ( \frac{1}{Z \alpha } +\frac{1}{2} \right ) \right ]-\frac{3}{2}+\frac{1}{t^2} \big \} \nonumber \\
&& \times \left [ e^{-2ct|r-x|} - e^{-2ct(r+x)} \right ]
\end{eqnarray}
and from the magnetic form-factor given by
\begin{eqnarray}
V_{SE}^{mg} (r) &=& \frac{i \alpha }{4 \pi c} \mbox{\boldmath$\gamma$} \cdot \mbox{\boldmath$\nabla$}_r \int_0^{\infty} d^3 x \ \rho(x) \nonumber \\
&& \times \left [ \left ( \int^{\infty}_{1}dt \frac{e^{-2tc R }}{Rt^2 \sqrt{t^2-1}}\right ) - \frac{1}{R} \right],
\end{eqnarray}
where $A(Z)=0.074+0.35Z \alpha $, $B(Z,r)=[1.071-1.97((Z-80) \alpha )^2 -2.128 ((Z-80) \alpha )^3+0.169 
((Z-80) \alpha )^4 ]cr/(cr+0.07(Z \alpha )^2 )$ and $R= | \textbf{r} - \textbf{x}|$. 

To incorporate the first order corrections due to the P,T-odd weak interactions, we express the total Hamiltonian of the atom as
\begin{eqnarray}
 H = H^{at} + \lambda H^{PT} ,
\end{eqnarray}
 where $H^{at}$ represents the atomic Hamiltonian; i.e. the DC Hamiltonian supplemented by higher order relativistic corrections if necessary 
 and $\lambda H^{PT}$ corresponds to either of the P,T-odd Hamiltonians given by Eqs. (\ref{nsmeq}) and (\ref{tpteq}). Here
 $\lambda$ can be $S$ or $C_{at}^T$ for the respective Hamiltonian. The atomic wave function can be expressed as
\begin{eqnarray}
 |\Psi_0 \rangle \approx |\Psi_0^{(0)} \rangle + \lambda |\Psi_0^{(1)} \rangle  ,
\end{eqnarray}
where nonlinear terms  in $\lambda$ have been neglected, and $| \Psi_0^{(0)} \rangle$ and $|\Psi_0^{(1)} \rangle$ are the wave functions of 
$H^{at}$ and its first order correction due to the P,T-odd interaction Hamiltonian, respectively. Hence Eq. (\ref{edmeq}) is approximated as
\begin{eqnarray}
 d_{at} & \simeq & 2 \lambda \frac{\langle \Psi_0^{(0)}|D|\Psi_0^{(1)} \rangle}{\langle \Psi_0^{(0)}|\Psi_0^{(0)} \rangle}.
 \label{eqed}
\end{eqnarray}
The actual quantity that is relevant for the calculation is
\begin{eqnarray}
 {\cal R} = d_{at} / {\lambda} &=& 2 \frac{\langle \Psi_0^{(0)}|D|\Psi_0^{(1)} \rangle}{\langle \Psi_0^{(0)}|\Psi_0^{(0)} \rangle}
\label{eqptt}
 \end{eqnarray}
 and it can be combined with the experimentally measured $d_{at}$ values to determine $\lambda$.

The first order perturbed wave function $|\Psi^{(1)} \rangle$ can be calculated by two different approaches. One is the sum-over-states
approach, where we express
\begin{eqnarray}
 |\Psi_0^{(1)} \rangle = \sum_{I \ne 0} |\Psi_I^{(0)} \rangle \frac{\langle \Psi_I^{(0)} | H^{PT}|\Psi_0^{(0)}\rangle}{E_I^{(0)} - E_0^{(0)}} ,
\end{eqnarray}
 where $|\Psi_I^{(0)} \rangle$s are the states other than $|\Psi_0^{(0)}\rangle$ of $H^{at}$ with the energies $E_I^{(0)}$ and $E_0^{(0)}$, which are the intermediate state
and the ground state energies respectively. The advantage of this approach is that one can include only the dominant contributions which come from the low-lying
states. However, this method cannot account for contributions from the core, the high-lying excited states including the
continuum which can be significant in heavy atomic systems. The other approach, which is often more desirable is to determine the first order perturbed
wave function by solving the following inhomogeneous equation
\begin{eqnarray}
(H^{at}-E_0^{(0)}) |\Psi_0^{(1)} \rangle &=& (E_0^{(1)}- H^{PT}) |\Psi_0^{(0)} \rangle \nonumber \\
                                         &=& - H^{PT} |\Psi_0^{(0)} \rangle ,
\label{eq4}
\end{eqnarray}
where the first order perturbed energy vanishes as $H^{PT}$ is an odd parity operator. It is well known that the accurate calculation of
$|\Psi_0^{(0)} \rangle$ in heavy atomic systems is challenging owing to the presence of the two-body Coulomb and Breit
interactions. So establishing suitable many-body methods to determine $|\Psi_0^{(0)} \rangle$ and its correction $|\Psi_0^{(1)} \rangle$
with the simultaneously inclusion of electromagnetic and weak interactions are of immense interest.

We discuss some of the all order relativistic many-body methods that have been employed to determine ${\cal R}$ in some of the
atoms of experimental interest such as $^{129}$Xe, $^{199}$Hg, $^{223}$Rn, and $^{225}$Ra. Earlier, only simpler approximated many-body methods
such as the relativistic third order many-body perturbation theory (MBPT(3) and the relativistic RPA were employed to determine these
quantities in $^{129}$Xe and $^{223}$Rn \cite{martensson,dzuba1,latha}. These methods cannot
accurately determine the values of ${\cal R}$ in other atoms like $^{199}$Hg and $^{225}$Ra, where pair-correlation effects contribute
significantly. Two calculations, where important classes of correlation effects were included using a perturbed relativistic coupled-cluster 
(PRCC) method \cite{lathalett} and a hybrid approach of configuration interaction with finite-order many-body perturbation theory (CI$+$MBPT) 
\cite{dzuba1} were employed to calculate the above quantities in $^{199}$Hg the results were fairly close. In the combined CI$+$MBPT
method, the initial single particle wave functions were determined using the $V^{N_c-2}$ potential with $N_c$ as the total number of electrons 
and the electron correlation effects are accounted for by dividing the electrons into valence and core electrons. For $^{225}$Ra, calculations 
using the CI$+$MBPT method with RPA corrections by Dzuba and coworkers \cite{dzuba1,dzuba_02} were performed. We shall discuss some of these methods
below. It is also worth mentioning here that the PRCC method and a relativistic coupled-cluster method that will be described subsequently are
similar; the difference between them lies in the treatment of the normalization of the wave function in the two theoretical approaches 
\cite{yashpal5}. 

To obtain both $|\Psi_0^{(0)} \rangle$ and $|\Psi_0^{(1)} \rangle$, we first calculate the Dirac-Hartree-Fock (DF) wave function 
$|\Phi_0 \rangle$ using the DF Hamiltonian
\begin{eqnarray}
H_{DF}^{at} &=& \sum_i  [c\mbox{\boldmath$\alpha$}_i\cdot \textbf{p}_i+(\beta_i -1)c^2+  
V_{nuc}(r_i) + U_{DF}(r_i)] \nonumber \\
      &=& \sum_i [h_0(r_i) + U_{DF}(r_i)],
\label{eq9}
\end{eqnarray}
with an average DF potential $U_{DF}(r)$, disregarding contributions from the residual interaction
\begin{eqnarray}
 V_{es}=\sum_{j>i}^N V(r_{ij}) -\sum_i U_{DF}(r_i).
 \label{eq10}
\end{eqnarray}
where $V(r_{ij})$ is(are) the two-body interaction (Coulomb or Coulomb with Breit) interaction operator(s).

The DF potential and the single particle wave function $|\phi_i^{(0)} \rangle$ of $|\Phi_0 \rangle$
are obtained by solving the following equations
\begin{eqnarray}
\langle \phi_i^{(0)}|U_{DF}|\phi_j^{(0)} \rangle &=&\sum_b^{N_c}[\langle \phi_i^{(0)} \phi_b^{(0)}|V(r_{ij}) |\phi_b^{(0)} \phi_j^{(0)}\rangle \nonumber \\
&& -\langle \phi_i^{(0)} \phi_b^{(0)}|V(r_{ij}) |\phi_j^{(0)} \phi_b^{(0)}\rangle]
\label{eq13}
\end{eqnarray}
and
\begin{eqnarray}
 (h_0+U_{DF})|\phi_i^{(0)}\rangle &=& \epsilon_i^{(0)}|\phi_i^{(0)} \rangle
 \label{eq12}
\end{eqnarray}
simultaneously in a self-consistent procedure. In the above expression, when the sum is considered up to $N_c-M$, for $M$ number of
valence electrons $M$, this particular kind of DF potential is referred to as the $V^{N_c-M}$ potential in the literature. 

We now focus on the similarities and differences between some of the many-body methods that have been 
widely employed to treat $V_{es}$ in order to calculate atomic properties. In particular, we present the pertinent details of certain 
methods that consider $V_{es}$ to all order. There are several ways to deal with this, but different approaches
will capture various correlation effects corresponding to approximations in the levels of particle-hole excitations. For a comprehensive
understanding of these theories, we take recourse to an approach based on the Bloch equation \cite{lindgren} in which one expresses
\begin{eqnarray}
|\Psi_0^{(0)} \rangle &=&  \Omega^{(0)} |\Phi_0 \rangle = \sum_k^n \Omega^{(k,0)} |\Phi_0 \rangle,
\end{eqnarray}
where $\Omega^{(0)}$ is known as the wave operator in the MBPT(n) method that accounts only up to $n$ (say) orders of Coulomb interactions 
and $k$ represents the order of $V_{es}$ associated with each wave operator in a perturbative expansion of $\Omega^{(0)}$. In the presence
of another external interaction, like the operator $H^{PT}$, the exact state can be written as
\begin{eqnarray}
 |\Psi_0 \rangle &=& \Omega |\Phi_0 \rangle = \sum_{\beta}^n \sum_{\delta}^m \Omega^{(\beta,\delta)} |\Phi_0 \rangle,
 \label{eq14}
\end{eqnarray}
where the perturbation expansion is described by $n$ orders of $V_{es}$ and $m$ orders of $H^{PT}$. For our
requirement of obtaining the first order wave function due to $H^{PT}$, we have
\begin{eqnarray}
 |\Psi_0^{(1)} \rangle  &=& \sum_{\beta}^n \Omega^{(\beta,1)} |\Phi_0 \rangle .
\end{eqnarray}

To obtain the solutions for the wave operators, we use the following generalized Bloch equations
\begin{eqnarray}
 [\Omega^{(\beta,0)},H_{DF}^{at} ] P &=& Q V_{es} \Omega^{(\beta-1,0)}P  \nonumber \\ && -
 \sum_{m=1 }^{\beta-1} \Omega^{(\beta-m,0)} P V_{es} \Omega^{(m-1,l)}P \ \  \  \  \
\end{eqnarray}
and
\begin{eqnarray}
 [\Omega^{(\beta,1)},H_{DF}^{at} ]P &=& QV_{es} \Omega^{(\beta-1,1)}P + Q D \Omega^{(\beta,0)}P 
\nonumber \\ && - \sum_{m=1 }^{\beta-1} \big ( \Omega^{(\beta-m,1)}
 P V_{es} \Omega^{(m-1,0)}P \nonumber \\ && - \Omega^{(\beta-m,1)}PD \Omega^{(m,0)}P \big ),
\label{eq44}
\end{eqnarray}
where $P= |\Phi_0 \rangle \langle \Phi_0 |$ and $Q=1-P$. It implies that $\Omega^{(0,0)}=1$, $\Omega^{(1,0)}=
\sum_I \frac{ \langle \Phi_I | V_{es} | \Phi_0 \rangle} { E_I^{DF} - E_0^{DF}} = 0$ and $\Omega^{(0,1)}= \sum_I 
\frac{ \langle \Phi_I | H^{PT} | \Phi_0 \rangle} { E_I^{DF} - E_0^{DF}}$. Here $|\Phi_I\rangle$ with DF energy $E_I^{DF}$ is an 
excited state with respect to $|\Phi_0\rangle$ and $E_0^{DF}$ is the sum of DF single particle energies.

In the case of the $V^{N_c-M}$ potential, it requires a slightly different formalism to account for the electron correlation effects.
In this approach, electrons are divided into a closed core and $M$ valence electrons which are expected to play
the major role in describing the dominant part of the electron correlation effects. The wave operator in such a scenario can be
expressed as 
\begin{eqnarray}
 \Omega =  1+ \chi_c + \chi_v ,
\end{eqnarray}
where $\chi_c$ and $\chi_v$ are the operators that are responsible for excitations within the closed-core (say $\vert \Phi_c \rangle$) and
among the valence orbitals (say $\vert \Phi_v \rangle$), respectively. It is necessary to solve a set of equations similar to those above
by expanding the wave operators as 
\begin{eqnarray}
 \chi_c = \sum_{\beta}^n \sum_{\delta}^m \chi_c^{(\beta,\delta)} 
 \end{eqnarray}
 and
 \begin{eqnarray}
\chi_v=\sum_{\beta}^n \sum_{\delta}^m \chi_v^{(\beta,\delta)} .
\end{eqnarray}
Core-valence correlations must also be taken into account in this approach. The other demerit of this approach
is that the orbitals and all the correlation effects are not treated on equal footing. In particular, the correlations among the
valence electrons are estimated ambiguously. This may not be appropriate for the heavier atoms when the core correlations are 
quite significant.

Below we discuss a few many-body methods starting with a common DF wave function $|\Phi_0 \rangle$ constructed using the $V^N$
potential. Later we shall present results from these methods to demonstrate the gradual inclusion of the electron correlation effects
from lower to higher order in a variety of all order perturbative methods.  

\subsubsection{The DF method}

Following Eqs. (\ref{edmeq}) and (\ref{eq44}), we can obtain the lowest order contribution to ${\cal R}$, the DF result, as 
\begin{eqnarray}
 {\cal R}  &=& 2 \langle \Phi_0| {\Omega^{(0,0)}}^{\dagger} D \Omega^{(0,1)} |\Phi_0 \rangle = 2 \langle \Phi_0| D \Omega^{(0,1)} |\Phi_0 \rangle \nonumber \\
& =& 2 \sum_I \frac{  \langle \Phi_0| D| \Phi_I \rangle \langle \Phi_I | H^{PT} | \Phi_0 \rangle} { E_I^{DF} - E_0^{DF}} .
\end{eqnarray}

\subsubsection{The MBPT(k) method}

In this  approximation, we assume $(k-1)$ orders of Coulomb and one order $H^{PT}$. Thus, it corresponds to
\begin{eqnarray}
{\cal R} &=& 2 \frac{\sum_{\beta=0}^{k-1} \langle \Phi_0| {\Omega^{(k-\beta,0)}}^{\dagger} D \Omega^{(\beta,1)} |\Phi_0 \rangle}
{ \sum_{\beta=0}^{k-1} \langle \Phi_0| {\Omega^{(k-\beta,0)}}^{\dagger} \Omega^{(\beta,0)} |\Phi_0 \rangle} 
\end{eqnarray}
This quantity can be expressed at the MBPT(2) as
\begin{eqnarray}
{\cal R} &=& \frac{2}{{\cal N}_2} \langle \Phi_0 | [\Omega^{(0,0)}+\Omega^{(1,0)}]^{\dagger} D [\Omega^{(0,1)}+\Omega^{(1,1)}]|\Phi_0 \rangle \nonumber \\
&=& \frac{2}{{\cal N}_2} \langle \Phi_0| D\Omega^{(0,1)} + D\Omega^{(1,1)} + {\Omega^{(1,0)}}^{\dagger} D\Omega^{(0,1)} \nonumber \\&&  +
{\Omega^{(1,0)}}^{\dagger} D\Omega^{(1,1)} |\Phi_0 \rangle  ,
\label{eq20}
\end{eqnarray}
and similarly in the MBPT(3) method it is
\begin{eqnarray}
 {\cal R}  &=& \frac{2}{{\cal N}_3} \langle \Phi_0 | [\Omega^{(0,0)}+\Omega^{(1,0)}+\Omega^{(2,0)}]^{\dagger} D \nonumber \\ && \times[\Omega^{(0,1)}+\Omega^{(1,1)}+\Omega^{(2,1)}]|\Phi_0 \rangle \nonumber \\
&=& \frac{2}{{\cal N}_3} \langle \Phi_0| D\Omega^{(0,1)} + D\Omega^{(1,1)}+D\Omega^{(2,1)} + {\Omega^{(1,0)}}^{\dagger} D\Omega^{(0,1)}  \nonumber \\ && +
{\Omega^{(1,0)}}^{\dagger} D\Omega^{(1,1)} +{\Omega^{(2,0)}}^{\dagger} D\Omega^{(0,1)}|\Phi_0 \rangle  ,
\label{eq21}
\end{eqnarray} 
with the respective normalization constants  ${\cal N}_2=\langle \Phi_0| 1 +{\Omega^{(1,0)}}^{\dagger} \Omega^{(1,0)} |\Phi_0 \rangle$ 
and ${\cal N}_3=\langle \Phi_0| 1 + {\Omega^{(1,0)}}^{\dagger} \Omega^{(1,0)} + {\Omega^{(1,0)}}^{\dagger} \Omega^{(2,0)} + {\Omega^{(2,0)}}^{\dagger} \Omega^{(1,0)}
+ {\Omega^{(2,0)}}^{\dagger} \Omega^{(2,0)}|\Phi_0 \rangle$. 

The above expressions clearly indicate that the complexity of the calculations grows steadily as the order of perturbation increases. We describe 
two all order perturbative methods to describe the electron correlation effects on the properties of the closed-shell atoms.

\subsubsection{The RPA method}

To arrive at the final working equation for the RPA method, we start by perturbing the DF orbitals and the single particle energies due 
to the perturbation $H^{PT}$. i.e. 
\begin{eqnarray}
  |\phi_i^{(0)}\rangle && \rightarrow |\phi_i^{(0)} \rangle+\lambda|\phi_i^{(1)}\rangle 
  \end{eqnarray}
 and 
 \begin{eqnarray}
 \epsilon_i^{(0)} &&\rightarrow \epsilon_i^{(0)} + \lambda \epsilon_i^{(1)} ,
 \label{eq22}
\end{eqnarray}
where $|\phi_i^{(1)} \rangle$ and $\epsilon_i^{(1)}$ are the first order corrections to the particle wave function and energy, respectively.
Owing to the fact that $H^{PT}$ is an odd parity operator, $\epsilon_i^{(1)}=0$. In the presence of a perturbation, the modified DF equation
for the single particle wave function yields
\begin{eqnarray}
(h_0+\lambda H^{PT})(|\phi_i^{(0)} \rangle+\lambda|\phi_i^{(1)}\rangle) &+& \sum_b^{N_c} (\langle \phi_b^{(0)}+ \lambda \phi_b^{(1)} |
 V(r_{ij}) \nonumber \\  |\phi_b^{(0)} + \lambda \phi_b^{(1)} \rangle |\phi_i^{(0)} &+& \lambda \phi_i^{(1)}\rangle \nonumber \\  -
 \langle \phi_b^{(0)} + \lambda  \phi_b^{(1)} | V(r_{ij}) |\phi_i^{(0)} &+& \lambda \phi_i^{(1)} \rangle |\phi_b^{(0)}  +\lambda \phi_b^{(1)} \rangle) \nonumber \\  
  &=& \epsilon_i^{(0)} (|\phi_i^{(0)} \rangle+\lambda|\phi_i^{(1)} \rangle).
 \label{eq23}
\end{eqnarray}
Collecting only the terms that are linear in $\lambda$, we get
\begin{eqnarray}
 (h_0+U_{DF}-\epsilon_i^{(0)} )|\phi_i^{(1)}\rangle= (-H^{PT} -U_{DF}^{(1)})|\phi_i^{(0)} \rangle,  
 \label{eq24}
\end{eqnarray}
where we use the notation $U_{DF}^{(1)}$ for
\begin{eqnarray}
 U_{DF}^{(1)} |\phi_i^{(0)}\rangle &=&\sum_b^{N_c} [ \langle \phi_b^{(0)}| V(r_{ij}) |\phi_b^{(1)}\rangle |\phi_i^{(0)} \rangle \nonumber \\
&& -\langle \phi_b^{(0)}| V(r_{ij}) |\phi_i^{(0)} \rangle |\phi_b^{(1)} \rangle \nonumber \\ 
&&  +\langle \phi_b^{(1)}| V(r_{ij}) |\phi_b^{(0)} \rangle |\phi_i^{(0)} \rangle \nonumber \\ && -\langle \phi_b^{(1)}| V(r_{ij}) |\phi_i^{(0)}\rangle 
 |\phi_b^{(0)} \rangle ] . 
 \label{eq25}
\end{eqnarray}

We express the single particle perturbed wave function in terms of the unperturbed single particle wave functions as
\begin{eqnarray}
 |\phi_i^{(1)}\rangle=\sum_{j \ne i} C_i^j |\phi_j^{(0)} \rangle,
 \label{eq26}
\end{eqnarray}
where $C_i^j$s are the expansion coefficients. In the RPA approach, we write
\begin{eqnarray}
 \sum_{j \ne i} C_i^i (h_0 + U_{DF} - \epsilon_j^{(0)} ) |\phi_j^{(0)} \rangle= (- H^{PT} - U_{DF}^{(1)}) |\phi_i^{(0)}\rangle,  
\end{eqnarray}
and solve this equation self-consistently to obtain the $C_i^j$ coefficients to all orders in the Coulomb interaction.

The RPA wave operator can be expressed as
\begin{eqnarray}
\Omega_{RPA}^{(1)} &=&  \sum_k^{\infty} \sum_{p,a} \Omega_{a \rightarrow p}^{(k, 1)} \nonumber \\
    &=& \sum_{\beta=1}^{\infty} \sum_{pq,ab} { \{} \frac{[\langle pb | V(r_{ij})  | aq \rangle 
- \langle pb | V(r_{ij}) | qa \rangle] \Omega_{b \rightarrow q}^{(\beta-1,1)} } {\epsilon_p - \epsilon_a}  \nonumber \\ 
&& + \frac{ \Omega_{b \rightarrow q}^{{(\beta-1,1)}^{\dagger}}[\langle pq | V(r_{ij}) | ab \rangle - \langle pq | V(r_{ij}) | ba \rangle] 
}{\epsilon_p-\epsilon_a} { \}},
\label{eq27}
\end{eqnarray} 
where $a \rightarrow p$ means replacement of an occupied orbital $a$ from $|\Phi_0 \rangle$ by a
virtual orbital $p$ which alternatively refers to a singly excited state with respect to $|\Phi_0 \rangle$.
It can be shown in the above formulation that the RPA method subsumes a certain class of singly excited
configurations corresponding to the core-polarization effects to all orders.

Using the above RPA wave operator, we evaluate ${\cal R}$ by
\begin{eqnarray}
 {\cal R}  &=& 2 \langle \Phi_0| {\Omega^{(0,0)}}^{\dagger} D \Omega_{RPA}^{(1)} |\Phi_0 \rangle \nonumber \\
         &=& 2 \langle \Phi_0| D \Omega_{RPA}^{(1)} |\Phi_0 \rangle .
\end{eqnarray}

\subsubsection{The RCC theory}
\label{sec44}

In the RCC method, we express the unperturbed atomic wave function as
\begin{eqnarray}
 | \Psi_0^{(0)} \rangle &=& \Omega_{RCC}^{(0)}|\Phi_0 \rangle = \sum_k^{\infty} \Omega_{RCC}^{(k,0)}|\Phi_0 \rangle  \nonumber \\ 
                        &=& e^{T^{(0)}} |\Phi_0 \rangle 
\label{eq32}
\end{eqnarray}
and the first order perturbed wave function as
\begin{eqnarray}
 | \Psi_0^{(1)} \rangle &=& \Omega_{RCC}^{(1)} |\Phi_0 \rangle  = \sum_k^{\infty} \Omega_{RCC}^{(k,1)}|\Phi_0 \rangle  \nonumber \\ 
                        &=& e^{T^{(0)}} T^{(1)} |\Phi_0 \rangle, 
\label{eq33}
\end{eqnarray} 
where $T^{(0)}$ and $T^{(1)}$ are the excitation operators from the reference state $|\Phi_0 \rangle$
that take care of contributions from $V_{es}$ and $V_{es}$ along with the perturbed $H^{PT}$ operator, respectively. 

The amplitudes of the excitation $T^{(0)}$ and $T^{(1)}$ operators are determined using the equations
\begin{eqnarray}
 \langle \Phi_0^{\tau}|\overline{H_N^{at}}|\Phi_0\rangle&=&0
 \label{eq36}
 \end{eqnarray}
and
\begin{eqnarray}
\langle \Phi_0^{\tau}|\overline{H_N^{at}}T^{(1)}|\Phi_0\rangle&=&-\langle \Phi_0^{\tau}|\overline{H}_N^{PT}|\Phi_0\rangle ,
  \label{eq37}
\end{eqnarray}
where the subscript $N$ represents normal ordered form of the Hamiltonian, $\overline{O}=(Oe^{T^{(0)}})_{con}$ with $con$
means only the connected terms and $| \Phi_0^{\tau} \rangle$ corresponds to the 
excited configurations with $\tau$ referring to level of excitations from $| \Phi_0 \rangle$. In our 
calculations, we only consider the singly and doubly excited configurations ($\tau=1,2$) by defining
\begin{eqnarray}
T^{(0)} &=& T_1^{(0)} + T_2^{(0)} \ \ \ \ \text{and} \ \ \ \ 
T^{(1)} = T_1^{(1)} + T_2^{(1)} ,
\label{eq35}
\end{eqnarray}
which is known as the CCSD method in the literature. When we consider the approximation 
$\overline{O} \simeq O + O T$, we refer it as the LCCSD method.

\begin{table*}[t]
\caption{Calculated values of ${\cal R}$ due to both T-PT (given as ${\cal R}^{TPT}$ in $\times 10^{-20} \langle \sigma\rangle |e|cm$) 
and NSM (given as ${\cal R}^{NSM}$ in $\times [10^{-17}/|e|fm^3]|e|cm$) interactions in the $^{129}$Xe and $^{223}$Rn noble gas atoms. 
The final recommended values with uncertainties are given as ``Best value" for the respective quantities.}
\begin{center}
\begin{tabular}{l | cccc | cccc }
\hline \hline
               &  \multicolumn{4}{c|}{$^{129}$Xe}  &   \multicolumn{4}{c}{$^{223}$Rn}   \\
\cline{2-5} \cline{6-9}  \\
  Method &     \multicolumn{2}{c}{This Work}   & \multicolumn{2}{c|}{Others} &   \multicolumn{2}{c}{This Work}   & \multicolumn{2}{c}{Others} \\
 \cline{2-3}  \cline{4-5} \cline{6-7} \cline{8-9} \\
   &  ${\cal R}^{TPT}$ & ${\cal R}^{NSM}$ & ${\cal R}^{TPT}$ & ${\cal R}^{NSM}$  &  ${\cal R}^{TPT}$ & ${\cal R}^{NSM}$ & ${\cal R}^{TPT}$ & ${\cal R}^{NSM}$  \\ \hline
DF     & 0.447 &  0.288   &  0.45 \cite{dzuba1}   & 0.29 \cite{dzuba1}    &   4.485   & 2.459  & 4.6 \cite{dzuba1} &  2.5 \cite{dzuba1}, 2.47 \cite{dzuba_02} \\
MBPT(2)& 0.405 &  0.266   &     &      & 3.927  &  2.356  & & \\    
MBPT(3)& 0.515  & 0.339   & 0.52 \cite{martensson}   &    & 4.137  &  2.398 & & \\       
RPA    & 0.562  & 0.375 & 0.57 \cite{dzuba1}, 0.564 \cite{latha}   & 0.38 \cite{dzuba1}   & 5.400  & 3.311  & 5.6 \cite{dzuba1} & 3.3 \cite{dzuba1}, 3.33 \cite{dzuba_02} \\
LCCSD  &  0.608     & 0.417      &   &   &    5.069     & 3.055 &  & \\
CCSD$^{(3)}$ & 0.501  & 0.336 &   &   &  4.947      &  2.925 & & \\
CCSD$^{(5)}$ & 0.489   & 0.334  &   &   &  4.851      &  2.890\\
CCSD$^{(\infty)}$ & 0.475  & 0.333  &    &  &  4.459 & 2.782 & & \\
\hline\\
Best value &  0.475(4) & 0.333(4) &  &  &  4.46(6) & 2.78(4) \\
\hline \hline
\end{tabular}
\end{center}  
\label{tab1}
\end{table*}

 We have adopted an optimal computational strategy by constructing the intermediate diagrams in the RCC method. In this approach, we 
 divide the effective $\overline{H_N^{at}}$ and $\overline{H}_N^{PT}$ operators containing the non-linear CC terms into effective 
one-body, two-body etc. terms using the Wick's generalized theorem \cite{lindgren}. The intermediate diagrams for the computation of the 
$T^{(0)}$ amplitudes are described at length in our previous work \cite{yashpal1,yashpal2}. We define intermediate diagrams
for the evaluation of the $T^{(1)}$ amplitudes in a slightly different way. As can be seen from Eq. (\ref{eq37}),
$\overline{H_N^{at}}$ contains all the non-linear terms while for solving Eq. (\ref{eq36}) it is
required to express as $\overline{H_N^{at}}=\overline{H_N^{at}}' \otimes T_{\tau}$. Thus the intermediate
diagrams in the latter case comprise terms from $\overline{H_N^{at}}'$ which require special scrutiny of
the diagrams to avoid repetition in the singles and doubles amplitude calculations of $T^{(0)}$. These effective diagrams 
are finally connected with the respective $T$ operators to 
obtain the amplitudes of the singles and doubles excitations. Contributions from the terms of $\overline{H}_N^{PT}$ are evaluated 
directly in the $T^{(1)}$ amplitude calculations.

${\cal R}$  is evaluated by
\begin{eqnarray}
 {\cal R} &=& 2 \frac{\langle\Phi_0 | e^{T^{\dagger (0)}} D e^{T^{(0)}} T^{(1)} | \Phi_0 \rangle }
                  {\langle\Phi_0 | e^{T^{\dagger (0)}} e^{T^{(0)}} | \Phi_0 \rangle }.
\end{eqnarray}
Since all the operators in the above expression are in normal order form and $e^{T^{\dagger (0)}} D e^{T^{(0)}}$ is a non-terminating series,
we can express $e^{T^{\dagger (0)}} D e^{T^{(0)}}= (e^{T^{\dagger (0)}} e^{T^{(0)}})_{cl} (e^{T^{\dagger (0)}} D e^{T^{(0)}})_{cc}$ where
the subscript $cl$ and $cc$ mean closed and closed with connected terms, respectively \cite{pal,bartlett}. We can then show that
\begin{eqnarray}  
{\cal R} &=& 2 \frac{\langle\Phi_0 | (e^{T^{\dagger (0)}} e^{T^{(0)}})_{cl} (e^{T^{\dagger (0)}} D e^{T^{(0)}} T^{(1)})_{cc} | \Phi_0 \rangle }
                  {\langle\Phi_0 | (e^{T^{\dagger (0)}} e^{T^{(0)}})_{cl} | \Phi_0 \rangle } \nonumber \\
  &=& 2 \frac{\langle\Phi_0 | (e^{T^{\dagger (0)}} e^{T^{(0)}})_{cl} |\Phi_0 \rangle \langle \Phi_0 | (e^{T^{\dagger (0)}} D e^{T^{(0)}} T^{(1)})_{cc} | \Phi_0 \rangle }
                  {\langle\Phi_0 | (e^{T^{\dagger (0)}} e^{T^{(0)}})_{cl} | \Phi_0 \rangle } \nonumber \\
  &=& 2 \langle\Phi_0 |(\overline{D}^{(0)} T^{(1)})_{cc}|\Phi_0 \rangle,
\label{eq38}
\end{eqnarray}
with $\overline{D}^{(0)} = e^{T^{\dagger{(0)}}}De^{T^{(0)}}$, which is a non-terminating series. Note that its
$(e^{T^{\dagger (0)}} e^{T^{(0)}}T^{(1)})_{cl}$ part will
vanish owing to odd-parity of $T^{(1)}$. In the LCCSD method, we get $\overline{D}^{(0)} = D + DT^{(0)} + T^{\dagger{(0)}} D + 
T^{\dagger{(0)}} D T^{(0)}$. To account for contributions from $\overline{D}^{(0)}$ in the CCSD method, we first evaluate terms from 
$\overline{D}^{(0)}$ that are very unique in the sense that they will not be repeated after connecting with another $T^{(0)}$ or 
$T^{\dagger (0)}$ operator. Then, the contributions from the other non-linear terms are considered by contracting with another $T^{(0)}$ 
and $T^{\dagger (0)}$ operators till self-consistent results were achieved. We present these contributions with $k$ numbers of $T^{(0)}$ and/or 
$T^{\dagger (0)}$ as the CCSD$^{(k)}$ method to demonstrate convergence of the results with $k \rightarrow \infty$.

 In order to estimate the dominant contributions from the neglected triple excitations in the CCSD method, we define
an excitation operator by appealing to perturbation theory in the RCC framework as following
\begin{eqnarray}
 T_3^{(0),pert}= \frac{1}{3!}\sum_{abc,pqr} 
 \frac{ ( \overline{H}_a T_2^{(0)})_{abc}^{pqr} }{\epsilon_a
 + \epsilon_b+\epsilon_c-\epsilon_p -\epsilon_q -\epsilon_r} ,
 \label{eq30}
\end{eqnarray}
where $\epsilon$'s are the energies of the occupied (denoted by $a,b,c$) and unoccupied (denoted by $p,q,r$) orbitals. From the differences 
between the results from the CCSD method and from the calculations carried out including the $T_3^{(0),pert}$ operator with $T^{(0)}$
in the expression given by Eq. (\ref{eq38}), we find typical order of magnitude estimates from the triple excitations. Note that the contributions of the counterparts of these
excitations coming through the $T^{(1)}$ RCC operators will be extremely small.

The multi-configuration Dirac-Fock (MCDF) method has recently been used to calculate ${\cal R}$ for diamagnetic atoms \cite{jaceak}. This is an extension of the DF 
method, in which the wave function of an atomic state is expressed as a linear combination of the wave functions corresponding to appropriate configurations which are
built from the occupied and virtual orbitals. The orbitals and the mixing coefficients are determined simultaneously in a self-consistent manner by using the variational principle.
The P,T-odd Hamiltonian has been considered as a first order perturbation and ${\cal R}$ is calculated by explicitly summing over many of the important odd parity 
intermediate atomic states.
   
\begin{table*}[t]
\caption{Calculated values of ${\cal R}$ due to both T-PT (given as ${\cal R}^{TPT}$ in $\times 10^{-20} \langle \sigma\rangle |e|cm$) 
and NSM (given as ${\cal R}^{NSM}$ in $\times [10^{-17}/|e|fm^3]|e|cm$) interactions in the $^{199}$Hg and $^{225}$Ra diamagnetic atoms. 
The final recommended values with uncertainties are given as ``Best value" for the respective quantities.}
\begin{center}
\begin{tabular}{l | cccc | cccc }
\hline \hline
               &  \multicolumn{4}{c|}{$^{199}$Hg}  &   \multicolumn{4}{c}{$^{225}$Ra}   \\
\cline{2-5} \cline{6-9}  \\
  Method &     \multicolumn{2}{c}{This Work}   & \multicolumn{2}{c|}{Others} &   \multicolumn{2}{c}{This Work}   & \multicolumn{2}{c}{Others} \\
 \cline{2-3}  \cline{4-5} \cline{6-7} \cline{8-9} \\
   &  ${\cal R}^{TPT}$ & ${\cal R}^{NSM}$ & ${\cal R}^{TPT}$ & ${\cal R}^{NSM}$  &  ${\cal R}^{TPT}$ & ${\cal R}^{NSM}$ & ${\cal R}^{TPT}$ & ${\cal R}^{NSM}$  \\ \hline
DF     & $-$2.39  &$-$1.20 & $-$2.0 \cite{martensson} & $-$1.19 \cite{dzuba_02}  & $-3.46$ & $-1.86$ & $-3.5$ \cite{dzuba1} & $-1.8$ \cite{dzuba1} \\
       &          &        &    $-2.4$ \cite{dzuba1}  &   $-1.2$ \cite{dzuba1} &         &         &                      &           \\
       &          &        &    $-7.29$ \cite{jaceak}$^a$ &   $-2.86$ \cite{jaceak}$^a$  &         &         &                      &            \\                      
MBPT(2)& $-$4.48  &$-$2.30  &   &          & $-11.00$ & $-5.48$ & \\    
MBPT(3)& $-$3.33  &$-$1.72   &   &         & $-10.59$ & $-5.30$ & \\       
RPA    & $-$5.89  &$-$2.94   & $-$6.0 \cite{martensson}  & $-$2.8 \cite{dzuba_02}  & $-16.66$ & $-8.12$ & $-17$ \cite{dzuba1}  & $-8.3$ \cite{dzuba1}  \\
       &          &          & $-5.9$ \cite{dzuba1}      & $-3.0$ \cite{dzuba1}  &          &         & $-16.59$ \cite{latha} & $-8.5$ \cite{dzuba_02} \\               
CI+MBPT&       &         &    $-$5.1 \cite{dzuba1} & $-$2.6 \cite{dzuba1} &  &  & $-18$ \cite{dzuba1} & $-8.8$ \cite{dzuba1} \\
PRCC   &       &         &    $-$4.3 \cite{lathalett} & $-$2.46 \cite{lathalett} & & \\ 
MCDF   &       &         &   $-4.84$ \cite{jaceak}$^a$   & $-2.22$ \cite{jaceak}$^a$             &  & \\
LCCSD   & $-$4.52  &$-$2.24   &   &                &  $-13.84$ & $-8.40$ & & \\
CCSD$^{(3)}$   &$-$3.82  &$-$2.00   &   &          &  $-10.40$ & $-6.94$ & & \\
CCSD$^{(5)}$   &$-$4.02  &$-$2.00   &   &          & $-10.01$ & $-6.79$ & & \\
CCSD$^{(\infty)}$ & $-3.38$ &  $-1.78$ &    &   & $-9.926$ & $-6.215$ \\
\hline \\
Best value  & $-3.4(5)$  & $-1.8(3)$ &  & & $-9.93(8)$  & $-6.22(6)$  &  \\
\hline \hline
\end{tabular}
\begin{tabular}{c}
$^a$ Only the {\it ab initio} results are cited for the comparison.\\
\end{tabular}
\end{center} 
\label{tab2}
\end{table*}

\subsection{Atomic results}

In Table \ref{tab1}, we present the calculated ${\cal R}$ values from T-PT and NSM interactions for $^{129}$Xe and $^{223}$Rn
noble gas atoms using the methods that we have described in this review, and also from previously reported calculations. From
a theoretical point of view, it would be instructive to compare the correlation trends for both the atoms as they 
belong to the same periodic table of elements. As can be seen from the results quoted from different methods with the lower 
to higher order approximations, the magnitudes first decrease, then increase and the final results increase marginally from
their DF values for both the atoms and for both the interactions. However, on close scrutiny suggests it is evident that correlation effects
are stronger in $^{223}$Rn due to its larger size. The previous calculations, referred to in the above table, carried out using
the DF, MBPT(3) and RPA methods \cite{dzuba1,martensson,dzuba_02,latha} cannot capture some
of the correlation effects in ${\cal R}$ for the ground states of heavy inert gases in an efficient manner. The RPA method ignores
pair correlation contributions, but takes into account the core-polarization effects to all orders. It is therefore not surprising
that the results of the RPA and CCSD methods differ significantly. In fact, there are large cancellations between the results from
the all order RPA and the all-order non-RPA contributions at the CCSD level. The importance of including non-RPA correlation effects can be realized from
the differences in the results between the MBPT(2) and MBPT(3) methods as the non-RPA contributions first start appearing at the
 MBPT(3) approximation in a perturbative theory framework. The MCDF method very often cannot capture the polarization effects arising
 from the deep core efficiently for practical reasons. The large differences seen among the results from the LCCSD and CCSD methods
 and among the results obtained  with various levels of truncation in the CCSD$^{(k)}$ calculations suggest that there are strong 
cancellations between the linear and non-linear RCC terms for estimating ${\cal R}$ values. More detailed discussions on these 
results can be found elsewhere \cite{yashpal3,yashpal4}. It is, therefore, imperative to use an all order approach like our CCSD 
method to capture both the RPA and non-RPA correlation contributions. To assess the accuracies of our CCSD results, we also
estimate order of magnitudes of the neglected effects, such as corrections due to the truncated basis in the construction of
atomic orbitals and higher level excitations (estimating from the leading order triply excitations). We provide recommended values along 
with the net uncertainties at the end of Table \ref{tab1} quoting as ``Best value''. These results in combination 
with the measured EDMs of the $^{129}$Xe and $^{223}$Rn atoms would provide best limits on $C_T$ and $S$ when they become available.

\begin{table}[t]
\caption{Breakdown of contributions to the ${\cal R}$ values from the CCSD method due to the T-PT interaction (in $\times 10^{-20} 
\langle \sigma\rangle |e|cm$) in the considered diamagnetic atoms.}
\begin{center}
\begin{tabular}{lcccc}
\hline \hline
CC term & $^{129}$Xe & $^{223}$Rn & $^{199}$Hg & $^{225}$Ra \\
\hline
 & & & & \\
$DT_1^{(1)}$                &  0.459  & 4.345   &  $-4.400$  & $-13.10$  \\
$T_1^{(0)\dagger}DT_1^{(1)}$&  $-0.001$ & 0.005   & 0.027   & $-0.100$  \\
$T_2^{(0)\dagger}DT_1^{(1)}$&  0.039   & 0.333   &  1.224  &  3.303 \\
$T_1^{(0)\dagger}DT_2^{(1)}$&  $-0.006$   & $-0.069$   &  $-0.058$  & $-0.086$  \\
$T_2^{(0)\dagger}DT_2^{(1)}$&  $-0.009$   & $-0.108$   &  0.107  &  0.778 \\
Extra                       & $-0.007$    & $-0.047$   &  $-0.28$  &  $-0.721$  \\
\hline \hline
\end{tabular} 
\end{center}
\label{tab3}
\end{table}

The diamagnetic atoms, $^{199}$Hg and $^{225}$Ra, are the two current leading candidates for EDM experiments. The electron correlation effects in
these two atoms are strong. The primary reason for this is that the leading ground state configuration which has two $s$  electrons, mixes fairly strongly with
low lying opposite configurations with $s$  and $p$ electrons and the correlation effects modifying them make substantial contributions.
We present ${\cal R}$ values in Table \ref{tab2}
from all the methods that we have discussed earlier in the same sequence as were given in Table \ref{tab1}. As can be seen,
the trends in the results are completely different from those in the noble gas atom discussed in the previous paragraph. Unlike
the noble gas atoms, the differences between the results from RPA and RCC methods are quite large. The final
results, especially for $^{225}$Ra, are significantly different from their corresponding DF results (refer \cite{yashpal5,yashpal6}
for more discussions). In fact, our DF and RPA results are in close agreement with
calculations reported earlier, but our final CCSD result for Ra is very different from that of the CI+MBPT
method. This makes a strong case for using a suitable relativistic
many-body theory that can treat both the core polarization and pair correlations to all orders and treat them on equal footing.
Adopting a procedure similar to the one discussed earlier, we also estimate uncertainties of the calculated ${\cal R}$ values of $^{199}$Hg
and $^{225}$Ra and quoted the ``Best values'' towards the bottom of Table \ref{tab2}. Moreover, we had evaluated dipole 
polarizabilities of the considered systems by replacing the P,T-odd interaction Hamiltonians by the dipole operator and had compared 
them against the available experimental values and other sophisticated calculations to gauge validity of our methods (more  
discussions are given in our previous works \cite{yashpal5,yashpal3,yashpal4,yashpal6}). In this view, we find our results are
the most accurate calculations to date due to a balanced treatment of all possible electron correlation effects exhibited by these atoms.

\begin{table}[t]
\caption{Breakdown of contributions to the ${\cal R}$ values from the CCSD method due to the NSM interaction (in $\times [10^{-17}/|e|fm^3]|e|cm$)
in the considered diamagnetic atoms.}
\begin{center}
\begin{tabular}{lcccc}
\hline \hline
CC term & $^{129}$Xe & $^{223}$Rn & $^{199}$Hg & $^{225}$Ra \\
\hline
 & & & & \\
$DT_1^{(1)}$                & 0.313   & 2.695    &  $-2.388$  & $-7.577$  \\
$T_1^{(0)\dagger}DT_1^{(1)}$& $-0.001$   & $-0.004$ & 0.018   &  0.008   \\
$T_2^{(0)\dagger}DT_1^{(1)}$& 0.023   & 0.134    & 0.607  &   1.557  \\
$T_1^{(0)\dagger}DT_2^{(1)}$& 0.0002   & $-0.006$    & 0.011   & 0.046    \\ 
$T_2^{(0)\dagger}DT_2^{(1)}$&  0.004  & 0.020    & $-0.026$   &   $-0.594$  \\
Extra                       & $-0.006$   & $-0.057$    & $-0.002$   & 0.345    \\
\hline \hline
\end{tabular} 
\end{center}
\label{tab4}
\end{table}

\subsection{Analyzing the CCSD results}

It is possible to get insights into the contributions from the singly excited and doubly excited configurations for the
calculation of ${\cal R}$ values using the expression given in Eq. (\ref{eq38}). The total sum of contributions from CCSD terms
associated with $T_1^{(1)}$ and $T_2^{(2)}$ represent contributions from the singly excited and doubly excited configurations,
 respectively. Unlike a CI method where configurations are explicitly selected in the calculations, the RCC operators generate
 all possible configurations automatically that are allowed. In Table \ref{tab3} and Table \ref{tab4}, we present contributions from various 
 CCSD terms to the ${\cal R}$ values due to the T-PT interaction and NSM respectively for all the atoms we have considered. The net
 contributions from the leading doubly excited odd parity configuration state functions. Though it appears as if the contribution
due to the leading singly excited odd parity configuration state functions, while the remaining terms represent for the 
contributions from the leading doubly excited odd parity configuration state functions. Though it appears as if the contributions 
from the non-linear terms, given as ``Extra'', in the table are small, but actually the major contributions from the non-linear RCC
terms have been included through the evaluation of the $T^{(0)}$ and $T^{(1)}$ amplitude equations. It is also worth mentioning here
is that in accordance with the description in in Sec. \ref{sec44}, corrections due to the normalization of the wave functions are not necessary
here.

 Now comparing the trends of contributions to ${\cal R}$ due to the T-PT interaction from all the atoms given in Table
\ref{tab3}, we find that the correlation trends for $^{129}$Xe and $^{223}$Rn are almost similar, but they are very different for
$^{199}$Hg and $^{225}$Ra. With respect to the DF values given in Tables \ref{tab1} and \ref{tab2}, the $DT_1^{(1)}$
contributions are very large in $^{225}$Ra than for other atoms. This means the correlation effects are very strong in $^{225}$Ra
and to account for these effects rigorously, it is imperative to use a powerful many-body method like our RCC theory. 
Though correlation trends between the $^{199}$Hg and $^{225}$Ra atoms are almost the same, but the strong correlation effects 
in the $^{225}$Ra atom suggest that the latter behaves more like an open-shell atom.

Since both the rank and parity of the P and T odd interaction Hamiltonians given by Eqs. (\ref{nsmeq}) and (\ref{tpteq}) are the same, one would
expect that the correlation trends for ${\cal R}$ values due to the T-PT interaction and due to the interaction of the atomic electrons
with the NSM to be similar. However,  a comparison between these values given in Tables \ref{tab3} and \ref{tab4} from the different CCSD
terms reveals that this is not the case. The reason may be due to an extra $r$ dependence appearing in Eq. (\ref{nsmeq}). One can also 
see that the trends for the contributions from the NSM interaction are different  for all the atoms.

\section{Experiments on EDMs of closed-shell atoms\label{sec:experiment}}

\begin{figure}[t]
\begin{center} 
\includegraphics[height=5.0cm,width=8.5cm]{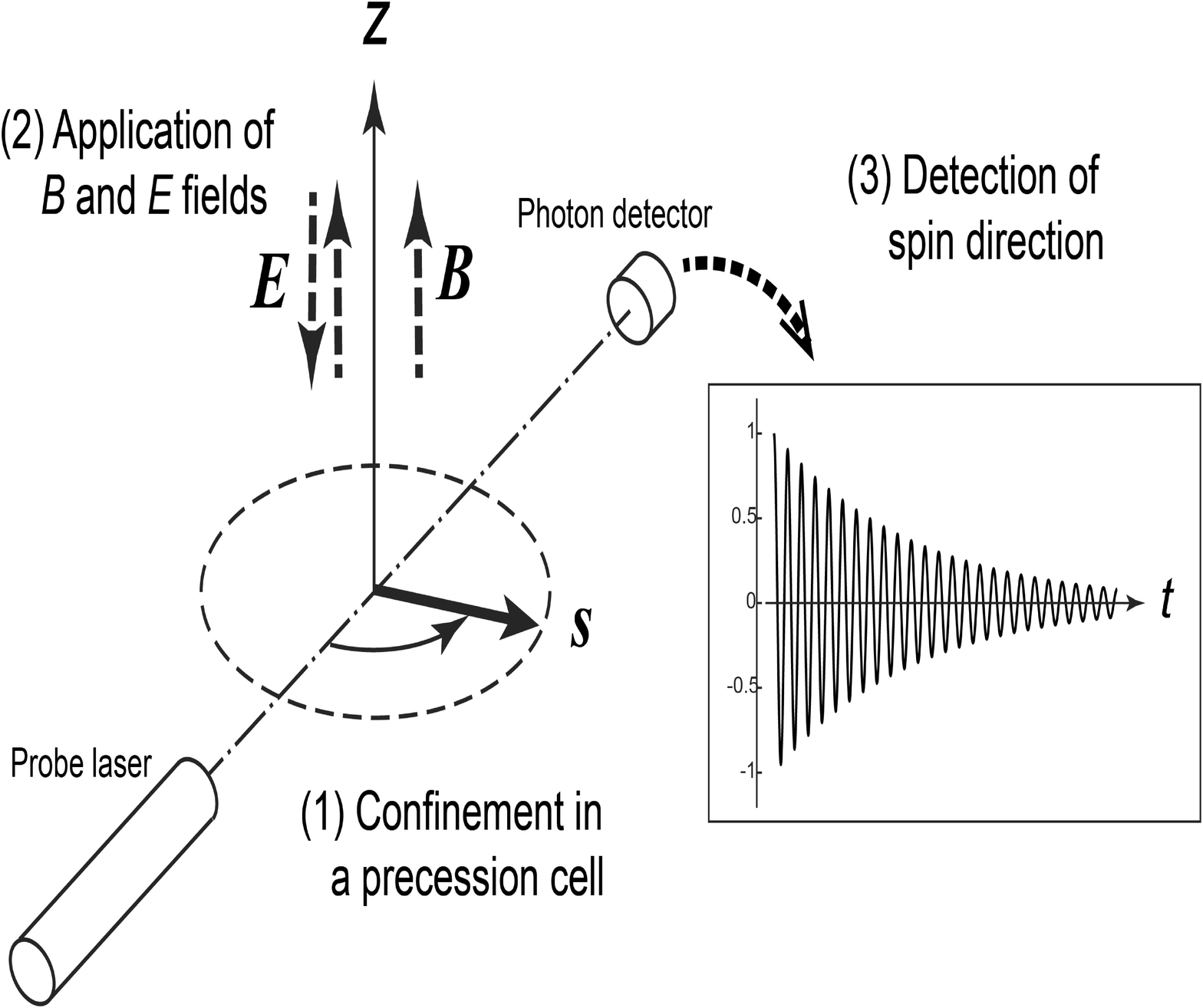}
\caption{\label{fig_setup} Geometry and concepts relevant to typical EDM measurements.}
\end{center}
\end{figure}

The EDM is a property of a spin carrying particle, and is detected through observation 
of the difference in energy between two spin states, {\it i.e.} with {\boldmath$s$} parallel and antiparallel to 
a static electric field {\boldmath$E$}. In a typical EDM measurement, the particle is placed under 
a static magnetic field {\boldmath$B$} so that the Zeeman energy splitting  between the magnetic 
substates $m = +1/2$ and  $-1/2$, or the spin precession frequency $\omega$, 
changes upon a reversal of  {\boldmath$E$}. Thus, the EDM $d$ is determined from a relation 
\begin{eqnarray}
 d_{at} = \frac{\hbar(\omega_{+}-\omega_{-})}{4E }
 \label{eq_Exp01}
\end{eqnarray}
where symbols $+$ and $-$ refer to the directions of the $E$ field parallel and antiparallel, 
respectively, to the $B$ field. In order to measure the precession frequencies $\omega_{\pm}$, 
typically the spin is polarized and is pointed toward a direction transverse to {\boldmath$B$} 
at time $t = 0$, from which the spin starts precession about the {\boldmath$B$} direction. 
Figure \ref{fig_setup} schematically illustrates geometry and concepts relevant to such an experiment: 
(1) confinement of polarized spins within some space ({\it e.g.} in a gas cell for 
the cases of Hg \cite{graner} and Xe \cite{rosenberry} atoms, 
in an optical dipole trap for the Ra case \cite{parker}, or in a 
beam path for the TlF case \cite{cho1_89,cho2_89}), (2) application of static $B$ and $E$ fields, and 
(3) detection of spin direction (typically through transmission of a circularly 
polarized laser light). To get a feeling on the typical scales for experimental 
settings and signals, one might recall that a size of EDM of $10^{-28}$ $e$ cm aimed 
to be detected poses to experimenter a requirement of frequency determination 
with a nHz precision under an applied $E$ field of 10 kV/cm and a stabilization
or monitoring of the $B$ field within 0.1 fT. 

With such demanding conditions required, any experiment which either have 
provided the present lowest limits or are newly proposed to revise them 
are equipped with their own advantageous features: Currently the best restricting 
limit on the EDM of a closed-shell atom has been obtained for $^{199}$Hg in \cite{griffith,graner}, 
in which a four-cell scheme is employed with the inner two cells placed under 
applied electric fields pointing opposite directions to each other and the two 
outer cells under a zero $E$ field.  The signal for the EDM is observed as 
a difference of the precession frequencies for the middle two cells, and the other 
combinations of the four cells are used to measure the averaged magnetic field and its gradient.  
Optical pumping is being used to spin-polarize 
the atoms orthogonal to the applied magnetic field, and the Faraday rotation of 
near-resonant light is observed to determine an electric-field-induced perturbation 
to the Larmor precession frequency carrying information on EDM.  As a result, they obtained 
$d( ^{199}\rm{Hg})=(-2.20 \pm 2.75(\rm{stat}) \pm 1.48 (\rm{sys})) \times 10^{-30}$ $e$cm 
corresponding to a new upper limit  
$|d( ^{199}\rm{Hg})| < 7.4 \times 10^{-30}$ $e$cm with 95\% confidence level \cite{graner}. 

The EDM of $^{129}$Xe has been studied in \cite{rosenberry} by taking advantage of 
high Xe polarization obtained using spin exchange optical pumping technique 
and of unlimitedly long spin coherence time realized by virtue of a spin maser, 
yielding the result 
$d(^{129}\mbox{Xe}) = [0.7\pm 3.3(\mbox{stat}) \pm 0.1(\mbox{syst})]\times10^{-27} e\mbox{cm}$.  
Also there are several currently ongoing developments on 
the $^{129}$Xe EDM: Tokyo Tech-RIKEN group introduces a new scheme for the 
spin-coherence time elongation, an external feedback spin oscillator \cite{inoue-xe,yoshimi,yoshimi_12} 
which works even at very low $B$ fields below $0.2 \mu\mbox{T}$.  
Heil {\it et al}. \cite{schmidt} take advantage 
of remarkably long transverse relaxation times which are only realized under 
extremely high homogeneity of their magnetic field.  Kuchler {\it et al}. \cite{fierlinger} 
proposes a completely new method for detecting the EDM, in which a 
rotating $E$ field is used instead of a static one.  

Still another way is at work to confront the demanding conditions posed to 
EDM experiments on closed-shell atoms: 
As discussed in Sect. 4 the nuclear octupole deformation/vibration, 
the parity-odd type collective degrees of freedom in the nucleus, can bring 
a large enhancement to the Schiff moment \cite{auerbach}.   
At present, the appearance of the octupole collectivity is known to be in rather 
limited regions of the nuclear chart, and the candidate nuclei currently 
attracting attention are radioactve isotopes:    
EDM measurements of the radioactive $ ^{225}$Ra atom have been carried out recently \cite{bishof,parker}.  
EDM of $^{225}$Ra nucleus is calculated to be 2-3 orders of magnitude larger than that of $^{199}$Hg 
\cite{Dobaczewski05,Spevak,dzuba_02}
due to octupole deformation. This means that 
even a one to two orders of magnitude less accurate determination 
of the EDM of  $^{225}$Ra atom than that of $^{199}$Hg 
is still advantageous for extracting the required physics. 
To measure the EDM of $^{225}$Ra atom, 
a cold-atom technique has been developed to detect  
spin precession of this atom held in an optical dipole trap 
and an upper limit has been set as $|d( ^{225}\rm{Ra})|
< 1.4 \times 10^{-23}$ $e$cm with 95\% confidence level \cite{bishof}. 
Likewise, an experiment to measure EDM of $^{223}$Rn is 
under development \cite{rand,tardiff}.  

\section{Discussion\label{sec:discussion}}

\subsection{EDMs of diamagnetic atoms and the underlying elementary particle physics}

By combining coefficients related to CP violating processes mentioned in the previous sections, we can write a general formula for the dependence of the EDMs of diamagnetic atoms on elementary 
level CP violation in the following manner
\begin{eqnarray}
d_{at}
&=& \sum_i K_{ i} x_i ,
\label{eq:atomicedmcoef}
\end{eqnarray}
where $x_i = d_e$, $d_u$, $d_d$, $d_s$, $d^c_u$, $d^c_d$, $d^c_s$, $w$, $C_{eq}^{T}$, 
$C_{eq}^{SP}$, $C_{eq}^{PS}$, and other CP-odd four-quark couplings ($i$ is the index of CP-odd effects considered) and $K_{i}$ is the ratio of the atomic EDM to the
CP violating coefficient.
We list these coefficients obtained from different levels of calculations in Table \ref{table:edmcoef}.

\begin{table*}[t]
\caption{\label{table:edmcoef}The coefficients relating to the EDMs of the diamagnetic atoms with the elementary level CP violating 
processes as given by Eq. (\ref{eq:atomicedmcoef}).
The renormalization point of quark and gluon level operators is taken as $\mu =1$ GeV. The $\theta$-term contribution is estimated with 
the single source assumption.}
\begin{center}
\begin{tabular}{lccl}
\hline \hline
$^{129}$Xe & &\\
\hline
$K_{\rm Xe}$ & Central value & Error bar & Largest sources of error \\
\hline
$K_{{\rm Xe}, \theta}$ & $7 \times 10^{-22}e$ cm & $O(50\%)$ & Hadronic level ($\theta$-term contribution to the nucleon EDM) \\
$K_{{\rm Xe}, d_e}$ & $10^{-3}$ & $O(100\%)$ & Atomic level (higher order contribution from hyperfine interaction) \\
$K_{{\rm Xe}, d_u}$ & $2.2 \times 10^{-6}$ & $O(30\%)$ & Nuclear level (nucleon EDM contribution to Schiff moment)  \\
$K_{{\rm Xe}, d_d}$ & $-8.5 \times 10^{-6}$ & $O(30\%)$ & Nuclear level (nucleon EDM contribution to Schiff moment) \\
$K_{{\rm Xe}, d^c_u}$ & $-6 \times 10^{-6} e $ & $O(70\%)$ & Hadronic level (light quark chromo-EDM contribution to nucleon EDM) \\
$K_{{\rm Xe}, d^c_d}$ & $-2 \times 10^{-5} e $ & $O(70\%)$ & Hadronic level (light quark chromo-EDM contribution to nucleon EDM) \\
$K_{{\rm Xe}, d^c_s}$ & $10^{-6} e $ & $O(100\%)$ & Hadronic level (strange quark chromo-EDM contribution to nucleon EDM) \\
$K_{{\rm Xe}, w}$ & $10^{-21}  [e \, {\rm GeV}^2 \, {\rm cm}]$ & $O(100\%)$ & Hadronic level (Weinberg operator contribution to $d_N$) \\
$K_{{\rm Xe}, C^{\rm SP}_{eu}}$ & $10^{-23} [e \, {\rm cm}]$ & $O(100\%)$ & Atomic level (higher order contribution from analytical expression) \\
$K_{{\rm Xe}, C^{\rm SP}_{ed}}$ & $10^{-23} [e \, {\rm cm}]$ & $O(100\%)$ & Atomic level (higher order contribution from analytical expression) \\
$K_{{\rm Xe}, C^{\rm SP}_{es}}$ & $10^{-25} [e \, {\rm cm}]$ & $O(100\%)$ & Atomic level (higher order contribution from analytical expression) \\
$K_{{\rm Xe}, C^{\rm T}_{eu}}$ & $-2.2 \times 10^{-22} [e \, {\rm cm}]$ & $O(30\%)$ & Nuclear level (nuclear spin matrix element) \\
$K_{{\rm Xe}, C^{\rm T}_{ed}}$ & $8.8 \times 10^{-22} [e \, {\rm cm}]$ & $O(30\%)$ & Nuclear level (nuclear spin matrix element) \\
$K_{{\rm Xe}, C^{\rm PS}_{eu}}$ & 
$-5.2 \times 10^{-22} [e \, {\rm cm}]$
& $O(60\%)$ & Hadronic level (nucleon pseudoscalar density) \\
$K_{{\rm Xe}, C^{\rm PS}_{ed}}$ & 
$5.5 \times 10^{-22} [e \, {\rm cm}]$
& $O(60\%)$ & Hadronic level (nucleon pseudoscalar density) \\
$K_{{\rm Xe}, C^{\rm PS}_{es}}$ & 
$-1.6 \times 10^{-23} [e \, {\rm cm}]$ 
& $O(60\%)$ & Hadronic level (nucleon pseudoscalar density) 
\\
$K_{{\rm Xe}, C_{LR}}$ & $10^{-19} [e \, {\rm GeV}^2 \, {\rm cm}]$ & $O(100\%)$ & Hadronic level (CP-odd four-quark interaction contribution to $\bar g_{\pi NN}^{(1)}$) \\
\hline \hline
& &\\
$^{199}$Hg & &\\
\hline
$K_{\rm Hg}$ & Central value & Error bar & Largest sources of error \\
\hline
$K_{{\rm Hg}, \theta}$ & $10^{-20}e$ cm & $O(100\%)$ & Nuclear level ($d_N$ and $\bar g_{\pi NN}^{(0)}$ contribution to Schiff moment) \\
$K_{{\rm Hg}, d_e}$ & $10^{-2}$ & $O(100\%)$ & Atomic level (higher order contribution from analytical expression) \\
$K_{{\rm Hg}, d_u}$ & $10^{-5}$ & $O(100\%)$ & Nuclear level (nucleon EDM contribution to Schiff moment) \\
$K_{{\rm Hg}, d_d}$ & $10^{-4}$ & $O(100\%)$ & Nuclear level (nucleon EDM contribution to Schiff moment) \\
$K_{{\rm Hg}, d^c_u}$ & $10^{-4} e $ & $O(100\%)$ & Nuclear level (CP-odd nuclear force contribution to Schiff moment) \\
$K_{{\rm Hg}, d^c_d}$ & $10^{-3} e $ & $O(100\%)$ & Nuclear level (CP-odd nuclear force contribution to Schiff moment) \\
$K_{{\rm Hg}, d^c_s}$ & $10^{-4} e $ & $O(100\%)$ & Hadronic level (strange quark chromo-EDM contribution to nucleon EDM) \\
$K_{{\rm Hg}, w}$ & $10^{-19} [e \, {\rm GeV}^2 \, {\rm cm}]$ & $O(100\%)$ & Hadronic level (Weinberg operator contribution to $d_N$) \\
$K_{{\rm Hg}, C^{\rm SP}_{eu}}$ & $10^{-21} [e \, {\rm cm}]$ & $O(100\%)$ & Atomic level (higher order contribution from analytical expression) \\
$K_{{\rm Hg}, C^{\rm SP}_{ed}}$ & $10^{-21} [e \, {\rm cm}]$ & $O(100\%)$ & Atomic level (higher order contribution from analytical expression) \\
$K_{{\rm Hg}, C^{\rm SP}_{es}}$ & $10^{-23} [e \, {\rm cm}]$ & $O(100\%)$ & Atomic level (higher order contribution from analytical expression) \\
$K_{{\rm Hg}, C^{\rm T}_{eu}}$ & $10^{-21} [e \, {\rm cm}]$ & $O(100\%)$ & Nuclear level (unknown nuclear spin matrix element) \\
$K_{{\rm Hg}, C^{\rm T}_{ed}}$ & $10^{-21} [e \, {\rm cm}]$ & $O(100\%)$ & Nuclear level (unknown nuclear spin matrix element) \\
$K_{{\rm Hg}, C^{\rm PS}_{eu}}$ & $10^{-21} [e \, {\rm cm}]$ & $O(100\%)$ & Nuclear level (unknown nuclear spin matrix element) \\
$K_{{\rm Hg}, C^{\rm PS}_{ed}}$ & $10^{-21} [e \, {\rm cm}]$ & $O(100\%)$ & Nuclear level (unknown nuclear spin matrix element) \\
$K_{{\rm Hg}, C^{\rm PS}_{es}}$ & $10^{-22} [e \, {\rm cm}]$ & $O(100\%)$ & Nuclear level (unknown nuclear spin matrix element) \\
$K_{{\rm Hg}, C_{LR}}$ & $10^{-18} [e \, {\rm GeV}^2 \, {\rm cm}]$ & $O(100\%)$ & Hadronic level (CP-odd four-quark interaction contribution to $\bar g_{\pi NN}^{(1)}$) \\
\hline \hline
 & & \\
$^{225}$Ra & &\\
\hline
$K_{\rm Ra}$ & Central value & Error bar & Largest sources of error \\
\hline
$K_{{\rm Ra}, \theta}$ & $-3 \times 10^{-17}e$ cm & $O(60\%)$ & Nuclear level ($\bar g_{\pi NN}^{(0)}$ contribution to Schiff moment) \\
$K_{{\rm Ra}, d_e}$ & $10^{-2}$ & $O(100\%)$ & Atomic level (higher order contribution from analytical expression) \\
$K_{{\rm Ra}, d_u}$ & $10^{-5}$ & $O(100\%)$ & Nuclear level (unknown nucleon EDM contribution to Schiff moment) \\
$K_{{\rm Ra}, d_d}$ & $10^{-4}$ & $O(100\%)$ & Nuclear level (unknown nucleon EDM contribution to Schiff moment) \\
$K_{{\rm Ra}, d^c_u}$ & $-6 e $ & $O(70\%)$ & Nuclear level (CP-odd nuclear force contribution to Schiff moment) \\
$K_{{\rm Ra}, d^c_d}$ & $6 e $ & $O(70\%)$ & Nuclear level (CP-odd nuclear force contribution to Schiff moment) \\
$K_{{\rm Ra}, d^c_s}$ & $10^{-4} e$ & $O(100\%)$ & Hadronic level (strange quark chromo-EDM contribution to nucleon EDM) \\
$K_{{\rm Ra}, w}$ & $10^{-19} [e \, {\rm GeV}^2 \, {\rm cm}]$ & $O(100\%)$ & Hadronic level (Weinberg operator contribution to $d_N$) \\
$K_{{\rm Ra}, C^{\rm SP}_{eu}}$ & $10^{-21} [e \, {\rm cm}]$ & $O(100\%)$ & Atomic level (higher order contribution from analytical expression) \\
$K_{{\rm Ra}, C^{\rm SP}_{ed}}$ & $10^{-21} [e \, {\rm cm}]$ & $O(100\%)$ & Atomic level (higher order contribution from analytical expression) \\
$K_{{\rm Ra}, C^{\rm SP}_{es}}$ & $10^{-23} [e \, {\rm cm}]$ & $O(100\%)$ & Atomic level (higher order contribution from analytical expression) \\
$K_{{\rm Ra}, C^{\rm T}_{eu}}$ & $10^{-21} [e \, {\rm cm}]$ & $O(100\%)$ & Nuclear level (unknown nuclear spin matrix element) \\
$K_{{\rm Ra}, C^{\rm T}_{ed}}$ & $10^{-21} [e \, {\rm cm}]$ & $O(100\%)$ & Nuclear level (unknown nuclear spin matrix element) \\
$K_{{\rm Ra}, C^{\rm PS}_{eu}}$ & $10^{-21} [e \, {\rm cm}]$ & $O(100\%)$ & Nuclear level (unknown nuclear spin matrix element) \\
$K_{{\rm Ra}, C^{\rm PS}_{ed}}$ & $10^{-21} [e \, {\rm cm}]$ & $O(100\%)$ & Nuclear level (unknown nuclear spin matrix element) \\
$K_{{\rm Ra}, C^{\rm PS}_{es}}$ & $10^{-23} [e \, {\rm cm}]$ & $O(100\%)$ & Nuclear level (unknown nuclear spin matrix element) \\
$K_{{\rm Ra}, C_{LR}}$ & $10^{-14} [e \, {\rm GeV}^2 \, {\rm cm}]$ & $O(100\%)$ & Hadronic level (CP-odd four-quark interaction contribution to $\bar g_{\pi NN}^{(1)}$) \\
\hline \hline

\end{tabular}
\end{center}
\end{table*}

We present first the ``conventional" way of obtaining constraints from the EDM experimental and theoretical results by assuming a single
source of CP violation at the atomic level. It has already been stated earlier that, currently, the most precise measurement of EDM of
diamagnetic atoms comes from $^{199}$Hg. Thus, by combining the EDM of this atom reported as $|d( ^{199}\rm{Hg})|< 7.4\times 10^{-30} |e|\,
\rm{cm}$ with 95\% confidence level \cite{graner} with the corresponding ${\cal R}$ value for the NSM and T-PT CP-odd electron-nucleus 
interaction using the CCSD method given in Table \ref{tab3}, we get
\begin{eqnarray}
C_{N}^T < \ 6.5 \times 10^{-10} 
\end{eqnarray}
and 
\begin{eqnarray}
S_{\rm Hg} < 4.1 \times 10^{-13} |e| \, {\rm fm}^3 
.
\end{eqnarray}

The NSM of $^{199}$Hg 
in terms of $\bar{g}^{(i)}_{\pi N N}$'s is given by 
\begin{eqnarray}
{\boldmath S}_{\rm Hg}
&=& 
\big [ (0.23 \pm 0.15) \bar{g}_{\pi N N}^{(0)} - (0.007 \pm 0.011) \bar{g}_{\pi N N}^{(1)} \big ]\mbox{ $|e|$ fm}^3
,
\nonumber\\
\label{eq:shggbar}
\end{eqnarray}
where we have adopted the central value of several mean field calculations from Ref. \cite{Ban10}, with the  standard deviation as the error bar.
From the above limit on the NSM, we can infer constraints on $\bar{g}^{(0)}_{\pi N N}$ as
\begin{eqnarray}
|\bar{g}_{\pi N N}^{(0)}| < \ 5.1 \times 10^{-12} 
.
\end{eqnarray}
Here we note that we have maximized the upper limit of $|\bar{g}_{\pi N N}^{(0)}|$ by minimizing the isoscalar coefficient of Eq. (\ref{eq:shggbar}) within the error bar.

Furthermore, using the relation of Eq. (\ref{eq:g0theta}),
we can extract the upper limits on $\bar{\theta}$ as 
\begin{eqnarray}
|\bar{\theta}| < \ 5.7 \times 10^{-10}
,
\end{eqnarray}
where we have again maximized the allowed region, given that the error bars of the coefficient of Eqs. (\ref{eq:g0theta}) and (\ref{eq:shggbar}) are uncorrelated, to obtain the most conservative constraint.
While inferring this limit, we have assumed that the PQ mechanism is not active. As can be noticed the above inferred limit of $\bar{\theta}$ 
from $^{199}$Hg and that is obtained from the combination of measured $d_n$ and EFT calculation in Eq. (\ref{eq:nedmconstraint}) are of same order. 

We emphasize here that, in giving each of the above limits, we have assumed that only one CP-odd coupling is finite, and the others are zero.
The above way of reasoning is expected to work well to constrain the $\theta$-term, since the $\theta$-term is the largest contribution to the atomic EDM under the current upper limit given by experiment, compared with the effects of other TeV scale new physics or the CP violation of the CKM matrix.
In these optimistic approximation, it would be useful to provide linear relationship between different atomic EDMs with $\bar \theta$ 
following Table \ref{table:edmcoef} as
\begin{itemize}
\item
EDM of $^{129}$Xe atom:
\begin{eqnarray}
d_{Xe}
&=&
(7.5 \pm 3.4) \times 10^{-22} \bar \theta \,e\, {\rm cm},
\end{eqnarray}

\item
EDM of $^{199}$Hg atom:
\begin{eqnarray}
d_{Hg}
&=&
(4.1 \pm 2.8)
\times 10^{-20} \bar \theta \,e\, {\rm cm},
\end{eqnarray}

and\\

\item
EDM of $^{225}$Ra atom:
\begin{eqnarray}
d_{Ra}
&=&
(-3.1 \pm 1.9) \times 10^{-17} \bar \theta \,e\, {\rm cm}.
\end{eqnarray}

\end{itemize}
It is to be noted that the $\theta$-term contribution to the atomic EDM has a sizable theoretical uncertainty, mainly coming from the nuclear structure calculation.
Moreover, the CP-odd interactions used in the calculation of the nuclear level coefficients, with restricted model space, may differ from the bare ones, and this fact may bring additional unknown systematics.

This single source assumption is however not valid when several sources contribute to the atomic EDM with similar orders of magnitude, which is the case for the majority of the TeV scale new physics (when the PQ mechanism is invoked and the $\theta$-term is unphysical).
As a concrete example, 
$d_{Xe}$ and $d_{Hg}$ have similar orders enhancement factors for the quark EDMs and chromo-EDMs (see Table \ref{table:edmcoef}), which is due to the absence of the enhancement or suppression of the contributions from  the nucleon EDMs and the CP-odd force between the nucleons. 
Another important case is the effect of P,CP-odd e-N interaction.
We see that the effect of the quark level $C^{\rm PS}_{eu}$ and $C^{\rm PS}_{ed}$ is comparable to that of $C^{\rm T}_{eq} (q=u,d)$, 
although T-PT interaction contributes larger than the S-PS interaction in the diamagnetic atoms.
This counterbalancing is due to the enhancement of the nucleon pseudoscalar charge [see Eqs. (\ref{eq:u5u}) and (\ref{eq:d5d})].
From these properties, we see that it is not possible to give limits by simply assuming a single elementary level process.
Rather, we have to accurately evaluate the contributions from each process to the final observable EDMs, as there could be a destructive interference between the quark EDM and the chromo-EDM, reducing the sensitivity and loosening the constraint.
This fact increases the error bars of the constraints on the CP violating parameters of the new physics generating the quark EDM and the chromo-EDM of the same order of magnitude. 
A typical example is the generic SUSY model, which generates the quark EDM and the chromo-EDM of the same order of magnitude.
This also concerns the general new physics candidates which generate the chromo-EDM or the Weinberg operator at the TeV scale, as the 
RGE renders quark EDM and chromo-EDM of the same order at the hadronic scale.

In relation to the quark chromo-EDM contribution, the EDM of $^{225}$Ra is a cleaner system, since the effect of the CP-odd nuclear force is 
much more enhanced compared to the nucleon EDM. As was pointed in Sec. \ref{ocmsec}, this enhancement is due to the nuclear octupole 
deformation. On the contrary, the NSM due to the intrinsic nucleon EDM is suppressed, and the upper limit on its 
contribution is suppressed by several orders of magnitude, when it is expressed in terms of $\bar g_{\pi NN}^{(0)}$. 
This is a very remarkable property, as the CP-odd nuclear force is singularly sensitive to the quark chromo-EDM, or the left-right 
symmetric type four-quark operator. 
Of course, this enhancement is only a nuclear level effect, and it does not prohibit the suppression of the quark chromo-EDM effect at the elementary level, which may counterbalance the effect of other sources.
In the analysis of the EDMs of diamagnetic atoms, careful inspections of the contributions from all possible important sources of CP violation are required.

\subsection{Implication for particle physics}

We now discuss the implications of our current knowledge of the EDMs of  diamagnetic atoms for different particle physics models.
The first case we consider is the SUSY class of models.
In the generic SUSY model, the EDM and the chromo-EDM of quarks are generated at the one-loop level.
In Sec. \ref{sec:susy}, we have seen that $d_q$ and $d^c_q$ are of the order of $10^{-25}e$ cm, at the TeV scale with typical parameters in TeV scale 
SUSY breaking scenarios.
The RGE can be calculated without large theoretical uncertainty, but the hadron and nuclear level evaluations involve a large error bar.
As can be seen in Table \ref{table:edmcoef}, the error might be enlarged due to the destructive interference for the EDMs of $^{129}$Xe and $^{199}$Hg.
If we consider the most conservative case, the best limit of the EDM of $^{199}$Hg cannot even constrain the SUSY CP phases ($\theta_\mu$ and $\theta_A$), related to $d_q$ and $d_q^c$ by Eqs. (\ref{eq:susy_qedm}) and (\ref{eq:susy_cedm}).
The EDM of $^{225}$Ra can overcome this problem, since its sensitivity to the chromo-EDM is enhanced.
In ordinary SUSY models, the right-handed current of light quarks is strongly suppressed due to the Yukawa couplings, so there is no possibility of destructive interference with the left-right four-quark interaction \cite{fabbrichesi,drees}. 

In the split SUSY scenarios, the leading CP violation is given by the quark and electron EDMs, as mentioned in Sec. \ref{sec:susy}.
In that case, the hadronic uncertainty is better controlled since the quark EDM contribution to the NSM is better known.
In this case however, we have to consider the interference between the electron and quark EDMs.
It is also important to note that $^{225}$Ra EDM is not sensitive to the split SUSY scenario, since the quark chromo-EDM is suppressed.
The same remarks apply to several R-parity violating models for which the leading contribution comes from the Barr-Zee type diagrams with heavy leptons in the inner loop.
In the baryon number violating R-parity violating scenarios, the right-handed quark current is also generated, which leads to the left-right type four-quark interaction and may interfere with the quark chromo-EDM \cite{rpvbetadecay2,jng}.

In the Higgs-doublet models, the leading process is the Barr-Zee type diagram of quarks and electrons, as discussed in Sec. \ref{sec:2HDM}.
Here the quark chromo-EDM gives the most important contribution, as the electromagnetic Barr-Zee type diagram is suppressed by $\alpha_{\rm em}$.
As for the one-loop level SUSY fermion EDM and chromo-EDM, this process generates after the renormalization group evolution down to the hadronic scale a quark EDM with the same order of magnitude.
The $^{225}$Ra EDM is again the most efficient way of probing it.
The Weinberg operator also contributes to the hadronic effective CP-odd interaction, but it is subleading since its Wilson coefficient at the TeV scale is smaller by about two orders of magnitude for the Higgs mass $m_H = 125$ GeV.
It is also additionally suppressed by the RGE, down to the hadronic scale, so that its final contribution to the nucleon EDM is smaller than that generated by the quark chromo-EDM by more than an order of magnitude.

In the left-right symmetric models, the leading CP violation is given by the left-right type four-quark interaction (see Sec. \ref{sec:LRSM}), which generates the isovector CP-odd $\pi$-N-N interaction at the hadron level, without largely mixing with other hadronic interactions (see Sec. \ref{sec:quarkRGE}).
This process can therefore be probed with all diamagnetic atoms.
If we can observe a clear hierarchy respecting the coefficients $a_1$ times the coefficients relating the NSM and the atomic EDM in the experimental values of the EDMs of $^{129}$Xe, $^{199}$Hg, and $^{225}$Ra, it is strongly probable that the left-right symmetric model is the source of CP violation.
We might think that this hierarchy can be mimicked by the electron EDM or the CP-odd e-N interactions, but the huge enhancement of the EDM of $^{225}$Ra is difficult to realize.
Of course, continuous efforts in determining the NSM and reducing theoretical uncertainties are desirable.

Regarding the leptoquark model, the EDM of diamagnetic atoms is singularly important, as it is sensitive to the tensor-type CP-odd e-N interaction.
Current limit of $^{199}$Hg EDM can exclude the mass of the leptoquark to the PeV level, assuming $O(0.1)$ couplings with $O(1)$ CP phase.

We also discuss below the case of vector like fermions without direct interaction with the SM fermions.
In this case the leading process is the Weinberg operator, which generates the quark EDM and chromo-EDM with similar orders of magnitude through the renormalization group evolution from the TeV scale to the hadronic scale.
We therefore have to accurately determine its contribution to the nucleon EDM as well as those from the quark EDM and the chromo-EDM.
Unfortunately, the accuracy is currently not high.
The isoscalar chromo-EDM generated by the Weinberg operator at the hadronic scale can be probed using the $^{225}$Ra through the CP-odd nuclear force.
We have to note that the Weinberg operator is also expected to contribute to the short range contact interaction of the CP-odd nuclear force.
This contribution is also currently unknown, and has to be determined to unveil the CP violation of vectorlike fermions.

Finally, we consider the SM contributions generated by the CP phase of the CKM matrix.
Here we have to compare the CP violation due to the CP-odd e-N interaction and that from the NSM.
The CP-odd e-N interaction contribution is estimated as $C^{\rm SP}_N \sim O(10^{-17})$ (see Sec. \ref{sec:cpve-n}).
Combing with the atomic level coefficients that are determined using the analytical relations, we obtain atomic EDMs less than 
$10^{-38}e $ cm. The NSM contribution was estimated in Ref. \cite{smnuclearedm} as
\begin{eqnarray}
d_{Xe}
&\sim &
10^{-36}e \, {\rm cm}
,
\\
d_{Hg}
&\sim &
10^{-35}e \, {\rm cm}
,
\\
d_{Ra}
&\sim &
10^{-32}e \, {\rm cm}
.
\end{eqnarray}

In the SM, the NSM is giving the largest contribution.
It is of course well below the current experimental sensitivity.
Here we note that we are using the same CP-odd $\pi$-N-N coupling to estimate the CP-odd e-N interaction (see Fig. \ref{fig:sm_e-n}) and the nuclear Schiff moment.
The hierarchy between them has therefore a smaller error bar than the values themselves.

\section{Summary and Outlook\label{sec:summary}}

The EDMs of diamagnetic atoms depend on the hadronic CP violation, CP-odd e-N interaction and the electron EDM. 
In particular they are sensitive to the isoscalar and isovector CP-odd $\pi$-N-N interactions.
The neutron EDM, which is sensitive to the hadronic CP violation, is also rather sensitive to isoscalar interactions.
Diamagnetic atoms are, on the contrary, sensitive to the isovector CP-odd $\pi$-N-N interaction.
Another remarkable point is that the EDMs of these atoms can probe the tensor-type CP-odd e-N interaction, which is singularly sensitive to 
the leptoquark model. For each microscopic CP violating process, there are other competitive or even more sensitive experimental probes,
such as the EDMs of paramagnetic atoms or $d_n$. However, as we have mentioned in the introduction, the BSMs which can generate CP violation
in several sectors at the same time cannot be constrained with only those singularly sensitive experimental observables.

The sensitivity of the EDM of diamagnetic atoms on elementary level CP violation is orthogonal with those of  other observables due 
to its dependence on a number of quantities, so it is very useful in constraining models that encompass a large parameter space. An 
excellent example is the analysis of SUSY models, which have a very large degree of freedom. Previous analyses often assumed that only 
a restricted number of parameters are active (the so-called ``single source dominance'' assertion) and the constraints on CP phases 
were given by the most sensitive experimental data on them. There may be cancellations if we consider several couplings and CP phases at
the same time. In such a scenario, the EDM of diamagnetic atoms can constrain the CP violation which spreads over several sectors, or 
disentangle the CP violating sources if a nonvanishing CP violation is found in some other experiments.

Among the several diamagnetic atoms that have so far 
been the subject of experimental EDM studies, $^{199}$Hg  has yielded the best result; the current upper limit of its EDM 
being $7.4 \times 10^{-30}$ $e$ cm at the 95 \% confidence level \cite{graner}.
This is a remarkably stringent limit, not only because it is nominally 
the lowest among the upper limits ever placed on the EDM of an elementary particle or a composite system,
but also because it holds promising possibilities for EDMs of other diamagnetic atoms, suggesting that they can be measured with
similar or even better accuracy. In fact, the detection sensitivity of the 
ongoing search for EDM in $^{225}$Ra is rapidly improving \cite{bishof,parker}. 
By virtue of the large enhancement expected for the Schiff moment in this quadrupole- 
and octupole-deformed nucleus, its sensitivity to the CP-violating sources will reach 
comparable or even superior levels to that of mercury. $^{223}$Rn would also 
be another promising candidate in the search for EDMs in diamagnetic atomic systems \cite{rand,tardiff}. We note here also that
important developments are taking place in the search for the EDM of $^{129}$Xe atom, which have been
undertaken by several groups taking advantage of the exceptionally long spin coherence 
times realizable for this species \cite{inoue-xe,fierlinger,schmidt}. 

It is evident from our discussion on the atomic calculations of the ratio of the EDMs of diamagnetic to different CP violating coupling
constants that significant progress has been made in this area during the past decade. This has been possible due to advances in the
RCC theory and the hybrid CI+MBPT method. In particular, the ability of the former method to
capture the strong electron effects in Ra is truly impressive. The errors in the calculations which is of the order of two to five percent for the diamagnetic atoms can be reduced further by
using the normal coupled-cluster and the extended coupled-cluster methods \cite{bishop}.

The NSM provides a very important contribution to the diamagnetic atomic EDM. 
In the beginning of its study, the Schiff moment might be enhanced due to the collective motion in the nucleus in a similar way 
as quadrupole vibrations enhance the quadrupole moments and transitions. 
It has been found later that this is not the case. On the contrary the single-particle estimate of the Schiff moment is even quenched due
 to the many-body effects. The exception might be the octupole deformation seen in actinide Ra regions, 
where the octupole deformation forms parity doublet states with spin 1/2 . 
It is the same for the nuclear EDMs coming 
from the intrinsic nucleon EDMs. Nuclear EDMs are also quenched 
due to the many-body correlations.
It is therefore important to incorporate  the many-body effects in the nuclear wave function. In this respect
the nuclear shell model is superior to the other mean field theories.

The theoretical uncertainty for the nuclear calculations is fairly large.
In addition to the sizeable error bar in the result of the calculation of the NSM, several nuclear level quantities, 
such as the nuclear spin matrix elements, are not known.
The nuclear spin matrix elements are useful in determining the CP-odd e-N interaction contribution to the EDMs of atoms.
Their evaluation is expected to be much easier than the NSM, so future work in this direction is very desirable.
Another open question is to relate the ``bare" CP-odd nuclear force to the effective CP-odd nuclear force which is relevant in theories with restricted model space.
This procedure is required in bridging from hadron to nuclear physics. The uncertainty in the evaluation of the hadron matrix elements is larger than those of all the 
quantities that are needed for the determination of the EDMs of diamagnetic atoms. It is very challenging to reduce it and it cannot be achieved without performing
large scale lattice QCD computations. Results for several quantities that contribute to the atomic EDMs such as the nucleon scalar densities and tensor changes
have been obtained recently. The most important quark level CP-odd quantity that needs to be evaluated is probably the quark chromo-EDM \cite{devries3}. It is currently being 
computed on lattices by several groups and new results are expected soon. The chiral EFT approach is also useful in controlling the theoretical uncertainties originating
in unknown hadronic effective interactions which are difficult to obtain on a lattice.

Given that our understanding of the challenging experimental and theoretical issues of the EDMs of diamagnetic atoms is steadily improving, one can be optimistic
about new and improved results in this field in the foreseeable future. This will not only deepen our knowledge of CP violation, but also provide important insights into
physics of BSM.

\section*{Acknowledgement}

We thank Professor T. Fukuyama for useful discussions. This work was supported partly by INSA-JSPS under project no. IA/INSA-JSPS 
Project/2013-2016/February 28,2013/4098. BKS acknowledges Dr. Y. Singh and Dr. D. K. Nandy for many useful discussions and
participating in development of RCC codes and use of PRL HPC cluster at Ahmedabad, India. The work of NY was completed due to 
support of the RSF grant  15-12-20008 and of Riken iTHES project.


\begin{thebibliography}{100}

\bibitem{SM1}
S. L. Glashow, Nucl. Phys. {\bf 22}, 579 (1961). http://dx.doi.org/10.1016/0029-5582(61)90469-2

\bibitem{SM2}
S. Weinberg, Phys. Rev. Lett. {\bf 17}, 616 (1966). https://doi.org/10.1103/PhysRevLett.17.616

\bibitem{SM3}
A. Salam, {\it Proceedings of the Eighth Nobel Symposium}, edited by N. Svartholm (New York, Wiley-Interscience, 1968).

\bibitem{higgsatlas}
G. Aad {\it et al.} (ATLAS Collaboration), Phys. Lett. B {\bf 716}, 1 (2012). http://dx.doi.org/10.1016/j.physletb.2012.08.020

\bibitem{higgscms}
S. Chatrchyan {\it et al.} (CMS Collaboration), Phys. Lett. B {\bf 716}, 30 (2012). http://dx.doi.org/10.1016/j.physletb.2012.08.021

\bibitem{sakharov}
A. D. Sakharov, Pisma Zh. Eksp. Teor. Fiz. {\bf 5}, 32 (1967) [JETP Lett. {\bf 5}, 24 (1967)].

\bibitem{khriplovich}
I. B. Khriplovich and S. K. Lamoreaux, {\it CP violation without strangeness: Electric dipole moments of particles, atoms, and molecules}, (Springer, Berlin, 1997).

\bibitem{bigibook}
I. I. Bigi and  A. I. Sanda, {\it CP Violation}, (Cambridge University Press, 2000).

\bibitem{roberts}
B. L. Roberts and W. J. Marciano, {\it Lepton Dipole Moments, Advanced series on Directions in High Energy Physics}, vol. 20, World Scientific, Singapore (2010).

\bibitem{christenson}
J. H. Christenson, J. W. Cronin, V. L. Fitch, and R. Turlay, Phys. Rev. Lett. {\bf 13}, 138 (1964). https://doi.org/10.1103/PhysRevLett.13.138

\bibitem{abe}
K. Abe {\it et al}., Phys. Rev. Lett. {\bf 87}, 091802 (2001). https://doi.org/10.1103/PhysRevLett.87.091802

\bibitem{aubert}
B. Aubert {\it et al}., Phys. Rev. Lett. {\bf 87}, 091801 (2001). https://doi.org/10.1103/PhysRevLett.87.091801

\bibitem{aaij}
R. Aaij {\it et al}., Phys. Rev. Lett. {\bf 110}, 221601 (2013). https://doi.org/10.1103/PhysRevLett.110.221601

\bibitem{alvarez}
E. \'{A}avarez and A. Szynkman, Mod. Phys. Lett. A {\bf 23}, 2085 (2008). http://dx.doi.org/10.1142/S021773230802728X

\bibitem{K-M}
M. Kobayashi and T. Maskawa, Prog. Theor. Phys. {\bf 49}, 652 (1973). https://doi.org/10.1143/PTP.49.652

\bibitem{Jarlskog}
C. Jarlskog, Phys. Rev. Lett. {\bf 55}, 1039 (1985). https://doi.org/10.1103/PhysRevLett.55.1039

\bibitem{farrar}
G. R. Farrar and M. E. Shaposhnikov, Phys. Rev. D {\bf 50}, 774 (1994). https://doi.org/10.1103/PhysRevD.50.774

\bibitem{huet}
P. Huet and E. Sather, Phys. Rev. D {\bf 51}, 379 (1995). https://doi.org/10.1103/PhysRevD.51.2827

\bibitem{dine}
M. Dine and A. Kusenko, Rev. Mod. Phys. {\bf 76}, 1 (2003). https://doi.org/10.1103/RevModPhys.76.1

\bibitem{purcell}
E. M. Purcell and N. F. Ramsey, Phys. Rev. {\bf 78}, 807 (1950). https://doi.org/10.1103/PhysRev.78.807

\bibitem{hgedm1987}
S. K. Lamoreaux, J. P. Jacobs, B. R. Heckel, F. J. Raab and N. Fortson, Phys. Rev. Lett. {\bf 59}, 2275 (1987). https://doi.org/10.1103/PhysRevLett.59.2275

\bibitem{rosenberry}
M. A. Rosenberry and T. E. Chupp, Phys. Rev. Lett. {\bf 86}, 22 (2001). https://doi.org/10.1103/PhysRevLett.86.22

\bibitem{regan}
B. C. Regan, E. D. Commins, C. J. Schmidt, and D. DeMille, Phys. Rev. Lett. {\bf 88}, 071805 (2002). https://doi.org/10.1103/PhysRevLett.88.071805

\bibitem{baker}
C. A. Baker {\it et al.}, Phys. Rev. Lett.  {\bf 97}, 131801 (2006). https://doi.org/10.1103/PhysRevLett.97.131801

\bibitem{griffith}
W. C. Griffith {\it et. al.}, Phys. Rev. Lett. {\bf 102}, 101601 (2009). https://doi.org/10.1103/PhysRevLett.102.101601

\bibitem{hudson} 
J. J. Hudson {\it et al}, Nature {\bf 473}, 493 (2011). doi:10.1038/nature10104

\bibitem{baron} 
J. Baron {\it et al} (ACME Collaboration), Science {\bf 343}, 269 (2014). DOI: 10.1126/science.1248213

\bibitem{bishof}
M. Bishof, R. H. Parker, K. G. Bailey, J. P. Greene, R. J. Holt, M. R. Kalita, W. Korsch, N. D. Lemke, Z.-T. Lu, P. Mueller, T. P. O’Connor, J. T. Singh and M. R. Dietrich, Phys. Rev. C {\bf 94}, 025501 (2016). https://doi.org/10.1103/PhysRevC.94.025501

\bibitem{graner}
B. Graner, Y. Chen, E. G. Lindahl, and B. R. Heckel, Phys. Rev. Lett. {\bf 116}, 161601 (2016). https://doi.org/10.1103/PhysRevLett.116.161601

\bibitem{Pendlebury}
J. M. Pendlebury {\it et al.}, Phys. Rev. D {\bf 92}, 092003 (2015). https://doi.org/10.1103/PhysRevD.92.092003

\bibitem{ramsey1}
N. F. Ramsey, Annu. Rev. Nucl. Part. Sci. {\bf 32}, 211 (1982). DOI: 10.1146/annurev.ns.32.120182.001235 

\bibitem{fortson}
N. Fortson, P. Sandars and S. Barr, Phys. Today {\bf 56}, 33 (2003). http://dx.doi.org/10.1063/1.1595052

\bibitem{luders}
G. L\"uders, Ann. Phys. (N.Y.) {\bf 281}, 1004 (2000). http://dx.doi.org/10.1006/aphy.2000.6027

\bibitem{bernreuther}
W. Bernreuther and M. Suzuki, Rev. Mod. Phys. {\bf 63}, 313 (1991). https://doi.org/10.1103/RevModPhys.63.313

\bibitem{barr}
S. M. Barr, Int. J. Mod. Phys. A {\bf 8}, 209 (1993). http://dx.doi.org/10.1142/S0217751X93000096

\bibitem{pospelovreview}
M. Pospelov and A. Ritz, Ann. Phys. (N.Y.) {\bf 318}, 119 (2005). http://dx.doi.org/10.1016/j.aop.2005.04.002

\bibitem{ramsey}
M. J. Ramsey-Musolf and S. Su, Phys. Rep. {\bf 456}, 1 (2008). http://dx.doi.org/10.1016/j.physrep.2007.10.001

\bibitem{weiss}
D. S. Weiss, Private communication.

\bibitem{heinzen}
D. Heinzen, Private communication.

\bibitem{harada}
K. Harada, T. Aoki, K. Kato, H. Kawamura, T. Inoue, T. Aoki, A. Uchiyama, K. Sakamoto, S. Ito, M. Itoh, T. Hayamizu, A. 
Hatakeyama, K. Hatanata, T. Wakasa and Y. Sakemi,  J. Phys. Conf. Ser. {\bf 691}, 012017 (2016). DOI: 1742-6596-691-1-012017

\bibitem{furukawa-xe}
T. Furukawa {\it et al}., J. Phys. Conf. Ser. {\bf 312}, 102005 (2011). DOI: 1742-6596-312-10-102005

\bibitem{inoue-xe}
T. Inoue {\it et al}., Hyperfine Interact. (Springer Netherlands) {\bf 220}, 59 (2013). DOI: 10.1007/s10751-012-0751-z

\bibitem{rand}
E. T. Rand {\it et al}., J. Phys. Conf. Ser. {\bf 312}, 102013 (2011). DOI: 1742-6596-312-10-102013

\bibitem{tardiff} 
E. R. Tardiff, {\it Towards a Measurement of the Electric Dipole Moment of $^{223}$Rn}, PhD thesis
submitted at the University of Michigan (2009). 

\bibitem{fierlinger}
F. Kuchler et al, Hyperfine Interact. {\bf 237}, 1 (2016). DOI: 10.1007/s10751-016-1302-9

\bibitem{schmidt}
W. Heil, C. Gemmel, S. Karpuk, Y. Sobolev, K. Tullney, F. Allmendinger, U. Schmidt, M.  Burghoff, W. Kilian, S. K.-Gr\"uneberg, 
A. Schnabel, F. Seifert, and L. Trahms, Annalen der Physik {\bf 525}, 539 (2013). DOI: 10.1002/andp.201300048

\bibitem{yoshimi}
A. Yoshimi {\it et al}., Phys. Lett. A {\bf 304}, 13 (2002). http://dx.doi.org/10.1016/S0375-9601(02)01324-5

\bibitem{heedmreview}
X.-G. He, B. H. J. McKellar and S. Pakvasa, Int. J. Mod. Phys. A {\bf 4}, 5011 (1989) [Erratum-ibid. A {\bf 6}, 1063 (1991)].  DOI: http://dx.doi.org/10.1142

\bibitem{ginges}
J. S. M. Ginges and V. V. Flambaum, Phys. Rep. {\bf 397}, 63 (2004). http://dx.doi.org/10.1016/j.physrep.2004.03.005

\bibitem{Fukuyama}
T. Fukuyama, Int. J. Mod. Phys. A {\bf 27}, 1230015 (2012). http://dx.doi.org/10.1142/S0217751X12300153

\bibitem{dzubareview}
V. A. Dzuba and V. V. Flambaum, Int. J. Mod. Phys. E {\bf 21}, 1230010 (2012). http://dx.doi.org/10.1142/S021830131230010X

\bibitem{engel}
J. Engel, M. J. Ramsey-Musolf and U. van Kolck, Prog. Part. Nucl. Phys. {\bf 71}, 21 (2013). http://dx.doi.org/10.1016/j.ppnp.2013.03.003

\bibitem{schiff}
L. I. Schiff, Phys. Rev. {\bf 132}, 2194 (1963). https://doi.org/10.1103/PhysRev.132.2194

\bibitem{GIM}
S. L. Glashow, J. Iliopoulos and L. Maiani, Phys. Rev. D {\bf 2}, 1285 (1970). https://doi.org/10.1103/PhysRevD.2.1285

\bibitem{donoghuesmedm}
J. Donoghue, Phys. Rev. D {\bf 18}, 1632 (1978).

\bibitem{shabalin1}
E. P. Shabalin, Yad. Fiz. {\bf 28}, 151 (1978) [Sov. J. Nucl. Phys. {\bf 28}, 75 (1978)].

\bibitem{shabalin2}
E. P. Shabalin, Yad. Fiz. {\bf 31}, (1980) 1665 [Sov. J. Nucl. Phys. {\bf 31}, 864 (1980)].


\bibitem{czarnecki}
A.~Czarnecki and B.~Krause,Phys.\ Rev.\ Lett.\  {\bf 78}, 4339 (1997).

\bibitem{pospelovsmelectronedm}
M. E. Pospelov and I. B. Khriplovich, Sov. J. Nucl. Phys. {\bf 53}, 638 (1991) [Yad. Fiz. {\bf 53}, 1030 (1991)].

\bibitem{booth}
M. J. Booth, arXiv:hep-ph/9301293.

\bibitem{pospelovsmatomicedm}
M. Pospelov and A. Ritz, Phys. Rev. D {\bf 89}, 056006 (2014).

\bibitem{Archambault}
J. P. Archambault, A. Czarnecki and M. E. Pospelov, Phys. Rev. D {\bf 70}, 073006 (2004).

\bibitem{hemfv1}
X.-G. He, C.-J. Lee, S.-F. Li and J. Tandean, Phys. Rev. D {\bf 89}, 091901 (2014).

\bibitem{hemfv2}
X.-G. He, C.-J. Lee, S.-F. Li and J. Tandean, JHEP {\bf 1408}, 019 (2014).

\bibitem{novales}
H. Novales-S\'{a}nchez, M. Salinas, J. J. Toscano and O. V\'{a}zquez-Hern\'{a}ndez, arXiv:1610.06649 [hep-ph].

\bibitem{ellissmtheta}
J. R. Ellis and M. K. Gaillard, Nucl. Phys. B {\bf 150}, 141 (1979).

\bibitem{khriplovichtheta}
I. B. Khriplovich, Phys. Lett. B {\bf 173}, 193 (1986); Yad. Fiz. {\bf 44}, 1019 (1986) [Sov. J. Nucl. Phys. {\bf 44}, 659 (1986)].

\bibitem{smweinbergop}
M. E. Pospelov, Phys. Lett. B {\bf 328}, 441 (1994).

\bibitem{ellissmedm}
J. Ellis, M. K. Gaillard and D. V. Nanopoulos, Nucl. Phys. B {\bf 109}, 213 (1976).

\bibitem{nanopoulossmedm}
D. V. Nanopoulos, A. Yildiz and P. H. Cox, Phys. Lett. B {\bf 87}, 53 (1979).

\bibitem{Deshpande}
N. G. Deshpande, G. Eilam and W. L. Spence, Phys. Lett. B {\bf 108}, 42 (1982).

\bibitem{gavelasmedm}
M. B. Gavela, A. Le Yaouanc, L. Oliver, O. Pene, J. C. Raynal and T. N. Pham, Phys. Lett. B {\bf 109}, 215 (1982).

\bibitem{smneutronedmkhriplovich}
I. B. Khriplovich and A. R. Zhitnitsky, Phys. Lett. B {\bf 109}, 490 (1982).

\bibitem{eeg}
J. O. Eeg and I. Picek, Nucl. Phys. B {\bf 244}, 77 (1984).

\bibitem{hamzaouismedm}
C. Hamzaoui and A. Barroso, Phys. Lett. B {\bf 154}, 202 (1985).

\bibitem{smneutronedmmckellar}
B. McKellar, S. R. Choudhury, X.-G. He and S. Pakvasa, Phys. Lett. B {\bf 197}, 556 (1987).

\bibitem{mannel}
T. Mannel and N. Uraltsev, Phys. Rev. D {\bf 85}, 096002 (2012). 

\bibitem{seng}
C.-Y. Seng, Phys. Rev. C {\bf 91}, 025502 (2015). 

\bibitem{barr-zee}
S. M. Barr and A. Zee, Phys. Rev. Lett. {\bf 65} (1990) 21 [Erratum Phys. Rev. Lett. {\bf 65}, 2920 (1990)].

\bibitem{weinbergop}
S. Weinberg, Phys. Rev. Lett. {\bf 63}, 2333 (1989).

\bibitem{dicus}
D. A. Dicus, Phys. Rev. D {\bf 41}, 999 (1990).

\bibitem{chang2hdm1}
D. Chang, W.-Y. Keung and T. C. Yuan, Phys. Lett. B {\bf 251}, 608 (1990).

\bibitem{leigh}
R. G. Leigh, S. Paban and R.-M. Xu, Nucl. Phys. B {\bf 352}, 45 (1991).

\bibitem{chang2hdm2}
D. Chang, W.-Y. Keung and T. C. Yuan, Phys. Rev. D {\bf 43}, R14 (1991).

\bibitem{mahanta}
U. Mahanta, Phys. Lett. B {\bf 281}, 320 (1992).

\bibitem{kao2hdm}
C. Kao and R.-M. Xu, Phys. Lett. B {\bf 296}, 435 (1992).

\bibitem{barger}
V. Barger, A. Das and C. Kao, Phys. Rev. D {\bf 55}, 7099 (1997).

\bibitem{bowser-chao}
D. Bowser-Chao, D. Chang and W.-Y. Keung, Phys. Rev. Lett. {\bf 79}, 1988 (1997).

\bibitem{buras2hdm}
A. J. Buras, G. Isidori and P. Paradisi, Phys. Lett. B {\bf 694}, 402 (2011).

\bibitem{brod}
J. Brod, U. Haisch and J. Zupan, JHEP {\bf 1311}, 180 (2013).

\bibitem{sinoue2hdm}
S. Inoue, M. J. Ramsey-Musolf and Y. Zhang, Phys. Rev. D {\bf 89}, 115023 (2014). 

\bibitem{cheung2hdm}
K. Cheung, J. S. Lee, E. Senaha and P.-Y. Tseng, JHEP {\bf 1406}, 149 (2014). 

\bibitem{jung2hdmedm}
M. Jung and A. Pich, JHEP {\bf 1404}, 076 (2014).

\bibitem{tabe}
T. Abe, J. Hisano, T. Kitahara and K. Tobioka, JHEP {\bf 1401}, 106 (2014). 

\bibitem{bian}
L. Bian and N. Chen, arXiv:1608.07975 [hep-ph].

\bibitem{chenedm}
C.-Y. Chen, S. Dawson and Y. Zhang, JHEP {\bf 1506}, 056 (2015).

\bibitem{eeghiggs}
J. O. Eeg, arXiv:1611.07778 [hep-ph].

\bibitem{barrmasiero}
S. M. Barr and A. Masiero, Phys. Rev. Lett. {\bf 58}, 187 (1987).

\bibitem{e-nint}
S. M. Barr, Phys. Rev. D {\bf 45}, 4148 (1992).

\bibitem{haber}
H. E. Haber and G. L. Kane, Phys. Rep. {\bf 117}, 75 (1985).

\bibitem{gunionmssm}
J. F. Gunion and H. E. Haber, Nucl. Phys. B {\bf 272}, 1 (1986).

\bibitem{martinmssm}
S. P. Martin, in {\it Perspectives on Supersymmetry II}, edited by G. L. Kane (World Scientific, Singapore, 2010), p. 1 [arXiv:hep-ph/9709356].

\bibitem{gabbiani}
F. Gabbiani, E. Gabrielli, A. Masiero and L. Silvestrini, Nucl. Phys. B {\bf 477}, 321 (1996).

\bibitem{ellismssmedm}
J. R. Ellis, S. Ferrera and D. V. Nanopoulos, Phys. Lett. B {\bf 114}, 231 (1982).

\bibitem{buchmullermssmedm}
W. Buchm\"{u}ller and D. Wyler, Phys. Lett. B {\bf 121}, 321 (1983).

\bibitem{polchinski}
J. Polchinski and M. B. Wise, Phys. Lett. B {\bf 125}, 393 (1983).

\bibitem{aguilamssmedm}
F. del Aguila, M. B. Gavela, J. A. Grifols and A. Mendez, Phys. Lett. B {\bf 126}, 71 (1983).

\bibitem{nanopoulosmssmedm}
D. V. Nanopoulos and M. Srednicki, Phys. Lett. B {\bf 128}, 61 (1983).

\bibitem{duganmssmedm}
M. Dugan, B. Grinstein and L. J. Hall, Nucl. Phys. B {\bf 255}, 413 (1985).

\bibitem{nathmssmedm}
P. Nath, Phys. Rev. Lett. {\bf 66}, 2565 (1991).

\bibitem{kizukurimssmedm1}
Y. Kizukuri and N. Oshimo, Phys. Rev. D {\bf 45}, 1806 (1992).

\bibitem{kizukurimssmedm2}
Y. Kizukuri and N. Oshimo, Phys. Rev. D {\bf 46}, 3025 (1992).

\bibitem{inuimssmedm}
T. Inui, Y. Mimura, N. Sakai and T. Sasaki, Nucl. Phys. B {\bf 449}, 49 (1995).

\bibitem{mssmreloaded}
J. R. Ellis, J. S. Lee and A. Pilaftsis, JHEP {\bf 0810}, 049 (2008).

\bibitem{bingli}
B. Li and C. E. M. Wagner, Phys. Rev. D {\bf 91}, 095019 (2015).

\bibitem{F-A}
T. Fukuyama and K. Asahi, Int. J. Mod. Phys. A {\bf 31}, 1650082 (2016).

\bibitem{west}
T. H. West, Phys. Rev. D {\bf 50}, 7025 (1994).

\bibitem{kadoyoshi}
T. Kadoyoshi and N. Oshimo, Phys. Rev. D {\bf 55}, 1481 (1997).

\bibitem{chang}
D. Chang, W.-Y. Keung and A. Pilaftsis, Phys. Rev. Lett. {\bf 82}, 900 (1999).

\bibitem{pilaftsis1}
A. Pilaftsis, Phys. Lett. B {\bf 471}, 174 (1999).

\bibitem{chang2}
D. Chang, W.-F. Chang and W.-Y. Keung, Phys. Lett. B {\bf 478}, 239 (2000).

\bibitem{mssmrainbow1}
A. Pilaftsis, Phys. Rev. D {\bf 62}, 016007 (2000).

\bibitem{demir}
D. Demir, O. Lebedev, K. A. Olive, M. Pospelov and A. Ritz,  Nucl. Phys. B {\bf 680}, 339 (2004). 

\bibitem{feng1}
T.-F. Feng, X.-Q. Li, L. Lin, J. Maalampi and X.-M. Zhang, Phys. Rev. D {\bf 71}, 056005 (2005).

\bibitem{feng2}
T.-F. Feng, X.-Q. Li, L. Lin, J. Maalampi and H.-S. Song, Phys. Rev. D {\bf 73}, 116001 (2006).

\bibitem{lihiggshiggsino}
Y. Li, S. Profumo and M. J. Ramsey-Musolf, Phys. Rev. D {\bf 78}, 075009 (2008).

\bibitem{mssmrainbow2}
N. Yamanaka, Phys. Rev. D {\bf 87}, 011701 (2013). 

\bibitem{carena}
M. Carena, J. Ellis, J. S. Lee, A. Pilaftsis and C. E. M. Wagner, JHEP {\bf 1602}, 123 (2016).

\bibitem{nakai}
Y. Nakai and M. Reece, arXiv:1612.08090 [hep-ph].

\bibitem{daimssm}
J. Dai, H. Dykstra, R. G. Leigh, S. Paban and D. Dicus, Phys. Lett. B {\bf 237}, 216 (1990).

\bibitem{arnowitt1}
R. Arnowitt, J. L. Lopez and D. V. Nanopoulos, Phys. Rev. D {\bf 42}, 2423 (1990).

\bibitem{arnowitt2}
R. Arnowitt, M. J. Duff and K. S. Stelle, Phys. Rev. D {\bf 43}, 3085 (1991).

\bibitem{fischler}
W. Fischler, S. Paban and S. D. Thomas, Phys. Lett. B {\bf 289}, 373 (1992).

\bibitem{falk}
T. Falk, K. A. Olive, M. Pospelov and R. Roiban, Nucl. Phys. B {\bf 560}, 3 (1999).

\bibitem{lebedevmssm}
O. Lebedev and M. Pospelov, Phys. Rev. Lett. {\bf 89}, 101801 (2002).

\bibitem{pilaftsishiggsmssm}
A. Pilaftsis, Nucl. Phys. B {\bf 644}, 263 (2002).

\bibitem{mssmunify1}
S. A. Abel, W. N. Cottingham and I. B. Whittingham, Phys. Lett. B {\bf 370}, 106 (1996).

\bibitem{mssmunify2}
R. Gariato and J. D. Wells, Phys. Rev. D {\bf 55}, 1611 (1997).

\bibitem{mssmunify3}
A. Romanino and A. Strumia, Nucl. Phys. B {\bf 490}, 3 (1997).

\bibitem{mssmunify4}
A. Bartl, T. Gajdosik, W. Porod, P. St\"{o}ckinger and H. Stremnitzer, Phys. Rev. D {\bf 60}, 073003 (1999).

\bibitem{mssmunify5}
M. Brhlik, L. L. Everett, G. L. Kane and J. D. Lykken, Phys. Rev. Lett. {\bf 83}, 2124 (1999).

\bibitem{mssmunify6}
E. Accomando, R. L. Arnowitt and B. Dutta, Phys. Rev. D {\bf 61}, 115003 (2000).

\bibitem{splitsusy1}
N. Arkani-Hamed and S. Dimopoulos, JHEP {\bf 0506}, 073 (2005).

\bibitem{splitsusy2}
N. Arkani-Hamed, S. Dimopoulos, G.F. Giudice and A. Romanino, Nucl. Phys. B {\bf 709}, 3 (2005).

\bibitem{splitsusy3}
D. Chang, W.-F. Chang and W.-Y. Keung, Phys. Rev. D {\bf 71}, 076006 (2005).

\bibitem{splitsusy4}
G. F. Giudice and A. Romanino, Phys. Lett. B {\bf 634}, 307 (2006).

\bibitem{dhuria}
M. Dhuria and A. Misra, Phys. Rev. D {\bf 90}, 085023 (2014).

\bibitem{sarellis}
S. A. R. Ellis and G. L. Kane, JHEP {\bf 1601}, 077 (2016). 

\bibitem{grossman}
Y. Grossman and M. P. Worah, Phys. Lett. B {\bf 395}, 241 (1997).

\bibitem{barbieri}
R. Barbieri and A. Strumia, Nucl. Phys. B {\bf 508}, 3 (1997).

\bibitem{belle}
K. Abe {\it et al.} (Belle Collaboration), Phys. Rev. Lett. {\bf 91}, 261602 (2003). 


\bibitem{hisanoshimizu1}
J. Hisano and Y. Shimizu, Phys. Lett. B {\bf 581}, 224 (2004).

\bibitem{hisanoshimizu2}
J. Hisano and Y. Shimizu, 
Phys. Rev. D {\bf 70}, 093001 (2004). 

\bibitem{endomssmflavor}
M. Endo, M. Kakizaki and M. Yamaguchi, Phys. Lett. B {\bf 583}, 186 (2004).

\bibitem{chomssmflavor}
G.-C. Cho, N. Haba and M. Honda, Mod. Phys. Lett. A {\bf 20}, 2969 (2005).

\bibitem{hisanonagai1}
J. Hisano, M. Nagai and P. Paradisi, Phys. Lett. B {\bf 642}, 510 (2006).

\bibitem{hisanonagai2}
J. Hisano, M. Nagai and P. Paradisi, Phys. Rev. D {\bf 78}, 075019 (2008). 
 
\bibitem{hisanonagai3}
J. Hisano, M. Nagai and P. Paradisi, Phys. Rev. D {\bf 80}, 095014 (2009).

\bibitem{Altmannshofer}
W. Altmannshofer, A. J. Buras and P. Paradisi, Phys. Lett. B {\bf 688}, 202 (2010).

\bibitem{nmssm}
M. Maniatis, Int. J. Mod. Phys. A {\bf 25}, 3505 (2010).

\bibitem{nmssmedm}
S. F. King, M. Muhlleitner, R. Nevzorov and K. Walz, Nucl. Phys. B {\bf 901}, 526 (2015).

\bibitem{blmssm1}
P. F. Perez, Phys. Lett. B {\bf 711}, 353 (2012).

\bibitem{blmssm2}
J. M. Arnold, P. F. Perez, B. Fornal and S. Spinner, Phys. Rev. D {\bf 85}, 115024 (2012).

\bibitem{blmssmedm}
S.-M. Zhao, T.-F. Feng, X.-J. Zhan, H.-B. Zhang and B. Yan, JHEP {\bf 1507}, 124 (2015).

\bibitem{rpvreview1}
G. Bhattacharyya, arXiv:hep-ph/9709395.

\bibitem{rpvreview2}
H. K. Dreiner, in {\it Perspectives on Supersymmetry II}, edited by G. L. Kane (World Scientific, Singapore, 1997), p. 565 [arXiv:hep-ph/9707435].

\bibitem{rpvreview3}
M. Chemtob, Prog. Part. Nucl. Phys. {\bf 54}, 71 (2005).

\bibitem{rpvreview4}
R. Barbier {\it et al.}, Phys. Rep. {\bf 420}, 1 (2005);


\bibitem{keum1}
Y. Y. Keum and Otto C. W. Kong, Phys. Rev. Lett. {\bf 86}, 393 (2001).

\bibitem{keum2}
Y. Y. Keum and Otto C. W. Kong, Phys. Rev. D {\bf 63}, 113012 (2001).

\bibitem{choirpv}
K. Choi, E. J. Chun and K. Hwang, Phys. Rev. D {\bf 63}, 013002 (2001).

\bibitem{chiourpv}
C.-C. Chiou, O. C. W. Kong and R. D. Vaidya, Phys. Rev. D {\bf 76}, 013003 (2007).

\bibitem{godbole}
R. M. Godbole, S. Pakvasa, S. D. Rindani and X. Tata, Phys. Rev. D {\bf 61}, 113003 (2000).

\bibitem{abelrpv}
S. A. Abel, A. Dedes and H. K. Dreiner, JHEP{\bf 05}, 13 (2000).

\bibitem{changrpv}
D. Chang, W.-F. Chang, M. Frank and W.-Y. Keung, Phys. Rev. D {\bf 62}, 095002 (2000).

\bibitem{rpvedm1}
N. Yamanaka, Phys. Rev. D {\bf 85}, 115012 (2012).

\bibitem{rpvedm2}
N. Yamanaka, Phys. Rev. D {\bf 86}, 075029 (2012). 

\bibitem{rpvedm3}
N. Yamanaka, T. Sato and T. Kubota, Phys. Rev. D {\bf 87}, 115011 (2013). 

\bibitem{rpvedm4}
N. Yamanaka, arXiv:1212.5800 [hep-ph].

\bibitem{rpv4f1}
P. Herczeg, Phys. Rev. D {\bf 61}, 095010 (2000).

\bibitem{faessler1}
A. Faessler, T. Gutsche, S. Kovalenko and V. E. Lyubovitskij, Phys. Rev. D {\bf 73}, 114023 (2006).

\bibitem{faessler2}
A. Faessler, T. Gutsche, S. Kovalenko and V. E. Lyubovitskij, Phys. Rev. D {\bf 74}, 074013 (2006).

\bibitem{susycancel1}
T. Ibrahim and P. Nath, Phys. Rev. D {\bf 57}, 478 (1998).

\bibitem{susycancel2}
T. Ibrahim and P. Nath, Phys. Lett. B {\bf 418}, 98 (1998).

\bibitem{susycancel3}
T. Ibrahim and P. Nath, Phys. Rev. D {\bf 58}, 111301 (1998).

\bibitem{susycancel4}
M. Brhlik, G. J. Good and G. L. Kane, Phys. Rev. D {\bf 59}, 115004 (1999).

\bibitem{susycancel5}
S. Pokorski, J. Rosiek and C. A. Savoy, Nucl. Phys. B {\bf 570}, 81 (2000).

\bibitem{susycancel6}
S. Abel, S. Khalil and O. Lebedev, Nucl. Phys. B {\bf 606}, 151 (2001).

\bibitem{susycancel7}
K. A. Olive, M. Pospelov, A. Ritz and Y. Santoso, Phys. Rev. D 72, 075001 (2005).

\bibitem{susycancel8}
S. Y. Ayazi and Y. Farzan, Phys. Rev. D {\bf 74}, 055008 (2006).

\bibitem{susycancel9}
S. Y. Ayazi and Y. Farzan, JHEP {\bf 0706}, 013 (2007).

\bibitem{susycancel10}
J. Ellis, J.S. Lee and A. Pilaftsis, JHEP {\bf 1010}, 049 (2010),

\bibitem{susycancel11}
Y. Li, S. Profumo and M. J. Ramsey-Musolf, JHEP {\bf 1008}, 062 (2010).

\bibitem{susycancel12}
J. Ellis, J.S. Lee and A. Pilaftsis, JHEP {\bf 1102}, 045 (2011).

\bibitem{susycancel13}
J. Ellis, J.S. Lee and A. Pilaftsis, arXiv:1009.1151 [math.OC].

\bibitem{yamanakabook}
N.~Yamanaka,
  ``Analysis of the Electric Dipole Moment in the R-parity Violating Supersymmetric Standard Model'',
  Springer, Berlin, Germany (2014).

\bibitem{linearprogramming}
N. Yamanaka, T. Sato and T. Kubota, 
JHEP {\bf 1412}, 110 (2014). 

\bibitem{left-right1}
R. N. Mohapatra and J. C. Pati, Phys. Rev. D {\bf 11}, 566 (1975).

\bibitem{left-right2}
R. N. Mohapatra and J. C. Pati, Phys. Rev. D {\bf 11}, 2558 (1975).

\bibitem{left-right3}
G. Senjanovic and R. N. Mohapatra, Phys. Rev. D {\bf 12}, 1502 (1975).

\bibitem{LRLHC1}
A. Maiezza and M. Nemev\u{s}ek, 
Phys. Rev. D {\bf 82}, 055022 (2010). 

\bibitem{CMSW'1}
S. Chatrchyan {\it et al.} (CMS Collaboration), JHEP {\bf 1405}, 108 (2014).

\bibitem{ATLASW'}
G. Aad {\it et al.} (ATLAS Collaboration), Phys. Lett. B {\bf 743}, 235 (2015).

\bibitem{CMSW'2}
V. Khachatryan {\it et al.} (CMS Collaboration), Phys. Lett. B {\bf 755}, 196 (2016). 

\bibitem{CMSW'3}
V. Khachatryan {\it et al.} (CMS Collaboration), JHEP {\bf 1602}, 122 (2016).

\bibitem{LRedm1}
G. Beall and A. Soni, Phys. Rev. Lett. {\bf 47}, 552 (1981).

\bibitem{LRedm2}
D. Chang, C. S. Li and T. C. Yuan, Phys. Rev. D {\bf 42}, 867 (1990).

\bibitem{LRedm3}
J.-M. Fr\`{e}re, J. Galand, A. Le Yaouanc, L. Oliver, O. P\`{e}ne and J.-C. Raynal, Phys. Rev. D {\bf 45}, 259 (1992).

\bibitem{LRzhang1}
Y. Zhang, H. An, X. Ji and R. N. Mohapatra, Phys. Rev. D {\bf 76}, 091301 (2007).

\bibitem{LRzhang2}
Y. Zhang, H. An, X. Ji and R. N. Mohapatra, Nucl. Phys. B {\bf 802}, 247 (2008). 

\bibitem{LRxu}
F. Xu, H. An and X. Ji, JHEP {\bf 1003}, 088 (2010).

\bibitem{eft6dim}
J. de Vries, E. Mereghetti, R.G.E. Timmermans and U. van Kolck, Annals Phys. {\bf 338}, 50 (2013). 

\bibitem{LRLHC2}
A. Maiezza and M. Nemev\u{s}ek, Phys. Rev. D {\bf 90}, 095002 (2014). 

\bibitem{dekens}
W. Dekens, J. de Vries, J. Bsaisou, W. Bernreuther, C. Hanhart, U.-G. Mei{\ss}ner, A. Nogga and A. Wirzba,  
JHEP {\bf 1407}, 069 (2014). 

\bibitem{sengisovector}
C.-Y. Seng, J. de Vries, E. Mereghetti, H. H. Patel and M. Ramsey-Musolf, Phys. Lett. B {\bf 736}, 147 (2014). 

\bibitem{Aguilar}
J. A. Aguilar-Saavedra, R. Benbrik, S. Heinemeyer, and M. P\'{e}rez-Victoria, Phys. Rev. D {\bf 88}, 094010 (2013). 
\bibitem{djouadi}
A. Djouadi and A. Lenz, Phys. Lett. B {\bf 715}, 310 (2012).

\bibitem{dimopouloscomposite}
S. Dimopoulos and J. Preskill, Nucl. Phys. B {\bf 199}, 206 (1982).

\bibitem{kaplancomposite}
D. B. Kaplan, Nucl. Phys. B {\bf 365}, 259 (1991).

\bibitem{peterson}
K. A. Peterson, Phys. Lett. B {\bf 255}, 567 (1991).

\bibitem{contino}
R. Contino, T. Kramer, M. Son and R. Sundrum, JHEP {\bf 0705}, 074 (2007). 

\bibitem{hewette6}
J. L. Hewett and T. G. Rizzo, Phys. Rep. {\bf 183}, 193 (1989).

\bibitem{vectorlikefcnc1}
F. del Aguila and M. J. Bowick, Nucl. Phys. B {\bf 224}, 107 (1983).

\bibitem{vectorlikefcnc2}
G. C. Branco and L. Lavoura, Nucl. Phys. B {\bf 278}, 738 (1986).

\bibitem{vectorlikefcnc3}
Y. Nir and D. J. Silverman, Phys. Rev. D {\bf 42}, 1477 (1990).

\bibitem{vectorlikefcnc4}
G. C. Branco, T. Morozumi, P. A. Parada and M. N. Rebelo, Phys. Rev. D {\bf 48}, 1167 (1993).

\bibitem{vectorlikefcnc5}
F. del Aguila, J. A. Aguilar-Saavedra, and G. C. Branco, Nucl. Phys. B {\bf 510}, 39 (1998).

\bibitem{vectorlikefcnc6}
G. Barenboim, F. J. Botella, G. C. Branco and O. Vives, Phys. Lett. B {\bf 422}, 277 (1998).

\bibitem{vectorlikefcnc7}
T. Appelquist, M. Piai and R. Shrock, Phys. Lett. B {\bf 593}, 175 (2004). 

\bibitem{vectorlikefcnc8}
T. Appelquist, M. Piai and R. Shrock, Phys. Lett. B {\bf 595}, 442 (2004). 

\bibitem{vectorlikefcnc9}
Y. Adachi, N. Kurahashi, C. S. Lim and N. Maru,  JHEP {\bf 1011}, 150 (2010). 

\bibitem{vectorlikefcnc10}
K. Ishiwata and M. B. Wise, Phys. Rev. D {\bf 88}, 055009 (2013). 

\bibitem{vectorlikefcnc11}
M. K\"{o}nig, M. Neubert and D. M. Straub, Eur. Phys. J. C {\bf 74}, 2945 (2014). 

\bibitem{liaockm}
Y. Liao and X. Li, Phys. Lett. B {\bf 503}, 301 (2001).

\bibitem{Iltan}
E. O. Iltan, Eur. Phys. J. C {\bf 51}, 689 (2007). 

\bibitem{chang5d}
W.-F. Chang and J. N. Ng, JHEP {\bf 0212}, 077 (2002).

\bibitem{agashe}
K. Agashe, G. Perez and A. Soni, Phys. Rev. D {\bf 71}, 016002 (2005).

\bibitem{nishiwaki}
C. S. Lim, N. Maru and K. Nishiwaki, Phys. Rev.D {\bf 81}, 076006 (2010). 

\bibitem{Kalinowski}
M. W. Kalinowski, Eur. Phys. J. C {\bf 74}, 2742 (2014).

\bibitem{fan}
J. Fan and M. Reece, JHEP {\bf 1306}, 004 (2013).

\bibitem{choi}
K. Choi, S. H. Im, H. Kim and D. Y. Mo, Phys. Lett. B {\bf 760}, 666 (2016). 

\bibitem{appelquist1}
T. Appelquist and G.-H. Wu, Phys. Rev. D {\bf 51}, 240 (1995). 

\bibitem{leptoquarkcms1}
V. Khachatryan {\it et al.} (CMS Collaboration), Phys. Lett. B {\bf 739}, 229 (2014).

\bibitem{leptoquarkcms2}
V. Khachatryan {\it et al.} (CMS Collaboration), JHEP {\bf 1507}, 042 (2015).

\bibitem{leptoquarkcms3}
V. Khachatryan {\it et al.} (CMS Collaboration), Phys. Rev. D {\bf 93}, 032004 (2016).

\bibitem{leptoquarkcms4}
V. Khachatryan {\it et al.} (CMS Collaboration), Phys. Rev. D {\bf 93}, 032005 (2016).

\bibitem{leptoquarkatlas1}
G. Aad {\it et al.} (ATLAS Collaboration), Eur. Phys. J. C {\bf 76}, 5 (2016).

\bibitem{leptoquarkatlas2}
M. Aaboud {\it et al.} (ATLAS Collaboration), New J. Phys. {\bf 18}, 093016 (2016).

\bibitem{davidson}
S. Davidson, D. Bailey and B. A. Campbell, Z. Phys. C {\bf 61}, 613 (1994).

\bibitem{geng}
C. Q. Geng, Z. Phys. C {\bf 48}, 279 (1990).

\bibitem{hee-Nedm}
X.-G. He, B. H. J. McKellar and S. Pakvasa, Phys. Lett. B {\bf 283}, 348 (1992).

\bibitem{herczegleptoquark}
P. Herczeg, Phys. Rev. D {\bf 68}, 116004 (2003).


\bibitem{dekens1}
W. Dekens and J. de Vries, JHEP {\bf 1305}, 149 (2013).

\bibitem{brod2}
W. Altmannshofer, J. Brod and M. Schmaltz, JHEP {\bf 1505}, 125 (2015).

\bibitem{cirigliano2}
V. Cirigliano, W. Dekens, J. de Vries and E. Mereghetti, Phys. Rev. D {\bf 94}, 016002 (2016).

\bibitem{cirigliano3}
V. Cirigliano, W. Dekens, J. de Vries and E. Mereghetti, Phys. Rev. D {\bf 94}, 034031 (2016).


\bibitem{tensorrenormalization1}
X. Artru and M. Mekhfi, Z. Phys. C {\bf 45}, 669 (1990).

\bibitem{braaten1}
E. Braaten, C. S. Li and T. C. Yuan, Phys. Rev. Lett. {\bf 64}, 1709 (1990).

\bibitem{boyd}
G. Boyd, A. K. Gupta, S. P. Trivedi and M. B. Wise, Phys. Lett. B {\bf 241}, 584 (1990).

\bibitem{braaten2}
E. Braaten, C. S. Li and T. C. Yuan, Phys. Rev. D {\bf 42}, 276 (1990).

\bibitem{dineweinbergop}
M. Dine and W. Fischler, Phys. Lett. B {\bf 242}, 239 (1990).

\bibitem{tensorrenormalization2}
V. Barone, Phys. Lett. B {\bf 409}, 499 (1997).

\bibitem{degrassi}
G. Degrassi and S. Marchetti, E. Franco and L. Silvestrini, JHEP {\bf 0511}, 044 (2005). 

\bibitem{yang}
J. Hisano, K. Tsumura and M. J. S. Yang,  Phys. Lett. B {\bf 713}, 473 (2012). 

\bibitem{pdg}
K. A. Olive {\it et al.} (Particle Data Group), Chin. Phys. C {\bf 40}, 100001 (2016). 

\bibitem{buras2}
G. Buchalla, A. J. Buras and M. E. Lautenbacher, Rev. Mod. Phys. {\bf 68}, 1125 (1996). 

\bibitem{buras1}
A. J. Buras, M. Jamin, M. E. Lautenbacher and P. H. Weisz, Nucl. Phys. B {\bf 370}, 69 (1992) [Erratum ibid. B {\bf 375}, 501 (1992)].

\bibitem{smnuclearedm}
N. Yamanaka and E. Hiyama, JHEP {\bf 1602} (2016) 067.

\bibitem{crewther}
R. J. Crewther, P. Di Vecchia, G. Veneziano and E. Witten, Phys. Lett. B {\bf 88}, 123 (1979) [Erratum ibid. B {\bf 91}, 487 (1980)].

\bibitem{pich}
A. Pich and E. de Rafael, Nucl. Phys. B {\bf 367}, 313 (1991).

\bibitem{borasoy}
B. Borasoy, Phys. Rev. D {\bf 61}, 114017 (2000).

\bibitem{dib}
C. Dib, A. Faessler, T. Gutsche, S. G. Kovalenko, J. Kuckei, V. E. Lyubovitskij and K. Pumsa-ard, J. Phys. G {\bf 32}, 547 (2006).

\bibitem{kuckei}
J. Kuckei, C. Dib, A. Faessler, T. Gutsche, S. G. Kovalenko, V. E. Lyubovitskij and K. Pumsa-ard, Phys. Atom. Nucl. {\bf 70}, 349 (2007).

\bibitem{narison}
S. Narison, Phys. Lett. B {\bf 666}, 455 (2008).

\bibitem{pospelovtheta1}
M. Pospelov and A. Ritz, Phys. Rev. Lett. {\bf 83}, 2526 (1999). 

\bibitem{pospelovtheta2}
M. Pospelov and A. Ritz, Nucl. Phys. B {\bf 573}, 177 (2000). 

\bibitem{nedmholography1}
D. K. Hong, H.-C. Kim, S. Siwach and H.-U. Yee,  JHEP {\bf 0711}, 036 (2007). 

\bibitem{nedmholography2}
L. Bartolini, F. Bigazzi, S. Bolognesi, A. L. Cotrone and A. Manenti, arXiv:1609.09513 [hep-ph].

\bibitem{nedmlattice1}
S. Aoki, A. Gocksch, A. V. Manohar and Stephen R. Sharpe, Phys. Rev. Lett. {\bf 65}, 1092 (1990).

\bibitem{shintani1}
E. Shintani {\it et al.}, Phys. Rev. D {\bf 72}, 014504 (2005).

\bibitem{nedmlattice3}
F. Berruto, T. Blum, K. Orginos and A. Soni, Phys. Rev. D {\bf 73}, 054509 (2006).

\bibitem{shintani2}
E. Shintani, S. Aoki and Y. Kuramashi, Phys. Rev. D {\bf 78}, 014503 (2008).

\bibitem{nedmlattice5}
F.-K. Guo, R. Horsley, U.-G. Meissner, Y. Nakamura, H. Perlt, P. E. L. Rakow, G. Schierholz, A. Schiller and J. M. Zanotti, 
Phys. Rev. Lett. {\bf 115}, 062001 (2015).

\bibitem{nedmlattice6}
A. Shindler, T. Luu and J. de Vries, Phys. Rev. D {\bf 92}, 094518 (2015). 

\bibitem{nedmetm}
C. Alexandrou, A. Athenodorou, M. Constantinou, K. Hadjiyiannakou, K. Jansen, G. Koutsou, K. Ottnad and M. Petschlies,  
Phys. Rev. D {\bf 93}, 074503 (2016). 

\bibitem{shintani3}
E. Shintani, T. Blum, T. Izubuchi and A. Soni, Phys. Rev. D {\bf 93}, 094503 (2016). 

\bibitem{ottnad}
K. Ottnad, B. Kubis, U.-G. Mei{\ss}ner and F.-K. Guo, Phys. Lett. B {\bf 687}, 42 (2010).

\bibitem{mereghetti2}
E. Mereghetti, W. H. Hockings and U. van Kolck, Annals Phys. {\bf 325}, 2363 (2010).

\bibitem{mereghetti1}
E. Mereghetti, J. de Vries, W. H. Hockings, C. M. Maekawa and U. van Kolck, Phys. Lett. B {\bf 696}, 97 (2011).

\bibitem{guo1}
F.-K. Guo and U.-G. Mei{\ss}ner, JHEP {\bf 1212}, 097 (2012).

\bibitem{devriessplitting}
J. de Vries, E. Mereghetti and A. Walker-Loud, Phys. Rev. C {\bf 92}, 045201 (2015). 

\bibitem{devriesreview}
J. de Vries and Ulf-G. Mei{\ss}ner, Int. J. Mod. Phys. E {\bf 25}, 1641008 (2016).

\bibitem{peccei}
R. D. Peccei and H. R. Quinn, Phys. Rev. Lett. {\bf 38}, 1440 (1977).

\bibitem{shifmantheta}
M. A. Shifman, A. I. Vainshtein and V. I. Zakharov, Nucl. Phys. B {\bf 166}, 493 (1980).

\bibitem{bigi}
I. Bigi and N. G. Ural'tsev, Zh. Eksp. Teor. Fiz. {\bf 100}, 363 (1991) [Sov. Phys. JETP {\bf 73}, 198 (1991)]; Nucl. Phys. B {\bf 353}, 321 (1991).

\bibitem{belayev1}
V. M. Belayev and I. B. Ioffe, Sov. Phys. JETP {\bf 100}, 493 (1982).

\bibitem{belayev2}
V. M. Belayev and Ya. I. Kogan, Sov. J. Nucl. Phys.  {\bf 40}, 659 (1984).

\bibitem{4-quark5}
H. An, X. Ji and F. Xu, JHEP {\bf 02}, 043 (2010).

\bibitem{Barton}
G. Barton and E. G.White, Phys. Rev. {\bf 184}, 1660 (1969).

\bibitem{haxton}
W. C. Haxton and E. M. Henley, Phys. Rev. Lett. {\bf 51}, 1937 (1983).

\bibitem{gorchtein}
M. Gorchtein, arXiv:0803.2906 [hep-ph].

\bibitem{blin}
T. Gutsche, A. N. Hiller Blin, S. Kovalenko, S. Kuleshov, V. E. Lyubovitskij, M. J. Vicente Vacas and A. Zhevlakov, arXiv:1612.02276 [hep-ph].

\bibitem{chiraleft3nucleon}
J. de Vries, R. Higa, C.-P. Liu, E. Mereghetti, I. Stetcu, R. G. E. Timmermans and U. van Kolck, Phys. Rev. C {\bf 84}, 065501 (2011).

\bibitem{liu}
C.-P. Liu and R. G. E. Timmermans, Phys. Rev. C {\bf 70}, 055501 (2004).

\bibitem{stetcu}
I. Stetcu, C.-P. Liu, J. L. Friar, A. C. Hayes and P. Navratil, Phys. Lett. B {\bf 665}, 168 (2008).

\bibitem{song}
Y.-H. Song, R. Lazauskas and V. Gudkov, {\it Phys. Rev. C} {\bf 87}, 015501 (2013).

\bibitem{bsaisou}
J. Bsaisou, J. de Vries, C. Hanhart, S. Liebig, U.-G. Mei{\ss}ner, D. Minossi, A. Nogga and A. Wirzba, JHEP {\bf 1503}, 104 (2015) [Erratum ibid. {\bf 1505}, 083 (2015)]. 

\bibitem{bsaisou2}
J. Bsaisou, U.-G. Mei{\ss}ner, A. Nogga and A. Wirzba, Annals Phys. {\bf 359}, 317 (2015).

\bibitem{yamanakanuclearedm}
N. Yamanaka and E. Hiyama, Phys. Rev. C {\bf 91}, 054005 (2015).

\bibitem{brambilla}
N. Brambilla {\it et al.}, Eur. Phys. J. C {\bf 74}, 2981 (2014).

\bibitem{collinslattice2016}
S. Collins, Talk given at the 34th International Symposium on Lattice Field Theory, 24-30 July 2016.

\bibitem{chengdashen}
T. P. Cheng and R. Dashen, Phys. Rev. Lett. {\bf 26}, 594 (1971).

\bibitem{gasser}
J. Gasser, Ann. Phys. {\bf 136}, 62 (1981).

\bibitem{gasserleutwyler1}
J. Gasser, H. Leutwyler and M. E. Sainio, Phys. Lett. B {\bf 213}, 85 (1988).

\bibitem{gasserleutwyler2}
J. Gasser, H. Leutwyler and M. E. Sainio, Phys. Lett. B {\bf 253}, 252 (1991).

\bibitem{gasserleutwyler3}
J. Gasser, H. Leutwyler and M. E. Sainio, Phys. Lett. B {\bf 253}, 260 (1991).

\bibitem{alarcon1}
J. M. Alarcon, J. Martin Camalich and J. A. Oller, Phys. Rev. D {\bf 85}, 051503 (2012).

\bibitem{alarcon2}
J. M. Alarcon, J. Martin Camalich and J. A. Oller, Ann. Phys. {\bf 336}, 413 (2013).

\bibitem{ren}
X.-L. Ren, L.-S. Geng and J. Meng, Phys. Rev. D {\bf 91}, 051502 (2015). 

\bibitem{Hoferichter}
M. Hoferichter, J. Ruiz de Elvira, B. Kubis and U.-G. Mei{\ss}ner, Phys. Rev. Lett. {\bf 115}, 092301 (2015). 

\bibitem{yao}
D.-L. Yao {\it et al.}, JHEP {\bf 1605}, 038 (2016).

\bibitem{ohki}
H. Ohki {\it et al.}, Phys. Rev. D {\bf 78}, 054502 (2008).

\bibitem{young1}
R. D. Young and A. W. Thomas, Phys. Rev. D {\bf 81}, 014503 (2010). 

\bibitem{durr1}
S. D\"{u}rr {\it et al.}, Phys. Rev. D {\bf 85}, 014509 (2012). 

\bibitem{dinter}
S. Dinter {\it et al.} (ETM Collaboration), JHEP {\bf 1208}, 037 (2012). 

\bibitem{qcdsf1}
G. S. Bali {\it et al.} (QCDSF Collaboration), Phys. Rev. D {\bf 85}, 054502 (2012). 

\bibitem{qcdsf2}
G. S. Bali {\it et al.} (QCDSF Collaboration), Nucl. Phys. B {\bf 866}, 1 (2013).

\bibitem{qcdsf-ukqcdsigmaterm}
R. Horsley {\it et al.} (QCDSF-UKQCD Collaboration), Phys. Rev. D {\bf 85}, 034506 (2012).

\bibitem{durr2}
S. Durr  {\it et al.}, Phys. Rev. Lett. {\bf 116}, 172001 (2016). 

\bibitem{rqcdsigmaterm}
G. S. Bali, S. Collins, D. Richtmann, A. Sch\"{a}fer, W. S\"{o}ldner, and A. Sternbeck, Phys. Rev. D {\bf 93}, 094504 (2016).

\bibitem{etmsigmaterm}
A. Abdel-Rehim, C. Alexandrou, M. Constantinou, K. Hadjiyiannakou, K. Jansen, Ch. Kallidonis, G. Koutsou, and A. Vaquero Avil\'{e}s-Casco, Phys. Rev. Lett. {\bf 116}, 252001 (2016).

\bibitem{chiqcdsigmaterm}
Y.-B. Yang {\it et al.}, Phys. Rev. D {\bf 94}, 054503 (2016). 

\bibitem{adler}
S. Adler, E. Colglazier, J. Healy, I. Karliner, J. Lieberman, Y. Ng and H. Tsao, Phys. Rev. D {\bf 11}, 3309 (1975).

\bibitem{rpvbetadecay1}
N. Yamanaka, T. Sato and T. Kubota, J. Phys. G {\bf 37}, 055104 (2010). 

\bibitem{rpvbetadecay2}
N. Yamanaka, T. Sato and T. Kubota, Phys. Rev. D {\bf 86}, 075032 (2012). 

\bibitem{gonzales-alonso}
M. Gonz\'{a}lez-Alonso and J. Martin Camalich, Phys. Rev. Lett. {\bf 112}, 042501 (2014). 

\bibitem{mohr}
P. J. Mohr, B. N. Taylor and D. B. Newell, Rev. Mod. Phys. {\bf 84}, 1527 (2012). 

\bibitem{thomas}
A. W. Thomas, X. G. Wang and R. D. Young, Phys. Rev. C {\bf 91}, 015209 (2015). 

\bibitem{linisovectoraxial}
H.-W. Lin, T. Blum, S. Ohta, S. Sasaki and T. Yamazaki, Phys. Rev. D {\bf 78}, 014505 (2008).

\bibitem{rbcukqcdisovectortensor}
Y. Aoki, T. Blum, H.-W. Lin, S. Ohta, S. Sasaki, R. Tweedie, J. Zanotti and T. Yamazaki, Phys. Rev. D {\bf 82}, 014501 (2010).

\bibitem{green}
J. R. Green, J. W. Negele, A. V. Pochinsky, S. N. Syritsyn, M. Engelhardt and S. Krieg, Phys. Rev. D {\bf 86}, 114509 (2012).

\bibitem{pndmeconnected}
T. Bhattacharya, S. D. Cohen, R. Gupta, A. Joseph, H.-W. Lin and B. Yoon, Phys. Rev. D {\bf 89}, 094502 (2014).

\bibitem{rqcdisovector}
G. S. Bali {\it et al.} (RQCD Collaboration), Phys. Rev. D {\bf 91}, 054501 (2015).

\bibitem{chiqcdisovector}
Y.-B. Yang, A. Alexandru, T. Draper, M. Gong and K.-F. Liu, Phys. Rev. D {\bf 93}, 034503 (2016).

\bibitem{pndmeisovector}
T. Bhattacharya {\it et al.} (PNDME Collaboration), Phys. Rev. D {\bf 94}, 054508 (2016). 

\bibitem{etmisovector}
A. Abdel-Rehim {\it et al.}, Phys. Rev. D {\bf 92}, 114513 (2015) [Erratum: Phys. Rev. D {\bf 93}, 039904 (2016)].

\bibitem{hatsudakunihiro1}
T. Hatsuda and T. Kunihiro, Z. Phys. C {\bf 51}, 49 (1991).

\bibitem{hatsudakunihiro2}
T. Hatsuda and T. Kunihiro, Nucl. Phys. B {\bf 387}, 715 (1992).

\bibitem{hatsudareview}
T. Hatsuda and T. Kunihiro, Phys. Rep. {\bf 247}, 221 (1994).

\bibitem{yamanakasde2}
N. Yamanaka, S. Imai, T. M. Doi and H. Suganuma, Phys. Rev. D {\bf 89}, 074017 (2014).

\bibitem{takeda}
K. Takeda, S. Aoki, S. Hashimoto, T. Kaneko, J. Noaki and T. Onogi, Phys. Rev. D {\bf 83}, 114506 (2011).

\bibitem{ohki2}
H. Ohki, K. Takeda, S. Aoki, S. Hashimoto, T. Kaneko, H. Matsufuru, J. Noaki and T. Onogi (JLQCD Collaboration), Phys. Rev. D {\bf 87}, 034509 (2013). 

\bibitem{etmdisconnected}
A.~Abdel-Rehim {\it et al.}, Phys. Rev. D {\bf 89}, 034501 (2014).

\bibitem{etm1}
C. Alexandrou, M. Constantinou, S. Dinter, V. Drach, K. Hadjiyiannakou, K. Jansen, G. Koutsou and A. Vaquero, 
Phys. Rev. D {\bf 91}, 094503 (2015). 

\bibitem{milc1}
D. Toussaint and W. Freeman (MILC Collaboration), Phys. Rev. Lett. {\bf 103}, 122002 (2009). 

\bibitem{engelhardt}
M. Engelhardt, Phys. Rev. D {\bf 86}, 114510 (2012). 

\bibitem{junnarkar}
P. M. Junnarkar and A. Walker-Loud, Phys. Rev. D {\bf 87}, 114510 (2013). 

\bibitem{milccharmcontent}
W. Freeman and D. Toussaint (MILC Collaboration), Phys. Rev. D {\bf 88}, 054503 (2013).

\bibitem{chiQCDcharmcontent}
M. Gong {\it et al.} ($\chi$QCD Collaboration), Phys. Rev. D {\bf 88}, 014503 (2013).

\bibitem{alarcon3}
J. M. Alarcon, L. S. Geng, J. Martin Camalich and J. A. Oller, Phys. Lett. B {\bf 730}, 342 (2014).

\bibitem{gubler1}
P. Gubler and K. Ohtani, Phys. Rev. D {\bf 90}, 094002 (2014). 

\bibitem{gubler2}
P. Gubler and W. Weise, Phys. Lett. B {\bf 751}, 396 (2015).

\bibitem{gubler3}
P. Gubler and W. Weise, Nucl. Phys. A {\bf 954}, 125 (2016).

\bibitem{wittenheavyquark}
E. Witten, Nucl. Phys. B {\bf 104}, 445 (1976).

\bibitem{shifmanheavyquark}
M. A. Shifman, A. I. Vainshtein, and V. I. Zakharov, Phys. Lett. B {\bf 78}, 443 (1978).

\bibitem{zhitnitskyheavyquark}
A. R. Zhitnitsky, Phys. Rev. D {\bf 55}, 3006 (1997).

\bibitem{franzheavyquark}
M. Franz, M. V. Polyakov, and K. Goeke, Phys. Rev. D {\bf 62}, 074024 (2000).

\bibitem{compass}
C. Adolph {\it et al.}, Phys. Lett. B {\bf 753}, 18 (2016).

\bibitem{hycheng}
H.-Y. Cheng, Phys. Lett. B {\bf 219}, 347 (1989).

\bibitem{cheng}
T. P. Cheng and L. F. Li, Phys. Rev. Lett. {\bf 62} (1989) 1441.

\bibitem{dienes}
K. R. Dienes, J. Kumar, B. Thomas, and D. Yaylali, Phys. Rev. D {\bf 90}, 015012 (2014).

\bibitem{qcdsfprotonspin}
G. S. Bali {\it at al.} (QCDSF Collaboration), Phys. Rev. Lett. {\bf 108}, 222001 (2012).

\bibitem{hycheng2}
H.-Y. Cheng and C.-W. Chiang, JHEP {\bf 1207}, 009 (2012).

\bibitem{scopel}
S. Scopel and H. Yu, arXiv:1701.02215 [hep-ph].

\bibitem{aokitensorcharge}
S. Aoki, M. Doui, T. Hatsuda and Y. Kuramashi, Phys. Rev. D {\bf 56}, 433 (1997).

\bibitem{pndmetensor1}
T. Bhattacharya, V. Cirigliano, R. Gupta, H.-W. Lin and B. Yoon, Phys. Rev. Lett. {\bf 115}, 212002 (2015).

\bibitem{pndmetensor2}
T. Bhattacharya, V. Cirigliano, S. D. Cohen, R. Gupta, A. Joseph, H.-W. Lin and B. Yoon, Phys. Rev. D {\bf 92}, 094511 (2015).

\bibitem{jlqcd4}
N. Yamanaka, H. Ohki, S. Hashimoto and T. Kaneko (JLQCD Collaboration),  PoS LATTICE2015, 121 (2016). 

\bibitem{yamanakasde1}
N. Yamanaka, T. M. Doi, S. Imai and H. Suganuma, Phys. Rev. D {\bf 88}, 074036 (2013).

\bibitem{pitschmann}
M. Pitschmann, C.-Y. Seng, C. D. Roberts and S. M. Schmidt, Phys. Rev. D {\bf 91}, 074004 (2015).

\bibitem{bacchetta}
A. Bacchetta, A. Courtoy, and M. Radici, JHEP {\bf 1303}, 119 (2013). 

\bibitem{anselmino}
M. Anselmino, M. Boglione, U. D'Alesio, S. Melis, F. Murgia, and A. Prokudin, Phys. Rev. D {\bf 87}, 094019 (2013). 

\bibitem{courtoy}
A. Courtoy, S. Bae{\ss}ler, M. Gonz\'{a}lez-Alonso, and S. Liuti,  Phys. Rev. Lett. {\bf 115}, 162001 (2015). 

\bibitem{radici}
M. Radici, A. Courtoy, A. Bacchetta and M. Guagnelli, JHEP {\bf 1505}, 123 (2015). 

\bibitem{kang}
Z.-B. Kang, A. Prokudin, P. Sun and F. Yuan, Phys. Rev. D {\bf 93}, 014009 (2016). 

\bibitem{yez}
Z. Ye {\it et al.}, Phys. Lett. B {\bf 767}, 91 (2017). 

\bibitem{choithetaedm}
K. Choi and J. Hong, Phys. Lett. B {\bf 259}, 340 (1991).

\bibitem{Chubukov}
D. V. Chubukov and L. N. Labzowsky, Phys. Rev. A {\bf 93}, 062503 (2016).

\bibitem{pospelov1}
M. Pospelov, Phys. Lett. B {\bf 530}, 123 (2002).

\bibitem{bsaisou1}
J. Bsaisou, C. Hanhart, S. Liebig, U.-G. Mei{\ss}ner, A. Nogga and A. Wirzba, Eur. Phys. J. A {\bf 49}, 31 (2013).

\bibitem{fuyuto}
K. Fuyuto, J. Hisano and N. Nagata, Phys. Rev. D {\bf 87}, 054018 (2013). 

\bibitem{dashencp}
R. Dashen, Phys. Rev. D {\bf 3}, 1879 (1971).

\bibitem{senghigherorder}
C.-Y. Seng and M. Ramsey-Musolf, arXiv:1611.08063 [hep-ph].

\bibitem{cirigliano}
V. Cirigliano, W. Dekens, J. de Vries and E. Mereghetti, Phys. Lett. B {\bf 767}, 1 (2017). 

\bibitem{4-quark1}
V. M. Khatsimovsky, I. B. Khriplovich and A. S. Yelkhovsky, Ann. Phys. (N.Y.) {\bf 186}, 1 (1988).

\bibitem{4-quark2}
X.-G. He and B. McKellar, Phys. Rev. D {\bf 47}, 4055 (1993).

\bibitem{4-quark3}
X.-G. He and B. McKellar, Phys. Lett. B {\bf 390}, 318 (1997).

\bibitem{4-quark4}
C. Hamzaoui and M. Pospelov, Phys. Rev. D {\bf 60}, 036003 (1999).

\bibitem{sushkov}
O. P. Sushkov, V. V. Flambaum and I. B. Khriplovich, Zh. Eksp. Teor. Fiz. {\bf 87}, 1521 (1984) [Sov. Phys. JETP {\bf 60}, 873 (1984)].

\bibitem{Flambaum86}
V. V. Flambaum, I. B. Khriplovich and O. P. Sushkov, Nucl. Phys. A {\bf 449}, 750 (1986).

\bibitem{hesmedm}
X.-G. He and B. McKellar, Phys. Rev. D {\bf 46}, 2131 (1992).

\bibitem{smdeuteronedm}
N. Yamanaka, arXiv:1605.00161 [nucl-th].

\bibitem{Khatsymovsky1}
V. M. Khatsymovsky and I. B. Khriplovich, Phys. Lett. B {\bf 296}, 219 (1992).

\bibitem{devries1}
J. de Vries, R. G. E. Timmermans, E. Mereghetti and U. van Kolck, Phys. Lett. B {\bf 695}, 268 (2011).

\bibitem{ucna}
B. Plaster {\it et al}. (UCNA Collaboration), Phys. Rev. C {\bf 86}, 055501 (2012).

\bibitem{pospelovchromoedm}
M. Pospelov and A. Ritz, Phys. Rev. D {\bf 63}, 073015 (2001). 

\bibitem{hisanochromoedm}
J. Hisano, J.-Y. Lee, N. Nagata and Y. Shimizu, Phys. Rev. D {\bf 85}, 114044 (2012). 

\bibitem{bhattacharya4}
T. Bhattacharya, V. Cirigliano, R. Gupta, E. Mereghetti and B. Yoon, Phys. Rev. D {\bf 92}, 114026 (2015). 

\bibitem{bhattacharya5}
T. Bhattacharya, V. Cirigliano, R. Gupta, E. Mereghetti and B. Yoon, PoS LATTICE2015, 238 (2015). 

\bibitem{bhattacharya6}
R. Gupta, T. Bhattacharya, V. Cirigliano and B. Yoon, arXiv:1701.04132 [hep-lat].

\bibitem{Abramczyk}
M. Abramczyk, S. Aoki, T. Blum, T. Izubuchi and H. Ohki and S. Syritsyn, arXiv:1701.07792 [hep-lat].

\bibitem{pospelovweinbergop}
D. Demir, M. Pospelov and A. Ritz, Phys. Rev. D {\bf 67}, 015007 (2003). 

\bibitem{pvcpvhamiltonian2}
V. P. Gudkov, X.-G. He and B. H. J. McKellar, Phys. Rev. C {\bf 47}, 2365 (1993). 

\bibitem{pvcpvhamiltonian3}
I. S. Towner and A. C. Hayes, Phys. Rev. C {\bf 49}, 2391 (1994). 

\bibitem{korkin}
I. B. Khriplovich and R. A. Korkin, Nucl. Phys. A {\bf 665} (2000) 365.

\bibitem{yamanakareview}
N. Yamanaka, arXiv:1609.04759 [nucl-th].

\bibitem{tiator}
L. Tiator, C. Bennhold and S. S. Kamalov, Nucl. Phys. A {\bf 580}, 455 (1994). 

\bibitem{Senkov2008}
R. A. Senkov, N. Auerbach, V. V. Flambaum and V. G. Zelevinsky, Phys. Rev. A {\bf 77}, 014101 (2008).

\bibitem{Flambaum12}
V. V. Flambaum and A. Kozlov, 
Phys. Rev. C \textbf{85}, 068502 (2012).

\bibitem{Baldo99}
For a review, see M. Baldo "Nuclear Methods and the Nuclear Equation of State, International Review of Nuclear"
Physics Vol. 8, pp 1, World Scientific 1999, M. Baldo ed.

\bibitem{Dmitriev03}
V. F. Dmitriev and R. A. Senkov 
Phys. Rev. Lett. \textbf{91}, 212303(2003).

\bibitem{Dmitriev05}
V.~F.~Dmitriev, R.~A.~Sen'kov, and N.~Auerbach,
Phys. Rev. C \textbf{71}, 035501 (2005).

\bibitem{Engel03}
J. Engel, M. Bender, J. Dobaczewski, J. H. de Jesus, and P. Olbratowski, Phys. Rev. C \textbf{68}, 025501 (2003).

\bibitem{sandars1}
 P. G. H. Sandars, Phys. Lett. {\bf 14}, 194 (1965).

\bibitem{sandars2}
 P. G. H. Sandars, Phys. Lett. {\bf 22}, 290 (1966).

\bibitem{flambaum2}
V. V. Flambaum, Yad. Fiz. {\bf 24}, 383 (1976) [Sov. J. Nucl. Phys. {\bf 24}, 199 (1976)].

\bibitem{Jesus05}
J.~H.~de~Jesus and J.~Engel,
Phys. Rev. C \textbf{72}, 045503 (2005).

\bibitem{Dobaczewski05}
J.~Dobaczewski and J.~Engel,
Phys. Rev. Lett. \textbf{94}, 232502 (2005).

\bibitem{Ban10}
S.~Ban, J~.Dobaczewski, J.~Engel, and A.~Shukla,
Phys. Rev. C \textbf{82}, 015501 (2010).

\bibitem{Auerbach06}
N.~Auerbach, V.~F.~Dmitriev, V.~V.~Flambaum, A.~Lisetskiy, R.~A.~Sen'kov, and V.~G.~Zelevinsky
Phys. Rev. C \textbf{74}, 025502 (2006).

\bibitem{Yoshinaga13}
N. Yoshinaga,  K. Higashiyama, R. Arai and E. Teruya, 
Phys. Rev. C \textbf{87},  044332 (2013); \textbf{89}, 069902 (2014). 

\bibitem{Higashi03} 
K. Higashiyama, N. Yoshinaga, and K. Tanabe, 
Phys. Rev. C \textbf{67}, 044305 (2003).

\bibitem{Yoshi04} 
N.~Yoshinaga and K.~Higashiyama, 
Phys. Rev. C \textbf{69}, 054309 (2004).

\bibitem{Higashi11} 
K. Higashiyama and N. Yoshinaga, 
Phys. Rev. C \textbf{83}, 034321 (2011); \textbf{89}, 049903 (2014).

\bibitem{Teruya16}
E. Teruya, N. Yoshinaga,  K. Higashiyama and K. Asahi, Submitted. 

\bibitem{Yoshinaga10}
N.~Yoshinaga, K.~Higashiyama, and R.~Arai, Prog. Theor. Phys. \textbf{124}, 1115 (2010).

\bibitem{Fujita12}
T. Fujita and S. Oshima, J. Phys. G \textbf{39}, 095106 (2012).

\bibitem{Yoshinaga14}
N. Yoshinaga,  K. Higashiyama, R. Arai and E. Teruya, 
Phys. Rev. C \textbf{89}, 045501 (2014).

\bibitem{auerbach}
N. Auerbach, V. V. Flambaum, and V. Spevak, Phys. Rev. Lett. {\bf 76}, 4316 (1996). https://doi.org/10.1103/PhysRevLett.76.4316

\bibitem{Spevak}
V. Spevak, N. Auerbach and V. V. Flambaum, Phys. Rev. C {\bf 56}, 1357 (1997). https://doi.org/10.1103/PhysRevC.56.1357

\bibitem{dzuba1}
V. A. Dzuba, V. V. Flambaum and S. G. Porsev, Phys. Rev. A {\bf 80}, 032120 (2009). https://doi.org/10.1103/PhysRevA.80.032120

\bibitem{flambaum1985}
V. V. Flambaum and I. B. Khriplovich, Zh. Eksp. Teor. Fiz. {\bf 89}, 1505 (1985) [Sov. Phys. JETP {\bf 62}, 872 (1985)].

\bibitem{flambaum}
V. V. Flambaum and J. S. M. Ginges, Phys. Rev. A {\bf 72}, 052115 (2005). https://doi.org/10.1103/PhysRevA.72.052115

\bibitem{bks-qed}
B. K. Sahoo, Phys. Rev. A {\bf 93}, 022503 (2016). https://doi.org/10.1103/PhysRevA.93.022503

\bibitem{martensson}
A. M. Martensson-Pendrill, Phys. Rev. Lett. {\bf 54}, 1153 (1985). https://doi.org/10.1103/PhysRevLett.54.1153

\bibitem{dzuba_02}
V. A. Dzuba, V. V. Flambaum, J. S. M. Ginges and M. G. Kozlov, Phys. Rev. A {\bf 66}, 012111 (2002). https://doi.org/10.1103/PhysRevA.66.012111

\bibitem{jaceak}
L. Radziute, G. Gaigalas, P. J\"onsson, and Jacek Bieron, Phys. Rev. A {\bf 93}, 062508 (2016). https://doi.org/10.1103/PhysRevA.93.062508

\bibitem{latha}
K. V. P. Latha and P. R. Amjith, Phys. Rev. A {\bf 87}, 022509 (2013). https://doi.org/10.1103/PhysRevA.87.022509

\bibitem{lathalett}
K. V. P. Latha, D. Angom, B. P. Das and D. Mukherjee, Phys. Rev. Lett. {\bf 103}, 083001 (2009); Erratum: Phys. Rev. Lett. {\bf 115}, 059902 (2015). https://doi.org/10.1103/PhysRevLett.103.083001

\bibitem{yashpal5}
Y. Singh and B. K. Sahoo, Phys. Rev. A {\bf 91}, 030501(R) (2015). https://doi.org/10.1103/PhysRevA.91.030501

\bibitem{lindgren}
I. Lindgren and J. Morrison, {\it Atomic Many-Body Theory}, Second Edition, Springer-Verlag, Berlin, Germany (1986).

\bibitem{yashpal1}
Y. Singh, B. K. Sahoo, and B. P. Das, Phys. Rev. A {\bf 88}, 062504 (2013). https://doi.org/10.1103/PhysRevA.88.062504

\bibitem{yashpal2}
Y. Singh and B. K. Sahoo, Phys. Rev. A {\bf 90}, 022511 (2014). https://doi.org/10.1103/PhysRevA.90.022511

\bibitem{pal}
S. Pal, M. D. Prasad and D. Mukherjee, Pramana {\bf 18}, 261 (1982).

\bibitem{bartlett}
I. Shavitt and R. J. Bartlett, {\it Many-Body Methods in Chemistry and Physics: MBPT and Coupled-Cluster Theory},

\bibitem{yashpal3}
Y. Singh, B. K. Sahoo and B. P. Das, Phys. Rev. A {\bf 89}, 030502(R) (2014). https://doi.org/10.1103/PhysRevA.89.030502

\bibitem{yashpal4}
B. K. Sahoo, Y. Singh and B. P. Das, Phys. Rev. A {\bf 90}, 050501(R) (2014). https://doi.org/10.1103/PhysRevA.90.050501

\bibitem{yashpal6}
Y. Singh and B. K. Sahoo, Phys. Rev. A {\bf 92}, 022502 (2015). https://doi.org/10.1103/PhysRevA.92.022502

\bibitem{parker}
R. H. Parker, et al.,  Phys. Rev. Lett. {\bf 114}, 233002 (2015). https://doi.org/10.1103/PhysRevLett.114.233002

\bibitem{cho1_89}
D. Cho, K. Sangster, E.A. Hinds, Phys. Rev. Lett. {\bf 63}, 2559 (1989). https://doi.org/10.1103/PhysRevLett.63.2559

\bibitem{cho2_89}
D. Cho, K. Sangster, E.A. Hinds, Phys. Rev. A {\bf 44}, 2783 (1991). https://doi.org/10.1103/PhysRevA.44.2783

\bibitem{yoshimi_12}
A. Yoshimi, T. Inoue, T. Furukawa, T. Nanao, K. Suzuki, M. Chikamori, M. Tsuchiya, H. Hayashi, M. Uchida, N. Hatakeyama, S. Kagami, Y. Ichikawa, H. Miyatake and K. Asahi, Phys. Lett. A {\bf 376}, 1924 (2012).  http://dx.doi.org/10.1016/j.physleta.2012.04.043

\bibitem{fabbrichesi}
E. Christova and M. Fabbrichesi, Phys. Lett. B {\bf 315}, 113 (1993). http://dx.doi.org/10.1016/0370-2693(93)90166-F

\bibitem{drees}
M. Drees and M. Rauch, Eur. Phys. J. C {\bf 29}, 573 (2003). DOI: 10.1140/epjc/s2003-01247-8

\bibitem{jng}
J. Ng and S. Tulin, Phys. Rev. D {\bf 85}, 033001 (2012). https://doi.org/10.1103/PhysRevD.85.033001

\bibitem{bishop}
R. F. Bishop, {\it Lecture Notes in Physics}, Microscopic Quantum Many-Body Theories and their Applications (Springer Berlin Heidelberg) 
{\bf 510}, 1 (1998). 

\bibitem{devries3}
J. de Vries, E. Mereghetti, C.-Y. Seng and A. Walker-Loud, Phys. Lett. B {\bf 766}, 254 (2017). http://dx.doi.org/10.1016/j.physletb.2017.01.017


\end{thebibliography}
\end{document}